\documentclass{article}

\usepackage{PRIMEarxiv}

\usepackage{graphicx}          

\usepackage{lipsum}
\usepackage{bm,bbm,amsfonts,amsmath}%
\usepackage{graphicx}
\usepackage{epstopdf}
\usepackage{algorithmic}
\usepackage{amssymb}  
\usepackage{booktabs} 
\usepackage[title]{appendix}
\usepackage{makecell}
\usepackage{tabularx}

\usepackage{ntheorem}
\newtheorem{theorem}{Theorem}

\newtheorem{corollary}{Corollary}[theorem]
\newtheorem{lemma}[theorem]{Lemma}
\newtheorem{definition}{Definition}[theorem]
\newtheorem{proposition}{Proposition}[theorem]

\newtheorem{assumption}{Assumption}

\newenvironment{proof}{\paragraph{Proof:}}{\hfill$\square$}

\usepackage{xcolor}
\definecolor{ao(english)}{rgb}{0.0, 0.5, 0.0}
\definecolor{americanrose}{rgb}{1.0, 0.01, 0.24}
\definecolor{cerisepink}{rgb}{0.93, 0.23, 0.51}
\definecolor{darkorchid}{rgb}{0.6, 0.2, 0.8}
\definecolor{applegreen}{rgb}{0.55, 0.71, 0.0}
\definecolor{brightpink}{rgb}{1.0, 0.0, 0.5}
\definecolor{azure(colorwheel)}{rgb}{0.0, 0.5, 1.0}
\definecolor{black-violet}{rgb}{0.54, 0.17, 0.89}
\definecolor{deepmagenta}{rgb}{0.8, 0.0, 0.8}
\definecolor{fashionfuchsia}{rgb}{0.96, 0.0, 0.63}
\definecolor{armygreen}{rgb}{0.29, 0.33, 0.13}
\definecolor{dogwoodrose}{rgb}{0.84, 0.09, 0.41}

\newtheorem{remark}{Remark}
\usepackage{xcolor}

\title{On Pricing Data Privacy: Endogenous Evolution, Optimal Stopping, and Incentive Compatibility
\thanks{\textit{\underline{Citation}}: 
\textbf{Authors. Title. Pages.... DOI:000000/11111.}} 
}

\author{
  Tao Zhang, Quanyan Zhu \\
  Electrical and Computer Engineering \\
  New York University \\
  \texttt{\{tz636, qz494\}@nyu.edu} \\
}

\begin{document}
\maketitle

\begin{abstract}

Privacy is an essential issue in data trading markets. This work uses a mechanism design approach to study the optimal market model to economize the value of privacy of personal data, using differential privacy. The buyer uses a finite number of randomized algorithms to get access to the owners' data in a sequential-composition manner, in which each randomized algorithm is differentially private. 
Each usage of a randomized algorithm is referred to as a period.
Motivated by the discovery of an individual's dual motives for privacy protection, we partition each data owner's preference over privacy protection into the intrinsic and the instrumental components, in which the instrumental preference arises endogenously from the data buyer's sequential usages of multiple private algorithms.
Due to the composability of differential privacy, there are inevitable privacy losses accumulated over periods.
Hence, we allow the owners to leave the market at the end of any period by making stopping decisions.
We define an instrumental kernel function to capture the instrumentalness of owners' preferences and model the formation of each owner's (both intrinsic and instrumental) preference over periods by taking into consideration of the composability of differential privacy and time-varying nature of privacy concerns.
Our desideratum is to study the buyer's design regime of optimal market models in dynamic environment when each owner makes coupled decisions of stopping and reporting of their preferences.
The buyer seeks to design a privacy allocation rule that dynamically specifies the degree of privacy protections and a payment rule to compensate the privacy losses of the owners.
The buyer additionally chooses a payment rule which is independent of owners' report of their preferences to influence the owners' stopping decisions.  We characterize the dynamic incentive compatibility and provide a design principle to construct the payment rules in terms of the privacy allocation rule.
Further, we relax the buyer's market design problem and provide a sufficient condition for an approximated dynamic incentive compatible market model.

\end{abstract}


\section{Introduction}

Big data is proving itself as the biggest promising opportunity for businesses, research communities, and governments since the Internet went mainstream about two decades ago.
Gigabytes, terabytes, and petabytes of industrial, commercial, and personal data rush into a great wave of opportunities.
Business leaders are seeking actionable methods to exploit the enormous value of data to promote financial gains by improving customer management, enhancing risk analysis, placing accurate marketing strategies, and so on.
Meanwhile, data marketization is attracting increasing attention in response to the valuable benefits of and the keen demand for data.
Designing effective data market models is critical to efficiently utilize the data by enabling data trading between data owners and data buyers. 
Commoditization of data in a digital market can incentivize data owners' participation through monetary benefits and thus enables the data buyers to access data of higher quality and larger quantity.
Furthermore, data marketization also provides opportunities to adapt governing regime and market standardization into the digital domain.
Efforts in researches of data market modelings including analysis and pricing data trading have been invested for, for example, financial data (e.g., \cite{admati1988selling,allen1990market,10.1093/rfs/15.4.987}), IoT data (e.g., \cite{niyato2016market,niyato2016economics,zhang2019optimal}), and medical data (\cite{mankopf2008system,tanner2017our}).

However, privacy issues naturally follow. 
It is critical to provide privacy protection for any form of data releasing in the data market.
Hence, privacy-preserving schemes should be an indispensable component of the data market model.
Due to the natural tradeoff between the privacy and the utility (accuracy) of data usages, however, private data releasing without any privacy loss in general unavoidably eliminates the useful value of data.
As a result, the privacy-preserving data market model has to take into account the privacy-utility tradeoff and provide incentives for both data owners and data buyers to participate.
Yet, the tradeoff is in general difficult to model explicitly and uncertain to both data owners and buyers.
Owners of the database with sensitive information often inevitably release more information than intended even under carefully crafted privacy protection \cite{hsu2014differential}.
For example, the ineffectiveness of anonymization has been shown in the literature that a small amount of auxiliary information is sufficient for an adversary to de-anonymize an individual in a database consisting of anonymized data about that individual's personal information (see, for example, \cite{narayanan2008robust,backstrom2007wherefore,jones2007know}). 
Another main challenge for privacy-preserving data releasing is due to the limited information about the adversary's knowledge and ability.  
Hence, rigorous quantification of privacy loss and its influence on data utility and the robustness to adversarial privacy pry with heterogeneous knowledge and prior information becomes important in the accurate design of monetization and incentivization.

To this end, differential privacy (\cite{dwork2011differential}) (also refers to $\epsilon$-differential privacy, with $\epsilon\in\mathbb{R}_{+}$) is widely used as the privacy notion for privacy-preserving data processing.
Differential privacy provides strong privacy guarantees such that whether an individual data point is in the database or not is near-indistinguishable based on the output information released by randomized processing of the database regardless of what auxiliary knowledge or side information is available to the adversary.
Differential privacy has been studied in a significant amount of work in noise-perturbed data-releasing mechanisms and data-learning algorithms such as empirical risk minimization (e.g., \cite{chaudhuri2011differentially}), statistical learning (e.g., \cite{dziugaite2018data}), and deep learning (e.g., \cite{abadi2016deep}).
The rigorous mathematical formulation of differential privacy provides an elegant framework to quantify individual privacy loss.
In particular, \textcolor{black}{the parameter $\epsilon$ can be used to quantify the degree of privacy by characterizing the upper bound of privacy loss that any individual data point can suffer.}
In addition, the parameter $\epsilon$ gracefully parameterizes the tradeoff between privacy and accuracy.
Basically, the privacy of data usage increases when $\epsilon$ decreases (as the upper bound of privacy loss decreases) at the expense of decreasing accuracy, or the accuracy of data usages increases when $\epsilon$ increases at the expense of decreasing privacy (as the upper bound of privacy loss increases).
\textcolor{black}{As a result, the change of $\epsilon$ can also be used to parameterize the changes in the accuracy of data usage when the degree of privacy protection changes.}
Hence, the design of privacy-preserving schemes in the data market can be characterized by the craft of $\epsilon$.

\textcolor{black}{
The owners' motives to protect the privacy of their data could be \textit{instrumental} \cite{lin2020valuing} which rises endogenously from the buyer's usage of their data.
For example, data owners can benefit from protecting the privacy of their personal data (e.g., by decreasing $\epsilon$) to prevent costs from data misuse, such as identity theft, spam, or adverse price discrimination \cite{odlyzko2004privacy}.
Their motives to protect the data privacy could also be \textit{intrinsic} \cite{gradwohl2018voting,lin2020valuing} which treats privacy protection as a human right and part of the code of social conduct, or owners' psychological desires regardless of the economic cost or benefit from the buyer's usage of their data.
For example, the data misuse may cause less tangible costs such as psychological discomfort or stigma \cite{stone1990privacy}.
The owners' motives to participate in allowing the buyer to get access to their data (i.e., motives to partially reveal private information) could also be instrumental and intrinsic.
As well-recognized in the economic theory of privacy \cite{acquisti2016economics}, individuals may benefit from sharing their personal information with other parties.
For example, individuals could receive more personalized offers that actually interest them by sharing personal information with service providers, get promotions and discount by joining merchant’s loyalty program, or get improved search results by sharing browsing activities with search engines.
Without revealing personal information (or with more privacy protection), these benefits could turn into opportunity cost for the individuals.
This captures the instrumental components of their motives to participate, while the intrinsic motive of participation describes that sharing private personal information could also be intrinsically rewarding \cite{tamir2012disclosing}.
}

\textcolor{black}{
Hence, the buyer's choice of $\epsilon$ is essentially an economic and social question (see, also, \cite{dwork2008differential}).
Choosing an acceptable $\epsilon$ may depend on the risk of privacy leakage that causes the economic and the psychological costs.
Suppose that an $\epsilon'$ has been chosen for a process of data usage.
In situations with very low risks, it may be tolerable to a higher value, $\epsilon'' = k\epsilon'$, for some $k=2$ or $3$, while in cases when the risk is very high, increasing $\epsilon'$ by a very small factor, e.g., $1.01$ may even be intolerable as the cost from privacy breach outweighs the gain from sharing data.
As a result, the proper choice of $\epsilon$ should take into consideration the data owners' motives of the privacy protection as well as their willingness of participation.
}

In this paper, we study the economy of data privacy and consider a privacy trading market.
The market consists of a finite group of data owners and one data buyer.
We consider that each data owner is a \textit{rational economic agent} who is aware of and internalize the economic value (positive and negative) of data privacy and the risk of privacy breaches.
Inspired by the discovery of individuals' dual motives for privacy protection, we use the notion of \textit{privacy preference} (preference) to summarize each owner's intrinsic and instrumental valuations and motivations--\textit{dual preference}--for privacy protection.
In accordance with Laudon \cite{laudon1997extensions}, we consider that the data owners hold an economic claim over their data (no matter whether the data is generated in some platforms owned by the buyer; e.g., searching history by using a search engine).
The data owners have the right to ask the buyer to protect their data by using differential privacy and require compensation for the inevitable privacy loss.
The buyer uses a finite number of randomized algorithms to get access to the owners' data in a sequential-composition manner, in which each randomized algorithm is differentially private. 
We refer to each usage of a randomized algorithm is as one time period.

In reality, data buyers necessarily request multiple periods of usages of the same data for different purposes.
This induces other challenges of privacy preservation design in the data market.
First, an owner's instrumental component of the preference can endogenously change over time due to, for example, leaning-by-doing, context-dependence of privacy, or influence from external factors; his intrinsic component of the preference can also be time-varying due to some exogenous impacts.
For instance, the realized privacy protection (i.e., the realization of $\epsilon$) in one data usage may influence their privacy preference at the next usage of data.
The same data owner may in some cases be severely concerned about, but under some other circumstances be indifferent to, privacy leakage \cite{acquisti2015privacy}.
Also, an owner's privacy preference tends to be influenced by external aspects that aim to activate or suppress privacy concerns. For example, fake news is spread to create illusions of a safe (resp. risky) cyber environment to encourage (resp. discourage) data sharing.
As a result, the endogenous instrumentalness, the change of context, or time-evolution of external influences may lead to dynamics of an owner's privacy preference. 
Second, the total privacy loss might be amplified and accumulative when the number of data usages increases due to the inevitable privacy loss in each individual usage.
For example, the composition of $k$ randomized algorithms, each of which is $\epsilon$-differentially private, is at least $k\epsilon$-differentially private \cite{dwork2006calibrating,kairouz2017composition}.
To mitigate the risk of losses due to privacy leakage, our model allows each data owner to terminate his participation at the end of each data usage if he cannot tolerate the expected loss by continuing to participate.

By conceptualizing the relationship between the owners and the buyer by a principal-multiagent model in a finite horizon, this paper proposes a theoretical framework for pricing differential privacy of data in a dynamic environment, where each owner privately possesses his privacy preference which is time-evolving due to the endogenous instrumentalness as well as exogenous evolution of the intrinsic privacy preference.
The buyer is the mechanism designer whose goal is to minimize the expected cost by choosing a privacy allocation rule that specifies the value of the privacy parameter $\epsilon_{t}$ in each period $t$ and a payment rule profile that determines payment to each owner in each period to compensate the privacy loss.
Both the specifications of the privacy parameter and payment require the owners to report their privacy preferences.

The proposed market model also highlights the owners' willingness and autonomy in trading their privacy by making a \textit{take-it-or-leave-it} offer the owners and entitling each owner to use a stopping rule to terminate his participation at the end of each data usage.
Due to the Revelation Principle, we restrict attention to direct mechanisms in which each owner truthfully reveals his privacy preference in each period by imposing \textit{dynamic incentive compatibility} constraints to the mechanism.
The autonomy raised by allowing stopping rules fundamentally complicates the characterizations of the dynamic incentive compatibility.

This work studies the design of a dynamic market for trading data privacy and focuses on the theoretical analysis of how to optimally design the mechanism rules and how the mechanism influences the owners' coupled decision makings of reporting and stopping.
%
%
%
The contributions of this paper are summarized as follows. 
\begin{itemize}
    \item[1.] We propose a dynamic market model for trading the privacy of data using differential privacy, based on the fundamental tradeoff of privacy and utility of data in differential privacy.
    We consider that each owner has a dual privacy preference and model the generation of each owner's privacy preference by introducing an instrumental kernel function which captures the endogenous instrumental component of the preference and treating the intrinsic component as an independent exogenous shock.
    The market model consists of a privacy allocation rule that specifies $\epsilon_{t}$ in each period $t$ and two payment rules to compensate the owners' privacy loss.
    Our model allows each owner to leave the market at the end of any period by using a stopping rule.
    The buyer also design a posted-price rule that is independent of owners' privacy preferences to influence the owners' stopping decisions.

    \item[2.] The owners' strategic interactions are modeled as a dynamic Bayesian game when each owner makes coupled decisions of reporting and stopping.
    We define a stopping problem for each owner when he dynamically chooses how to report his privacy preference  to the buyer. 
    A new notion of dynamic incentive compatibility (DIC) is defined based on the Bellman equation, which captures robustness of the model to the coupled deviations from truthful reporting decisions and optimal stopping behaviors.

    \item[3.] We characterize the DIC and transform the owners' stopping decision into a threshold-based rule under a monotonicity assumption about owners' instrumentalness.
    A theoretical design regime is established by formulating the preference-dependent payment rules in terms of the privacy allocation rule and the preference-independent posted-privacy rule in terms of the privacy allocation rule and the threshold function.
    \item[4.] Based on the design regime, we relax the buyer's optimal market design problem from a four decision rule profiles and two constraint sets to a problem of determining the privacy allocation rule profile and the threshold function profile with a single constraint set.
    A notion of approximated DIC is defined to address the inevitable violations of DIC when the mechanism design problem is solved approximately.
    %
    %
\end{itemize}

\subsection*{Organization}

The rest of this paper is organized as follows.
In Section \ref{sec:related_work}, we provide related works.
Section \ref{sec:background_differential_privacy} provides background of differential privacy that is necessary for the formulations of our dynamic market model.
In Section \ref{sec:brief_static_privacy_trading}, we describe a one-stage static market model in which the buyer only uses one differentially-private algorithm to access to the owners' data.
Section \ref{sec:model} formally describes the model of the dynamic market of data privacy with differential privacy.
We model the decision makings of the owners by a dynamic Bayesian game and formulate the buyer's mechanism design problem.
Also, we construct a stopping time rule for each owner to make stopping decisions.
A new nontion of dynamic incentive compatibility is then defined.
In Section \ref{sec:characterization}, we characterize the dynamic incentive compatibility by obtaining theoretical design regimes.
In Section \ref{sec:relaxed_market_design}, we relax the data buyer's optimal mechanism design problem based on the theoretical results obtained in Section \ref{sec:characterization}.
Section \ref{sec:conclusion} concludes the paper.
A summary of main notations is given in Table \ref{table:notations}.

\section{Related Work}\label{sec:related_work}

\textcolor{black}{
Our market model considers the fundamental tradeoff of privacy and utility of differential privacy that is characterized by the privacy parameter $\epsilon$. 
This tradeoff coincides with the tradeoff of privacy and discovered in economics theoretically and empirically.
Dating back to 1970s and 1980s, the pioneering works produced by Chicago School scholars (e.g., Posner \cite{posner1981economics,posner1978right} and Stigler \cite{stigler1980introduction}) have studied the the economic tradeoff of privacy protection and the utility (value or damage) that individuals and society may incur.
These works highlight the economic value of individuals' privacy in terms of the cost that protecting such privacy may induce to other market participants \cite{posner1993blackmail} or the damage that revealing such privacy may create to the individuals themselves \cite{stigler1980introduction}.
In particular, Posner \cite{posner1981economics,posner1978right} has argued that protection of privacy raises inefficiency (in terms of, e.g., increased cost or reduced welfare) in the marketplace due to the concealing of potential payoff-relevant information.
Stigler \cite{stigler1980introduction} has made a similar argument that regulatory interventions of privacy protections would ultimately lead to inefficient use of economic resources and productive factors, or unfair reward allocations due to the removal of relevant personal information.
%
%
These economic tradeoffs of privacy and utility come from the parties' conflicts of interests in the cost and the benefit, respectively, from protecting and using the private information that contains individuals' negative traits.
}

\textcolor{black}{
These economic tradeoffs of privacy and utility come from the parties' conflicts of interests in the cost and the benefit, respectively, from protecting and using the private information that contains individuals' negative traits.
The privacy-utility tradeoff could also from the conflicts between the information holder's own utilities that can be generated from revealing private personal information and the cost from the (usually adverse) usage of such information by other parties.
Varian \cite{varian2002economic} has observed that there could be individual cost from privacy protection and customers may rationally want to share their personal information with other parties to receive, for example, personalized services and offers that actually benefits the customers. 
However, at the same time, the customers may want to limit the amount of personal information to be known by others. 
This is because personal information may be taken advantage of by others, which may lead to spam and adverse price discrimination.
For example, Odlyzko \cite{odlyzko2004privacy} has studied privacy and price discrimination and found that knowing more personal information about buyers' willingness to pay promote the sellers' ability to price discriminate (see, also, \cite{chen1997paying,jeong2009commitment}).
}

\textcolor{black}{Whereas the aforementioned works focus on articulating economic arguments about economics of privacy, there is literature on studying economic value of privacy in formal economic models.
Fudenberg and Tirole \cite{fudenberg2000customer} has studied a duopoly model in which customers of one firm decides to remain loyal to the firm or to defect to a rival. They have shown that it is better off for a firm to offer discounts to the competitor's customers because these customers' purchase history and their preference for the competitor's product (i.e., the private information of the customers) can provide economic value would outweigh the cost of offering discounts.
In a similar vein, Chen and Zhang \cite{chen2009dynamic} has studied a dynamic model of target pricing. They have demonstrated a strategy for the firms referred to as \textit{price for information}, with which the firms price their product less aggressively to attract customers, such that the firms can learn more about their customers (by tracking their private information).
In our model, the buyer's willingness to compensate the owners for their privacy loss can be interpreted in the manner where the owners' (privacy of) data contributes economic value for the buyer.}
%
%
%
%
%

\textcolor{black}{
Our work is also related to the literature on market models which studies how the economic value of participants' private information can influence the decision makings.
%
%
Bergemann and Bonatti \cite{bergemann2015selling} have studied a model of advertising platform in which advertisers costly acquire user-pertinent information from a data provider in order to know customers' type (by forming posterior beliefs), then purchase advertising space.
Hagiu and Jullien \cite{hagiu2011intermediaries} have studied how intermediaries of two-sided markets can use information of customers' characteristics to affect matching between customers and firms. 
In \cite{board2018competitive}, Board and Lu have considered a market consisting of buyers and studied how market outcomes changes as the amount of consumer information possessed by the buyers varies.
There are also works studying how limiting the accessible of information (i.e., increasing privacy protection) can influences the market outcomes of intermediary gatekeepers \cite{baye2001information,wathieu2002privacy,conitzer2012hide}.
For example, Conitzer et al. \cite{conitzer2012hide} have demonstrated that allowing users to freely anonymize (i.e., protect privacy) can be profit-maximizing for both the gatekeeper and firms.
}

\textcolor{black}{
As remarked in \cite{acquisti2016economics}, privacy issues exist in widely diverse contexts.
In this work, we restrict attention to the privacy issues of algorithmic data usages. 
By recognizing the value of data privacy in terms of the utility that can be extracted from the data, we treat the privacy of the data as a good and aim to design a market model for selling the privacy, based on the fundamental tradeoff of privacy-utility of differential privacy characterized by the privacy parameter $\epsilon$.
Our contributions lie in the engineering part--\textit{the design regime}--of the market design.
Our model can contribute as an additional component to a variety of economic models when the process of extracting knowledge from private data is algorithmic.
}

There is literature on the interactions of differential privacy and mechanism design.
\cite{ghosh2015selling} have initiated the study of private data markets.
They have treated differential privacy of data as a commodity and applied traditional static mechanism design approaches to model one-query private data trading as a variant of a multi-unit procurement auction.
They have considered the cost of privacy loss as each owner's private information and studied truthful mechanism in which each owner is incentivized to truthfully release his private information.
Works following \cite{ghosh2015selling} include
\cite{fleischer2012approximately,dandekar2011privacy,ligett2012take,roth2012conducting,aperjis2012market}, which have studied how to determine $\epsilon$ through auctions.
Other literature of studying how rational agents evaluate differential privacy loss and choose $\epsilon$ includes, e.g., \cite{nissim2012privacy,hsu2014differential,chen2016truthful,xiao2013privacy}.
Authors of \cite{hsu2014differential} have proposed a framework to choose differential privacy parameters through a simple static economic model with complete information of data owners and buyers based on quantities that can be estimated in practice
%

There is also related work in dynamic settings.
Authors of \cite{li2014theory} have studied an orthogonal problem to \cite{ghosh2015selling}: owners' valuations are public knowledge and there are multiple queries of data usages. 
Besides the accuracy of query outputs, \cite{li2014theory} have also considered unbiasedness.
Their model allows the data buyers to get an arbitrary number of queries and provides arbitrage-free pricing scheme for the buyers that is balanced by taking into account the compensation for privacy loss and the profits from data usages.
Other line of work in dynamic setting concerns optimal pricing in a time-evolving environment.
There is literature considering posted price models that do not require truthful revealing of private information (e.g., \cite{singla2013truthful}) and models that require incentive compatibility (e.g., \cite{amin2013learning}).
Authors of \cite{xu2016dynamic} have proposed a dynamic privacy pricing framework in a market where a data buyer repeatedly buys data from a group of data owners, whose valuations of privacy are randomly drawn from an unknown distribution.
They have treated each candidate price as one arm and modeled a multi-armed bandit problem to dynamically adjust the prices to compensate the data owners.

In contrast, we consider a dynamic market framework, in which each owner can learn and update new his valuation of privacy (his private information).
We use mechanism design approaches to dynamically set the value of $\epsilon$ and the price of privacy as a compensation for each owner's privacy loss at each period through a dynamic optimization problem that minimizes the buyer's cost by taking into account the incentive compatibility, individual rationality, and the buyer's accuracy requirement.
Our model offers a flexible commitment and allows each owner to leave the market by adopting a stopping rule once his pre-determined privacy budget is exceeded.

There is a significant amount of work on dynamic mechanism design problems. 
The literature on dynamic mechanism designs can be divided into two classes. Those are (1) mechanisms with dynamic population and static private information and (2) mechanisms with dynamic private information and static population.
Authors of \cite{parkes2004mdp} have studied a sequential allocation problems when the participating population is dynamic. 
In particular, their model has considered the environment when each self-interested agent arrives and departs dynamically overtime.
The information possessed by each agent is static and includes the arrival and the departure time as well as her valuation about allocation outcomes.
Other works consider this class of dynamic settings include, e.g., \cite{pai2008optimal,gallien2006dynamic,gershkov2009dynamic,said2012auctions,pai2013optimal,board2016revenue}.
Orthogonal to the dynamic population mechanisms, there are other works considering mechanisms, in which the underlying model is dynamic due to the time-evolution of agents' private information.
There is a large number of works lying in this category that studies for example, the dynamic pivot mechanisms (e.g., \cite{bergemann2010dynamic,kakade2013optimal}), dynamic team mechanisms (e.g., \cite{bapna2005efficient,athey2013efficient,nazerzadeh2013dynamic}), and more generally (e.g., \cite{pavan2014dynamic,zhang2019incentive}).
\cite{athey2013efficient} have considered a dynamic team problem and proposed a balanced team mechanism to implement dynamic efficiency with a balanced budget.
Each agent observes private signals over time and decisions are made periodically.
Their mechanism provides each agent an incentive payment in each period, which equals to the expected present value of the other agents' payoffs induced by this agent's current period report, to establish an equilibrium in truthful strategies.

The theoretical framework of our mechanism model lies in the interaction of mechanism design with dynamic population and with time-evolving private information.
In particular, each data owner's private information (i.e., valuation of privacy) changes over time and the population is dynamic due to the stopping time rule adopted by each data owner.
Unlike the aforementioned works with dynamic population, we do not consider the arrival of new data owners and the departure time is determined by the stopping rule (depends on the owner's valuation and the privacy guarantees) and is not treated as private information.

\section{Preliminaries}

This section introduces some preliminaries. We summarize the concept of differential privacy in Section \ref{sec:background_differential_privacy} and describe the basic one-stage framework of our data privacy trading model in Section \ref{sec:brief_static_privacy_trading}.

\subsection{Differential Privacy}\label{sec:background_differential_privacy}

In this section, we review basic concepts in differential privacy to properly support the contributions of this paper.

Let $\mathcal{D}\equiv\{D_{1}, D_{2},\dots, D_{n}\}$, where each $D_{k}\in \mathbb{D}$ is a single data point, denote a database consisting of $n$ data points.
Let $\mathcal{A}: \mathbb{D}^{n} \mapsto \mathcal{S}$ denote a \textit{randomized} algorithm such that $\mathcal{A}(\mathcal{D})\in \mathcal{S}$ is the output of the algorithm with $\mathcal{D}$ as the input data.
The following definition defines indistinguishability of any algorithm.

\begin{definition}\label{def:indistinguishability}
Let $\mathcal{D}\in \mathbb{D}^{n}$ and $\mathcal{D}'\in \mathbb{D}^{n}$ be any two databases. We say the randomized algorithm $\mathcal{A}$ is \textit{$\epsilon$-indistinguishable} (or indistinguishable) for these two databases if, for $\epsilon\geq 0$,
\begin{equation}\label{eq:def_indistinguishability}
    P_{r}(\mathcal{A}(\mathcal{D})\in \mathcal{S})\leq \exp(\epsilon) P_{r}(\mathcal{A}(\mathcal{D}')\in \mathcal{S}).
\end{equation}
\end{definition}

Basically, a higher degree of indistinguishability (i.e., smaller $\epsilon$) implies a higher degree of privacy.
Let $\mathcal{D}'\equiv\{D'_{1}, \dots, D'_{n}\}$ be another database that differs from $\mathcal{D}$ in one data point, i.e., $D_{k}\neq D'_{k}$ and $D_{j}=D'_{j}$, for all $j\neq k$. In other words, the Hamming Distance, which is defined as $\text{HD}(\mathcal{D}, \mathcal{D}')=\sum_{i=1}^{n}\mathbf{1}\{D_{i} \neq D'_{i}\}$, is $1$.
The notion of differential privacy is developed in \cite{dwork2011differential}.
Specifically, the algorithm $\mathcal{A}$ is differentially private if the probability likelihood of $\mathcal{A}(\mathcal{D})\in \mathcal{S}$ is close to the probability likelihood of $\mathcal{A}(\mathcal{D}')\in \mathcal{S}$.
We refer to a data point that contains private information (whose privacy needs to be protected) but is unknown by the adversary as \textit{sensitive} data point.
Basically, differential privacy captures the indistinguishability of the algorithm in the \textit{worst-case scenario}, in which the adversary knows every data points other than a single sensitive $D_{k}$, and guarantees that any single data point does not influence the distribution of algorithm outcome by much.
Thus, the adversary cannot obtain much information about the sensitive data point by observing the distributions of the outcomes of the algorithm.
Definition \ref{def:differential_privacy} formally describes the concept of differential privacy.

\begin{definition}\label{def:differential_privacy} ($\epsilon$-Differential Privacy.)
A randomized algorithm $\mathcal{A}: \mathbb{D}^{n} \mapsto \mathcal{S}$ is $\epsilon$-differentially private if for any pair of database $\mathcal{D}$ and $\mathcal{D}'$ with $\text{HD}(\mathcal{D}, \mathcal{D}')=1$,
\begin{equation}\label{eq:differential_privacy_def}
    P_{r}(\mathcal{A}(\mathcal{D})\in \mathcal{S} ) \leq \exp(\epsilon)P_{r}( \mathcal{A}(\mathcal{D}'\in \mathcal{S})),
\end{equation}
where $\epsilon\in\mathbb{R}_{+}$.
\end{definition}

Differential privacy is a strong privacy notion that protects any single sensitive data point in the worst-case scenario.
In particular, any $\epsilon$-differentially private algorithm $\mathcal{A}$ that is robust to the adversary who targets on knowing the $k$-th data point $D_{k}$ of the input database $\mathcal{D}$ is also robust to any other adversaries who have different target data points $D_{j}\in \mathcal{D}$, for any $j\neq k$.
It is difficult to know what information the adversary could have about the target database.
By considering the worst-case scenario, differential privacy makes no assumptions about the knowledge set of the adversary.
The standard randomization approach for promoting differential privacy is perturbation with Laplacian noise (see, e.g., \cite{dwork2006calibrating,chaudhuri2011differentially,zhang2016dynamic}).

Next, we consider a different scenario, i.e., there are $m>1$ sensitive data points and the adversary knows all other $n-m$ data points except these $m$ points.
Let $\mathcal{D}^{m}$ be any database such that $\text{HD}(\mathcal{D}^{m}, \mathcal{D}) = m$.
The following corollary directly follows Definition \ref{def:differential_privacy} (see, e.g.,  \cite{dwork2006calibrating,ghosh2015selling}).
\begin{corollary}\label{corollary:differential_privacy_multiple_points}
Let $\mathcal{A}$ be any $\epsilon$-differentially private algorithm defined in Definition \ref{def:differential_privacy}.
Let $\mathcal{D}$ and $\mathcal{D}^{m}$ be any two databases with $\text{HD}(\mathcal{D}^{m}, \mathcal{D}) = m$. Then the following holds:
\begin{equation}\label{eq:corollary_differential_privacy}
    P_{r}(\mathcal{A}(\mathcal{D})\in \mathcal{S}) \leq \exp(m\times\epsilon )P_{r}(\mathcal{A}(\mathcal{D}^{m})\in \mathcal{S} ).
\end{equation}
Let any randomized algorithm $\mathcal{A}$ satisfying (\ref{eq:corollary_differential_privacy}) be named as $m\epsilon$-indistinguishable.
\end{corollary}
%

\begin{proof}
Let $\mathcal{D}^{-1,0}=\mathcal{D}$, $\mathcal{D}^{0,1}$, $\mathcal{D}^{1,2}$, $\dots,$ $\mathcal{D}^{m-1,m}=\mathcal{D}^{m}$ be any sequence of databases such that each pair $\mathcal{D}^{k-1,k}$ and $\mathcal{D}^{k,k+1}$ have $\text{HD}(\mathcal{D}^{k-1,k}, \mathcal{D}^{k,k+1})=1$, for all $0\leq k \leq m-1$
Then, we have
$$
\begin{aligned}
\frac{P_{r}(\mathcal{A}(\mathcal{D})\in \mathcal{S})   }{ P_{r}(\mathcal{A}(\mathcal{D}^{m})\in \mathcal{S})}=&\prod_{k=0}^{m-1} \frac{P_{r}(\mathcal{A}(\mathcal{D}^{k-1,k})\in \mathcal{S})   }{ P_{r}(\mathcal{A}(\mathcal{D}^{k,k+1})\in\mathcal{S})}\\
\leq& \prod_{k=0}^{m-1}\exp(\epsilon)=\exp(m\times \epsilon).
\end{aligned}
$$
\end{proof}

\begin{figure*}[tbh]
\centering
\includegraphics[width=0.6\linewidth]{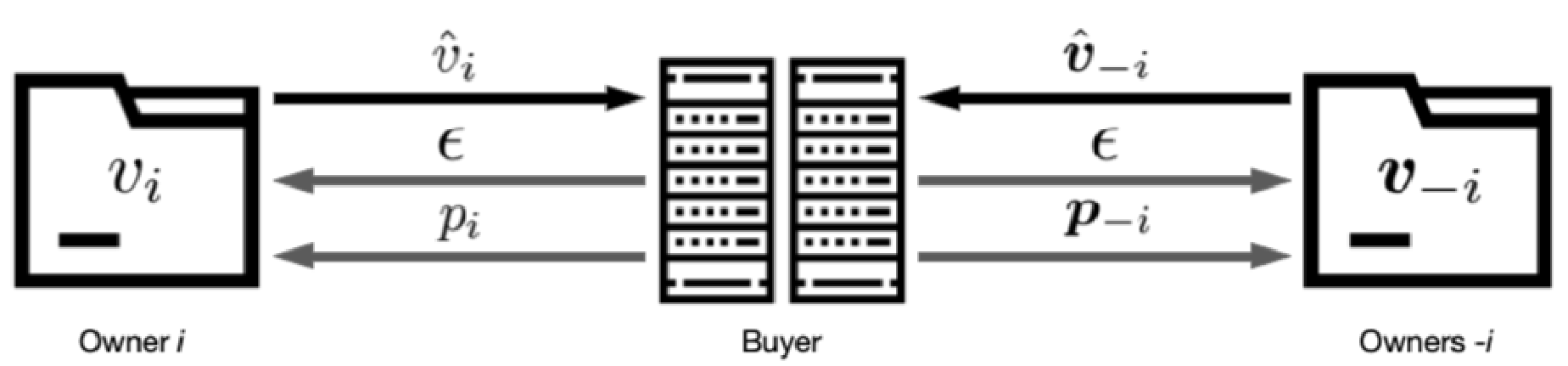}
\caption{One-stage market model for data privacy: Each owner $i$ with preference $v_{i}$ reports $\hat{v}_{i}$ to the buyer. Given the report profile $\hat{\bm{v}}=\{\hat{v}_{i},  \hat{\bm{v}}_{-i}\}$,  the buyer specifies a privacy allocation $\epsilon = \sigma(\hat{\bm{v}})$ to protect the privacy of all owners' data and payment $p_{i}= \beta_{i}(\hat{\bm{v}})$ to each owner $i$.}
\label{fig:privacy_market_one_stage}
\end{figure*}

\textcolor{black}{
In the \textit{non-worst-case scenario} when there are $m$ sensitive data points, i.e., when the adversary does not know $m$ data points in the private database, Corollary \ref{corollary:differential_privacy_multiple_points} states that if an algorithm $\mathcal{A}$ is $\epsilon$-indistinguishable for any pairs of databases $D^{m}$ and $D$ with $HD(D^{m}, D)=m$, then it is $\frac{\epsilon}{m}$-differentially private.
}

%
%

Another important feature of differential privacy is its \textit{(sequential) composability}.
In particular, composition of $k$ randomized algorithms that get access to the same database, each of which is $\epsilon$-differentially private, is at least $k\epsilon$-differentially private \cite{dwork2006calibrating,dwork2006our,kairouz2017composition}.
More generally, let $\mathcal{A}^{k}=\{\mathcal{A}_{t}\}_{t=1}^{k}$ denoted a composition of $k$ randomized algorithms that uses the same database $\mathcal{D}$, in which each $\mathcal{A}_{t}$ be the $\epsilon_{t}$-differentially private algorithm at the $t$-th order in the sequential composition, for some $\epsilon_{t}\in \mathbb{R}_{+}$.
Let $\mathcal{A}^{k}(D) = \{\mathcal{A}_{t}(D)\}$ denote the corresponding sequence of outputs. 
Then, the following holds, for any pair $\mathcal{D}, \mathcal{D}^{1}\in \mathbb{D}^{n}$ with $\text{HD}(\mathcal{D}, \mathcal{D}^{1})=1$,
\begin{equation}\label{eq:introduction_composition}
    P_{r}(\mathcal{A}^{k}(\mathcal{D}) \in \prod\limits_{t=1}^{k}\mathcal{S}_{t})\leq \exp(\sum_{t=1}^{k} \epsilon_{t})P_{r}(\mathcal{A}^{k}(\mathcal{D}^{1}) \in \prod\limits_{t=1}^{k}\mathcal{S}_{t}).
\end{equation}
%

\subsection{One-Stage Data Privacy Market Using Differential Privacy }\label{sec:brief_static_privacy_trading}

\textcolor{black}{
In this section, we define a single-stage data privacy market model and describe some basic concepts of mechanism design.
The \textit{static} in this work is twofold: \textit{(i)} owners' preference over privacy protection is static and \textit{(ii)} the buyer uses only \textit{one randomized algorithm} whose privacy parameter $\epsilon$ is fixed.
Here, the buyer can access to the owners' data multiple times by using the same randomized algorithm.
}

\textcolor{black}{
With respect to Fig. \ref{fig:privacy_market_one_stage}, we define the static one-stage market model and describe some basic concepts of mechanism design.
Consider a static market model consisting of two parties: those are \textit{(1)} $n$ data owners (\textit{owner}, he), denoted as $\mathbb{I}\equiv [n]$ and \textit{(2)} a data buyer (\textit{buyer}, she);
a generic owner is indexed by $i\in\mathbb{I}$.
Each owner $i$ possesses a private data point $D_{i}\in \mathbb{D}$ and the buyer wants to purchase the data from the owners to constitute a data base $\mathcal{D}=\{D_{i}\}_{i\in \mathbb{I}}$.
}

\textcolor{black}{
\textit{\textbf{Dual preference of privacy.}}
The motive of the owners' wanting their privacy protected in the market is based on their \textit{preference of privacy} (privacy preference, preference).
The privacy preference parameterizes an owner's \textit{cost of privacy loss}.
We consider a \textit{dual privacy preference} which consists of two components \cite{lin2020valuing}.
The \textit{intrinsic} part represents an owner's \textit{``taste"} of privacy which is utility primitive and is independent of how the data is used by the buyer.
The \textit{instrumental} part, on the other hand, endogenously depends on the buyer's usage of data and captures the owner's anticipated economic loss from  potential privacy leakage by participating in the buyer's market.
%
In this work, we consider that each owner $i$ privately observes his privacy preference, denoted by $v_{i}\in V_{i}$, for all $i\in I$, where $V_{i}$ is a compact set of privacy preferences of owner $i$.
We assume that each owner $i$'s privacy preference has prior probability distribution denoted by $\overline{K}_{i}\in \Delta(V_{i})$, for all $i\in\mathbb{I}$.
}

\textcolor{black}{
The buyer promises the owners to protect the privacy of data, using differential privacy, by taking into account the owners' privacy preferences.
Since the buyer uses the data from all the owners together, she takes advantage of the robustness of differential privacy and address \textit{global privacy protection} for $\mathcal{D}$ which is parameterized by a scalar $\epsilon\in \mathcal{E}\subseteq\mathbb{R}_{+}$.
We refer to $\epsilon$ as \textit{privacy allocation}.
However, the privacy loss is inevitable even if the data usage is differentially private.
Hence, the buyer additionally specifies a payment to compensate the privacy loss of each owner.  
This payment captures the price of privacy that is lost in the buyer's differentially private data usage.
Since the economic behaviors in this market is due to the owners' privacy concern, we refer to this market as \textit{data privacy market.}
}

\textcolor{black}{
However, the preference is the private information of each owner and the buyer can only know about it through the \textit{message}, $m_{i}\in M_{i}$, reported by each owner $i$.
We restrict attention to \textit{direct mechanism} in which each owner $i$ reveals his privacy preference; i.e., $\mathcal{M}_{i}=V_{i}$, for all $i\in\mathbb{I}$.
Hence, each owner may use such informational advantage to manipulate the market---\textit{adverse selection}--- due to the buyer's not knowing his true preference; i.e., each owner may find that it is his benefit to misreport his true privacy preference.
Owner $i$ uses a \textit{reporting strategy}, $\bar{\chi}_{i}:V_{i} \mapsto  V_{i}$, to report his privacy preference. Let $\hat{v}_{i}$ denote a typical report from owner $i$; i.e., $\hat{v}_{i} = \bar{\chi}_{i}(v_{i})$.
Owner $i$ misreports his preference if $\hat{v}_{i}\neq v_{i}$.
}

\textcolor{black}{
The buyer first collects the reported preference $\bm{\hat{v}}\equiv(\hat{v}_{i})_{i\in \mathbb{I}}$ and determines an $\epsilon\in \mathcal{E}$ that specifies the differential privacy protection of her data usage.
The buyer uses an \textit{privacy assignment rule} (assignment rule), $\bar{\sigma}:\bm{V}\mapsto \mathcal{E}$, to choose a privacy allocation $\epsilon$ when a report profile $\bm{\hat{v}}$ is collected; i.e., $\epsilon = \bar{\sigma}( \bm{\hat{v}})$.
The buyer uses a \textit{pricing rule}, $\bar{\beta}_{i}: V_{i}\times \bm{V}_{-i}\mapsto \mathcal{P} \subseteq \mathbb{R}$, to specify a payment $p_{i}$ to each owner $i$ based on their reports $\bm{\hat{v}}=(\hat{v}_{i}, \bm{\hat{v}}_{-i})$: $p_{i} = \bar{\beta}_{i}(\hat{v}_{i}, \bm{\hat{v}}_{-i})$.
We focus on the setting when the buyer's specifications of privacy protection $\epsilon$ and payments $p\equiv(p_{i})_{i\in I}$ based only on the (owner's reported) privacy preference.
This setting captures that the privacy concern is only from the owner and is exogenous to the buyer.
}

\textcolor{black}{
When owner $i$'s privacy preference is $v_{i}\in V_{i}$ and the buyer uses $\epsilon\in \mathcal{E}$ to protect the data privacy, owner $i$'s privacy loss is given by \cite{ghosh2015selling}:
\begin{equation}\label{eq:flow_loss}
    \begin{aligned}
        \ell(v_{i}, \epsilon) \equiv v_{i}\big[\exp(\epsilon )-1\big].
    \end{aligned}
\end{equation}
The loss function $\ell$ is increasing in $\epsilon$, i.e., the larger (resp. smaller) $\epsilon$ is, the less (resp. more) private the data usage becomes; when $\epsilon\rightarrow 0$, there is no private loss, i.e., $\lim_{\epsilon\rightarrow 0}\ell(v_{i}, \epsilon)=0$.
Since $v_{i}$ and $\epsilon$ are finite, $\ell(v_{i}, \epsilon)$ is bounded; i.e., $|\ell(v_{i}, \epsilon)|< \infty$, for all $v_{i}\in V_{i}$ and $\epsilon\in \mathcal{E}$.
On the other hand, the buyer suffers losses of utility that she can extract from the data due to differential privacy protection.
There is fundamental tradeoff between owners' privacy and the buyer's utility from the data: larger (resp. smaller) $\epsilon_{t}$ gives the buyer more (resp. less) utility from the data and the owners more (resp. less) privacy loss from participation.
One possible formulation of of the buyer's \textit{utility loss} is:
\begin{equation}
    \alpha(\epsilon)\equiv L\exp(-\epsilon),
\end{equation}
where $L\in \mathbb{R}_{++}$ represents the maximum utility loss when $\epsilon\rightarrow 0$.
}

\textcolor{black}{
The buyer's allocation rule and the payment rule profile, $<\bar{\sigma}, \bm{\bar{\beta}}>$, constitute a \textit{mechanism}, which causes strategic interaction of each owner $i$'s with other owners.
Since the privacy preference $v_{i}$ is a private information of each owner $i$, the mechanism $<\bar{\sigma}, \bm{\bar{\beta}}>$ induces a \textit{Bayesian game}.
%
%
By $\overline{\mathcal{M}}$, we denote the one-shot static market model for trading data privacy:
\begin{equation}\label{eq:def_one_shot_static_market}
    \begin{aligned}
    \overline{\mathcal{M}}\equiv \Big\{\{<\bar{\sigma}, \bm{\bar{\beta}}>\} , \{\bm{V},\bm{\bar{K}}\}, \ell, \alpha\Big|\mathcal{D}, \mathbb{I}\Big\}.
    \end{aligned}
\end{equation}
The model $\overline{\mathcal{M}}$ is common knowledge.
Each owner $i$ chooses an optimal $\chi_{i}$ that is a best response to his opponents' optimal strategy $\bm{\chi}_{-i}$ in $\overline{\mathcal{M}}$:
\begin{equation*}
    \begin{aligned}
    \chi_{i}\in &\arg\max\limits_{\chi'_{i}}\mathbb{E}_{ \bm{\tilde{v}}_{-i} \sim \bm{\bar{K}}_{-i}  }\Big[\\
    &\bar{\beta}_{i}(\chi'_{i}(v_{i}), \bm{\chi}_{-i}(\bm{\tilde{v}}_{-i}) ) - \ell\Big( v_{i},  \bar{\sigma}\big(\chi'_{i}(v_{i}),  \bm{\bar{\chi}}_{-i}(\bm{\tilde{v}}_{-i})\big) \Big) \Big].
    \end{aligned}
\end{equation*}
}
%

\textcolor{black}{
\textit{Incentive compatibility} and \textit{individual rationality} are two important constraints in mechanism design problems, which incentivizes owners to \textit{truthfully reveal} their private privacy preferences and motivates them to participate in the market, respectively.
In this work, we consider Bayesian incentive compatibility defined as follows.
\begin{definition}[Bayesian Incentive Compatibility]
The mechanism $<\bar{\sigma}, \bm{\bar{\beta}}>$ is Bayesian incentive compatible (BIC) if truthful reporting is each owner $i$'s best response to other owners' truthful reporting: for all $i\in \mathbb{I}$, $v_{i}, \hat{v}_{i}\in V_{i}$
\begin{equation}
    \begin{aligned}
  \mathbb{E}_{ \bm{\tilde{v}}_{-i} \sim \bm{K}_{-i}  }&\Big[ \bar{\beta}_{i}(v_{i}, \bm{\tilde{v}}_{-i} ) - \ell\big( v_{i},  \bar{\sigma}(v_{i}, \bm{\tilde{v}}_{-i}) \big) \Big]\\
   &\geq \mathbb{E}_{ \bm{\tilde{v}}_{-i} \sim \bm{K}_{-i}  }\Big[\bar{\beta}_{i}(\hat{v}_{i}, \bm{\tilde{v}}_{-i} ) - \ell\big( v_{i},  \bar{\sigma}(\hat{v}_{i}, \bm{\tilde{v}}_{-i}) \big)\Big].
    \end{aligned}
\end{equation}
\end{definition}
}

\textcolor{black}{
In general, we have three types of individual rationality.
\begin{definition}[Individual Rationality]
The mechanism $<\sigma, \beta>$ is ex-ante individually ratioal (EAIR) if each owner's ex-ante expected payoff is non-negative; i.e., for all $i\in \mathbb{I}$,
\begin{equation}
    \begin{aligned}
    \mathbb{E}_{ \bm{\tilde{v}} \sim \bm{K}  }\Big[ \bar{\beta}_{i}(v_{i}, \bm{\tilde{v}}_{-i} ) - \ell\big( v_{i},  \bar{\sigma}(v_{i}, \bm{\tilde{v}}_{-i})\big) \Big] \geq 0.
    \end{aligned}
\end{equation}
The mechanism is ex-interim individually rational (EIIR) if each owner's interim expected payoff is non-negative; i.e., for all $i\in \mathbb{I}$, $v_{i}\in V_{i}$,
\begin{equation}
    \begin{aligned}
    \mathbb{E}_{ \bm{\tilde{v}}_{-i} \sim \bm{K}_{-i}  }\Big[ \bar{\beta}_{i}(v_{i}, \bm{\tilde{v}}_{-i} ) - \ell\big( v_{i},  \bar{\sigma}(v_{i}, \bm{\tilde{v}}_{-i})\big) \Big] \geq 0.
    \end{aligned}
\end{equation}
The mechanism is ex-post individually rational (EPIR) if each owner's ex-post payoff is non-negative; i.e., for all $i\in \mathbb{I}$, $(v_{i}, \bm{v}_{-i})\in \bm{V}$,
\begin{equation}
    \begin{aligned}
     \bar{\beta}_{i}(v_{i}, \bm{v}_{-i} ) - \ell\big( v_{i},  \bar{\sigma}(v_{i}, \bm{v}_{-i})\big)\geq 0.
    \end{aligned}
\end{equation}
\end{definition}
}

\textcolor{black}{
The weakest form of individual rationality is the EAIR which implies that no individual owner wishes to decline to participate in the market before he knows his own preference and only have expectations over all the realizations of his preference and other owners' preferences and thus the resulting privacy allocations and payments.
The EIIR is the constraint of participation such that no individual owner wishes to leave the market after he observes his preference but does not know others' preferences. Hence, his participation decision is based on his expectations over all the realizations of his opponents' preferences and the resulting privacy allocations and payments.
The strongest form of individual rationality is the EPIR which is that no individual owner wishes to decline his participation in the market after all preferences have been revealed.
With EPIR, the privacy allocation and the payments are completely specified, regardless of the realizations of the owners' preferences; but the allocations and the payments are verifiable by the owners according to the preferences.
}

\textcolor{black}{
By fixing $\epsilon$, the privacy guarantee using $\epsilon$-differential privacy is time-invariant if the data $\mathcal{D}$ is repeatedly used by the same $\epsilon$-differentially private algorithm $\mathcal{A}$.
However, the owners' costs of privacy (captured by their privacy preferences) might be time-varying \cite{abowd2019economic}.
Moreover, due to the composability of differential privacy, if the buyer uses a sequence of independent algorithms to access to $\mathcal{D}$ over time, then the privacy loss accumulates.
As a result, multiple-time access to $\mathcal{D}$ requires the buyer to dynamically adjust privacy protection and decide the price of data privacy.
To this end, we propose a dynamic mechanism model to periodically allocate differential privacy protection \big(i.e., choosing $\epsilon_{t} \in \mathcal{E}$ for each time $t$) and specify the price of privacy (i.e., providing payment $\bm{p}_{t}\equiv (p_{i,t})_{i\in \mathbb{I}}$\big) to compensate privacy loss.
}

\begin{figure*}[tbh]
\centering
\includegraphics[width=0.7\textwidth]{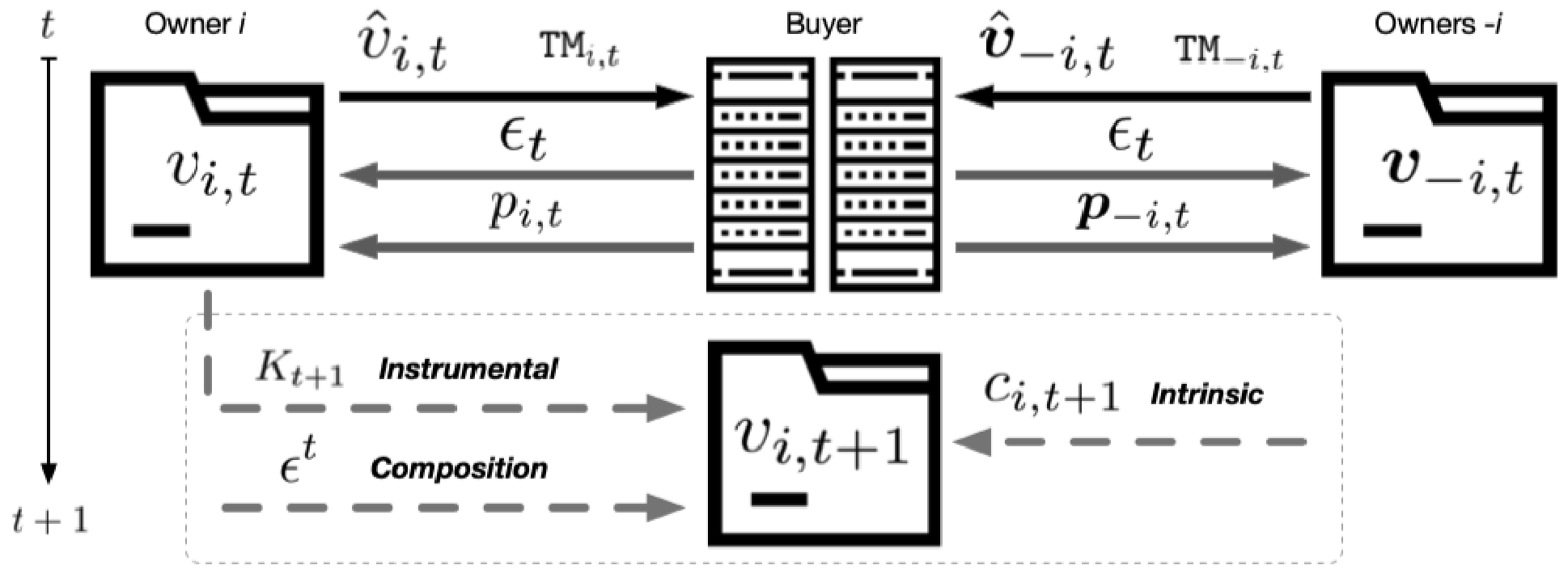}
\caption{Market model for data privacy: At each period $t$, each owner $i$ reports his preference $v_{i,t}$ as $\hat{v}_{i,t} = \sigma_{i,t}(v_{i,t}, h_{t})$ and his stopping decision $\mathtt{TM}_{i,t}\in \{0,1\}$ to the buyer.
Given the report profile $\hat{\bm{v}}_{t}=\{\hat{v}_{i,t}, \hat{\bm{v}}_{-i,t}\}$, the buyer specifies the privacy allocation $\epsilon_{t} = \sigma_{t}(\hat{\bm{v}}_{t})$ and payment $p_{i,t} = \beta_{i,t}(\hat{v}_{t})$ or $p_{i,t} = \theta_{i,t}(\hat{v}_{t})+ \rho_{i}(t)$ if $\mathtt{TM}_{i,t}=0$ or $\mathtt{TM}_{i,t}=1$, respectively.
The Markovian time-evolution of each owner $i$'s preference depends on his last-period preference $v_{i,t}$, the accumulated privacy loss characterized by $\epsilon^{t}$ due to the composition of differential privacy, current intrinsic preference $c_{i,t}$, and the instrumental kernel $K_{t+1}$: $v_{i,t+1} = K_{i,t+1}( v_{i,t}, \epsilon^{t}; c_{i,t+1})$.}
\label{fig:privacy_market}
\end{figure*}

\section{Dynamic Data Privacy Market Model}\label{sec:model}

In this section, we extend the one-stage market model described in Section \ref{sec:brief_static_privacy_trading} to a dynamic model in which the buyer uses a finite number of randomized algorithms to use the owners' data in a sequentially composition manner.
We refer to one usage of data by one randomized algorithm as one \textit{period of time}.
Hence, the dynamic model is finite-horizon in which time is discrete, denoted by $\mathbb{T}\equiv\{0,1,\dots, T\}$, with $0\leq T < \infty$.
Upon participation in the ex-ante stage, each owner $i$ updates his privacy preference to $v_{i,t}$ at the beginning of each period $t$.
Then, the owner reports his preference to the buyer.
We use $v_{i,t}$ and $\hat{v}_{i,t}$ to denote a generic preference and its report of owner $i$ in period $t$, respectively.
Each single-period data usage induces a privacy loss. Due to the composability of differential privacy, the privacy loss accumulates over time in an additive fashion (see, (\ref{eq:introduction_composition})).
Our dynamic model enables the owners to respond to accumulated privacy by allowing them to leave the market at the end of each period.
Let $\mathtt{TM}_{i,t}\in\{0,1\}$ denote owner $i$'s period-$t$ decision of stopping, in which $\mathtt{TM}_{i,t}=1$ and $\mathtt{TM}_{i,t}=0$ represent stopping and not stopping in $t$, respectively.
Once owner $i$ has chosen $\mathtt{TM}_{i,t}=1$, he cannot return to the market in any period $\tau > t$.

By extending the static model $\overline{\mathcal{M}}$ in (\ref{eq:def_one_shot_static_market}) in the dynamic environment, we denote each period-$t$ model by the following tuple: for all $t\in \mathbb{T}$,
\begin{equation}
    \begin{aligned}
    \mathcal{M}_{t}\equiv \Big\{\{<\sigma_{t}, \bm{\beta}_{t}, \bm{\theta}_{t}, \bm{\rho}(t)>\} ,  \{\bm{V}_{t},\bm{K}_{t}\} , \ell, \alpha\Big|\mathcal{D}, \mathbb{I}\Big\}.
    \end{aligned}
\end{equation}

Each element of $\mathcal{M}_{t}$ is elaborated in the following subsections.

\subsubsection{Non-anticipating Mechanism}

The buyer provides a \textit{take-it-or-leave-it} offer to the owners at the beginning of the initial period.
After each owner $i$ observes his initial privacy preference (preference) $v_{i,0}\in V_{i}$, he decides whether to participate in the dynamic market.
Starting from the initial period, the buyer dynamically chooses a privacy allocation. 
A generic privacy allocation in period $t$ is denoted as $\epsilon_{t}$.
Let $h_{t} \equiv (\hat{v}_{s}, \epsilon_{s})_{s\in \mathtt{T}_{0,t-1}}\in H_{t}\equiv \bm{V}^{t-1}\times \mathcal{E}^{t-1}$ denote the \textit{history} up to $t$.
We assume that the buyer discloses the reports of all the agents at the end of each period. 
Hence, the history $h_{t}$ is publicly observed.
%
%
%
%
%
%
The offer is composed of \textit{nonanticipating rules}, $<\sigma, \bm{\beta},$ $\bm{\theta}, \bm{\rho}>$ $=\{<\sigma_{t}, \bm{\beta}_{t}, \bm{\theta}_{t}, \bm{\rho}(t)>\}_{t\in\mathbb{T}}$. 
The privacy allocation rule profile $\sigma$ is nonanticipating if $\epsilon_{t}$ specified by $\sigma_{t}$ depends only on owners' current report $\hat{\bm{v}}_{t}$ and the history, $h_{t}$, up to $t$; i.e., $\sigma_{i,t}(\hat{v}, h_{t}) = \epsilon_{t}$. 
The payment rules $\bm{\beta},$ $\bm{\theta}$ are nonanticipating in the same way.

Upon participation, in each period $t$, owner $i$ updates his preference to $v_{i,t}\in V_{i,t}$ and then reports $\hat{v}_{i,t}\in V_{i,t}$ and sends the terminating message $\mathtt{TM}_{i,t}\in\{0,1\}$ to the buyer.
Then, the buyer chooses $\epsilon_{t} = \sigma_{t}(\bm{\hat{v}}_{t}, h_{t})$ and specifies a payment $p_{i,t}\in \mathcal{P}\subset \mathbb{R}$ to each owner $i$.

To cope with the owners' leave option, the buyer uses a \textit{switching payment} policy such that
\begin{equation}
    \begin{aligned}
    p_{i,t} =\begin{cases}
    \beta_{i,t}(\hat{v}_{t}, h_{t}), &\text{ if } \mathtt{TM}_{i,t}=0\\
    \theta_{i,t}(\hat{v}_{t}, h_{t}) + \rho_{i}(t), &\text{ if } \mathtt{TM}_{i,t}=1.
    \end{cases}
    \end{aligned}
\end{equation}
Specifically, owner $i$ receives a payment specified by $\beta_{i,t}(\hat{v}_{t}, h_{t})$ if he decides to continue to trade in the next period (i.e., $\mathtt{TM}_{i,t}=0$) while he receives a payment specified by $\theta_{i,t}(\hat{v}_{t}, h_{t}) + \rho_{i}(t)$ if he activates his leave option (i.e., $\mathtt{TM}_{i,t}=1$).
Here, $\rho_{i}:\mathbb{T}\mapsto \mathbb{R}$ specifies a monetary transfer that is independent of each owner's period-$t$ report. 
We refer to $\theta_{i,t}$ and $\rho_{i}$ as the \textit{pause} and the \textit{posted} (payment) rules, respectively.

\subsubsection{Preference Dynamics}

In the static model, the roles of the intrinsic and the instrumental components of each owner's privacy preference are not distinguished from each other. 
In the dynamic model, however, we leverage the endogenous nature of the instrumental component and articulate the roles of the intrinsic and the instrumental components in the formation of each owner's time-evolving privacy preference.
In the rest of the paper, we use intrinsic (instrumental) preference and intrinsic (instrumental) component interchangeably.

\textcolor{black}{
In this work, we assume that each owner $i$'s intrinsic preference, denoted as $c_{i,t}\in \mathcal{C}_{i}$, is a move by Nature, which is drawn independently in each period $t$ according to a common prior $K^{\mathcal{C}}_{i}(\cdot)\in\Delta(\mathcal{C}_{i})$, for all $i\in\mathbb{I}$, $t\in \mathbb{T}$.
We assume that owners' intrinsic preference are idiosyncratic and are independent of each other.
We restrict attention to when the instrumental component of each owner's privacy preference is due to the buyer's dynamic privacy allocations and is independent of the monetary transfers.
This setting coincides with the definition of the instrumentalness in that the instrumental preference over privacy is due to the concern of the anticipated privacy loss rather than the monetary value from participating in the buyer's market.
}

\textcolor{black}{
\begin{definition}[Instrumentalness]
The instrumentalness of the privacy preference is modeled by a collection of \textit{instrumental kernels} (kernels), $K\equiv<K_{t}>_{ t\in \mathbb{T}}$, in which $K_{t}: V^{t-1}_{i}\times\mathcal{E}^{t-1}\times \mathcal{C}_{i}  \mapsto V_{i}$, such that, the random variable $\tilde{v}_{i,t} \sim K_{i,t}( v^{t-1}_{i}, \epsilon^{t-1}; \tilde{c}_{i,t}) \in \Delta(V_{i,t})$ if $\tilde{c}_{i,t}\sim K^{\mathcal{C}}_{i}(\cdot)\in \Delta(\mathcal{C}_{i})$ and a realization $v_{i,t} = K_{i,t}( v^{t-1}_{i}, \epsilon^{t-1}; c_{i,t})$, for all $i\in\mathbb{I}$, $t\in \mathbb{T}$, when his history of preference is $v^{t-1}_{i}$, the history of privacy allocation is $\epsilon^{t-1}$, and his period-$t$ intrinsic preference is $c_{i,t}$. 
\end{definition}
}

\textcolor{black}{
Hence, the intrinsic component of owner $i$'s period-$t$ preference $v_{i,t}$ is $c_{i,t}$ and the instrumental component is given by the dependence on the past participation (i.e., $\{v^{t-1}_{i}, \epsilon^{t-1}\}$) through the kernel. 
The randomness of the intrinsic component and the endogenous nature of the instrumental component forms the dynamics of owners' preferences when the buyer requests multiple accesses to the owners' data by a sequence of independent algorithms.
Given $K^{\mathcal{C}}$, each kernel $K_{t}$ serves as a transition probability from period-$t-1$ preference to period-$t$ preference.
Here, we assume that conditional on $\{v^{t-1}_{i}, \epsilon^{t-1}\}$, owners' preferences in period $t$ are independent from each other.
This coincides with the assumption that owners' intrinsic components are idiosyncratic so that we can rule out aggregate random elements that involved in the dynamics of the owners' preferences that are common to all owners.
}

\textcolor{black}{
\begin{remark}
We can use the notion of independent \textit{shock}, denoted by $\mathtt{s}_{i,t}\in \mathcal{S}$, to define the dynamics of the owners' preferences.
With a slight abuse of notation, let $\widehat{K}_{t}: V^{t-1}_{i}\times\mathcal{E}^{t-1}\times \mathcal{C}_{i}  \mapsto \Delta(V_{i})$ denote the transition function of each owner $i$'s period-$t$ preference, such that $\tilde{v}_{i,t}$ is distributed according to $\widehat{K}_{t}(v^{t-1}_{i}, \epsilon^{t-1}_{i};c_{i,t})$ when the owner's history of preference is $v^{t-1}_{i}$, the history of privacy allocation is $\epsilon^{t-1}$, and his period-$t$ intrinsic preference is $c_{i,t}$.
Then, there exists a representation $O^{V}_{t}: V^{t-1}_{i}\times\mathcal{E}^{t-1}\times \mathcal{C}_{i} \times \mathcal{S}  \mapsto V_{i,t}$, such that, if $\tilde{\mathtt{s}}_{i,t}$ is distributed according to a distribution $W\in \Delta(\mathcal{S})$, then $\tilde{v}_{i,t} \sim \widehat{K}_{t}(v^{t-1}_{i}, \epsilon^{t-1}; c_{i,t}| \tilde{\mathtt{s}}_{i,t})$ and $v_{i,t} = O^{V}_{t}(v^{t-1}_{i}, \epsilon^{t-1}; c_{i,t}| \mathtt{s}_{i,t})$. 
Such representation $O^{V}_{t}$ exists for any transition function $\widehat{K}_{t}$ (see, e.g., \cite{esHo2007optimal}).
Another way to involve the independent shocks in the preference dynamics is via the generations of the owners' intrinsic components.
Specifically, there exists a representation $O^{\mathcal{C}}_{i,t}: \mathcal{S}\mapsto \mathcal{C}_{i}$, such that, if $\tilde{\mathtt{s}}_{i,t} \sim W$, then $\tilde{c}_{i,t} \sim K^{\mathcal{C}}_{i}$ and $c_{i,t} = O^{\mathcal{C}}_{i,t}(\mathtt{s}_{i,t})$.
In this work, we restrict attention to the latter case; i.e., the randomness of each owner's preference is due to the randomness of the intrinsic component. However, the results of this work can be easily extended to the former case.
\end{remark}
}

\textcolor{black}{
We consider that the owners have the same kernel in each period.
This is due to a reasonable assumption that the owners experience the same endogenous influence (i.e. instrumentalness) from participating in the same market.
The dependence of $v_{i,t}$ on the history $v^{t-1}_{i}$ can be relaxed by considering a Markovian setting.
}

\textcolor{black}{
\begin{definition}[Markovian Instrumentalness]
The instrumentalness of each owner's preference is Markovian if period-$t$ preference $v_{i,t}$ depends on past participation $\{v^{t-1}_{i}, \epsilon^{t-1}\}$ only through $\{v_{i,t-1}, \epsilon^{t-1}\}$ for all $i\in\mathbb{I}$, $t\in\mathbb{T}$; i.e., $\tilde{v}_{i,t} \sim K_{t}(v_{i,t-1}, \epsilon^{t-1}; \tilde{c}_{i,t})$ if $\tilde{c}_{i,t}\sim K^{\mathcal{C}}_{i}$ and $v_{i,t} = K_{t}(v_{i,t-1}, \epsilon^{t-1}; c_{i,t})$.
\end{definition}
}
\textcolor{black}{
With the Markovian instrumentalness, histories leading to the same $\{v_{i,t-1}, \epsilon^{t-1}\}$ have the same endogenous effect on owner $i$'s formation of preference in period $t$.
For the rest of the paper, we focus on the Markovian instrumentalness.
The realization of the preference and its intrinsic are the private information of each owner.
We assume that each owner is aware of his preferences through, e.g., data privacy audition, and all owners' preferences are evaluated according to the same publicly known standard.
Let $F_{i,t}(\cdot|v_{i,t-1}, \epsilon^{t-1})$ and $f_{i,t}(\cdot|v_{i,t-1}, \epsilon^{t-1})$ denote the cumulative distribution function (cdf) probability density function (pdf) of owner $i$ in period $t$, respectively, corresponding to the kernel $K_{t}$, for all $i\in\mathbb{I}$, $t\in\mathbb{T}$.
}

%
%

\subsection{Dynamic Bayesian Game}\label{sec:dynamic_bayesian_game}

We first suppose that agents are not allowed to leave the market until the end of the last period $T$ of the buyer's commitment.
The proposed market model couples the owners through their reports by the mechanism rules $\sigma$, $\bm{\beta}$, and $\bm{\theta}$.
Since each owner's private information is dynamic, the market model induces a dynamic Bayesian game among the owners.
The action of each owner $i$ in each period $t$ is to choose a report $\hat{v}_{i,t}$.
The reporting strategy of each owner $i$ is a collection $\chi_{i}\equiv(\chi_{i,t})_{t\in\mathbb{T}}$.
Because the owners' preferences in each period are independent, we restrict attention to \textit{non-anticipating reporting strategies} in the same way as the mechanism rules.
Specifically, agent $i$'s period-$t$ reporting strategy $\chi_{i,t}:V_{i,t} \times H_{t}\mapsto V_{i,t}$ determines a report $\hat{v}_{i,t}$ that depends on current preference $v_{i,t}$, past reports and privacy allocations, but not on past true preferences; i.e., $\hat{v}_{i,t} = \chi_{i,t}(v_{i,t}, h_{t})$.
%

A well-known equilibrium concept for such game is the \textit{perfect Bayesian equilibrium} (PBE).
Following the Revelation Principle, it is without loss of generality to focus on direct mechanisms in which owners truthfully report their preferences to the buyer.
In particular, we study PBE in truthful reporting strategy profile ($\chi_{i,t}(v_{i,t},h_{i,t}) = v_{i,t}$, for all $i\in\mathbb{I}$, $t\in \mathbb{T}$, $v_{i,t}\in V_{i,t}$, $h_{t}\in H_{t}$) for the underlying dynamic Bayesian game, in which each owner assigns probability $1$ to the event that all other owners report truthfully.
%
%
The general specification of PBE also requires players to form beliefs about unobserved payoff-relevant information according to Bayes' law.
Our non-anticipating strategies only conditions on previous reports and privacy allocations and this information is publicly observed. As a result, each owner does not need to form posterior beliefs about the past reports of his opponents or the past privacy allocations.
Since truthful PBE assumes other owners to report truthfully, the public history $h_{t}$ contains the history of owners' past true preferences.
Thus, the beliefs about the contemporaneous preferences of other owners can be formed according to the kernel $K_{-i,t}$ based on the public history $h_{t}$ (through $(v_{-i,t-1}, \epsilon_{t-1})$ in the Markovian environment).
%

To elaborate truthful PBE, we define some notations.
Given the privacy allocation $\epsilon_{t}\in \mathcal{E}$ and his preference $v_{i,t}\in V_{i,t}$, owner $i$ obtains a (monetary) \textit{one-stage privacy loss}, specifies by a \textit{loss function} $\ell_{i,t}: V_{i,t}\times \mathcal{E}\mapsto \mathbb{R}$ \cite{ghosh2015selling}:
\begin{equation}\label{eq:flow_loss_0}
    \begin{aligned}
        \ell(v_{i,t}, \epsilon_{t}) \equiv v_{i,t}\big[\exp(\epsilon_{t})-1\big].
    \end{aligned}
\end{equation}
The loss function $\ell$ is increasing in $\epsilon_{t}$; i.e., the larger (resp. smaller) $\epsilon_{t}$ is, the less (resp. more) private the data becomes (hence, more privacy loss is suffered by owner $i$).
When $\epsilon_{t}\rightarrow 0$, there is no private loss, i.e., $\lim_{\epsilon_{t}\rightarrow 0}\ell(v_{i,t}, \epsilon_{t})=0$.
Since $v_{i,t}$ and $\epsilon_{t}$ are finite, $\ell(v_{i,t}, \epsilon_{t})$ is bounded, i.e., $|\ell(v_{i,t}, \epsilon_{t})|< \infty$, for all $v_{i,t}\in V_{i,t}$ and $\epsilon_{t}\in \mathcal{E}$.

Given the payment $p_{i,t}$, owner $i$'s period-$t$ \textit{payoff} is defined as follows:
\begin{equation}\label{eq:owner_flow_payoff}
    z_{i,t }(v_{i,t }, \epsilon_{t}, p_{i,t}) \equiv -\ell(v_{i,t}, \epsilon_{t}) + p_{i,t}.
\end{equation}
Since $\ell$ is bounded and $p_{i,t}$ is finite, the payoff $z_{i,t}(v_{i,t}, \epsilon_{t}, p_{i,t})$ is bounded, i.e., $|z_{i,t}(v_{i,t},$ $\epsilon_{t}, p_{i,t})|<\infty$, for all $v_{i,t}\in V_{i}$, $\epsilon_{t}\in \mathcal{E}$, and $p_{i,t}\in \mathcal{P}$.

According to Ionescu Tulcea theorem (see, e.g., \cite{hernandez2012discrete}), the kernels $K$, the allocation rule $\sigma$, and the owners' reporting strategy profile $\chi$ define a unique probability measure $P[\sigma, \chi]$ on $V^{T}$.
%
%
Similarly, any history $h_{t}\in H_{t}$, current preference $v_{i,t}$, and $<K^{T}_{t}, \sigma^{T}_{t}, \chi^{T}_{i,t}>$ uniquely define a probability measure $P[\sigma, \chi]|v_{i,t},h_{t}$ on $V^{T}_{t+1}$.
We denote the expectation operators corresponding to $P[\sigma, \chi]$ and $P[\sigma, \chi]|v_{i,t},h_{t}$, respectively, as $\mathbf{E}^{\sigma}_{\chi}[\cdot]$ and $\mathbb{E}^{\sigma}_{\chi}[\cdot|v_{i,t}, h_{t}]$.
For the ease of notation, we suppress the reporting strategy in the expectation if owners are truthful: $\mathbb{E}^{\sigma}$ if $\chi$ is truthful and $\mathbb{E}^{\sigma}_{\chi_{i}}$ if owners other than $i$ are truthful.
Also, we may omit the the preference and history in the expectation operator when the conditioning event is obvious: $\mathbb{E}^{\sigma}_{\chi_{i}}[\cdot] = \mathbb{E}^{\sigma}_{\chi_{i}}[\cdot|v_{i,t}, h_{t}]$.

Let $\Gamma\equiv<\sigma, \bm{\beta},\bm{\theta},\bm{\rho}>$ denote the mechanism rule profile.
Suppose that each agent $i$ sends $\mathtt{TM}_{i,t}=0$ for all $i\in\mathcal{N}$, $t\in\mathbb{T}$.
We use the notation $J^{\Gamma}_{i,t}$ to denote the period-$t$ \textit{ex-interim payoff} function of owner $i$. Let, for all $i\in\mathbb{I}$, $t\in\mathbb{T}$, $\tau\in\mathbb{T}_{t}$,
\begin{equation}\label{eq:owner_interim_payoff}
    \begin{aligned}
    &J^{\Gamma}_{i,t}(v_{i,t}, h_{t}, \tau; \chi)\equiv \sum\limits_{k=0}^{t-1}z_{i,k}(v_{i,k}, \epsilon_{k}, p_{i,k})\\
    &+\mathbb{E}^{\sigma}_{\chi}\Big[ z_{i,t}(v_{i,t}, \tilde{\epsilon}_{t}, \tilde{p}_{i,t}) + \sum_{s=t+1}^{\tau }z_{i,s}(\tilde{v}_{i,s}, \tilde{\epsilon}_{s}, \tilde{p}_{i,s})    \Big],
    \end{aligned}
\end{equation}
with $\sum^{\tau}_{s=t+1}z_{i,s}(\cdot) = 0$ when $t=\tau$.
Then, $J^{\Gamma}_{i,t}(v_{i,t}, h_{t},$ $T; \chi)$ is owner $i$'s period-$t$ ex-interim payoff, when owner $i$'s current preference is $v_{i,t}\in V_{i,t}$, history is $h_{t}\in H_{t}$, and owners reporting strategy profile is $\chi$.

Let $\chi^{*}_{i}$ denote the truthful reporting strategy of each owner $i$.
We impose the \textit{ex-interim incentive compatibility} (IC) constraints which guarantees that each owner is better off reporting truthfully when other owners adopt the truthful strategies after observing his each-period preference under the market.
That is,
\begin{equation}\tag{$\mathtt{IC}_{i,t}$}\label{eq:d_ex_interim_IC}
    \begin{aligned}
    J^{\Gamma}_{i,t}(v_{i,t}, h_{t}, T; \chi^{*}) \geq J^{\Gamma}_{i,t}(v_{i,t}, h_{t}, T; \chi_{i}, \chi^{*}_{-i}).
    \end{aligned}
\end{equation}
We say a mechanism $\Gamma$ is IC if it satisfies (\ref{eq:d_ex_interim_IC}), for all $i\in \mathbb{I}$, $t\in\mathbb{T}$.
Any IC mechanism induces a PBE in truthful reporting strategies, in which the \textit{Bayesian} part refers to the assumption that each owner assigns probability $1$ to other owners' truthful reporting.
For each of notation, we omit the notations of truthful reporting strategies; e.g., $J^{\Gamma}_{i,t}(\cdot; \chi_{i}) = J^{\Gamma}_{i,t}(\cdot; \chi_{i}, \chi^{*}_{-i})$.

We next use the notion $J^{\Gamma}_{i}$ to denote the \textit{ex-ante expected payoff} function of owner $i$, when the reporting strategy profile of the owners is $\chi$. 
Define, for all $i\in\mathbb{I}$, $\tau\in\mathbb{T}$,
%
%
\begin{equation}\label{eq:owner_ex_ante_payoff}
    \begin{aligned}
    & J^{\Gamma}_{i}(\chi, \tau) \equiv \mathbf{E}^{\sigma}_{\chi}\Big[ J^{\Gamma}_{i,0}(\tilde{v}_{i,0}, \tau; \chi)  \Big].
    \end{aligned}
\end{equation}
Then, $J^{\Gamma}_{i}(\chi, T)$ is owner $i$'s ex-ante expected payoff.
For the ease of notation, when owners' reporting strategy profile is truthful, we suppress it; e.g., $J^{\Gamma}_{i}(\tau) = J^{\Gamma}_{i}(\chi^{*}, \tau)$ when $\chi^{*}$ is truthful.

Besides the incentive compatibility constraint, the buyer also wants the owners have incentive to participate in the market.
This is captured by the constraint of \textit{individual rationality}. 
In general, there are three notions of individual rationality and we describe each of them in the dynamic model as follows.

An IC mechanism is \textit{ex-ante individual rational} (EAIR) if 
\begin{equation}\tag{$\mathtt{EAIR}_{i}$}\label{eq:eair}
    \begin{aligned}
    J^{\Gamma}_{i}(T)\geq 0.
    \end{aligned}
\end{equation}
That is, the EAIR ensures that by participating in the IC market, each owner's expected (in ex-ante stage, i.e., before any preference is realized) privacy loss would be compensated enough by the expected payment to purchase their privacy in terms of non-zero expected payoff evaluated in the ex-ante stage.

An IT mechanism is \textit{ex-interim individual rational} (EIIR) if, for all $i\in\mathbb{I}$, $v_{i,t}\in V_{i,t}$, $h_{t}\in H_{t}$,
\begin{equation}\tag{$\mathtt{EIIR}_{i,T}$}\label{eq:eiir}
    \begin{aligned}
    J^{\Gamma}_{i,t}(v_{i,t}, h_{t}, T) -  J^{\Gamma}_{i,t}(v_{i,t}, h_{t}, t-1)\geq 0.
    \end{aligned}
\end{equation}
Here, the EIIR ensures that upon observing their each-period preference (but before reporting), each owner's expected current and the expected future privacy losses would be compensated enough by the expected payment in terms of non-zero ex-interim payoff to-go (including the expected current-period one-stage payoff).
%


Our notion of stopping decision is related to the notion of EIIR (i.e., (\ref{eq:eiir}), for all $i\in\mathbb{I}, t\in\mathbb{T}$).
However, there is a key difference between the two notions.
Specifically, the model with EIIR can be interpreted as enabling the owners to leave the market by checking whether their ex-interim expected payoff-to-go is non-negative.
The stopping decision, on the other hand, is an action available to each owner in addition to the actions of reporting.
Thus, as he can plan future reporting while making current reporting decision, each owner can also plan his stopping decision in future if he does not decide to leave at the end of the current period.
In the next section, we elaborate the optimal stopping decision for each owners while the incentive compatibility is guaranteed in the dynamic environment.

\subsection{Optimal Stopping}\label{sec:optimal_stopping_rule}

The payoff functions in Section \ref{sec:dynamic_bayesian_game} are defined when each owner $i$ neither chooses $\mathtt{TM}_{i,t}=1$ in each period $t$ nor plans to choose $\mathtt{TM}_{i,\tau}=1$ in period $t$ for any $\tau\in\mathbb{T}_{t+1}$.
When owners are allowed to leave at the end of each period, there is a weak decrease in the population size over time due to realizations of $\{\mathtt{TM}_{i,t} = 1\}$.
Furthermore, the planned future $\{\mathtt{TM}_{i,\tau}=1\}$, for $\tau\in\mathbb{T}_{t+1}$, requires each owner to predict future population change.
This is because owner $j$'s (planned) $\mathtt{TM}_{j,\tau_{j}}=1$, for $\tau_{j}\in\mathbb{T}_{t}$, $j\neq i$, would eliminate his expected generations of preferences after $\tau$; hence, the probability measure over future preferences as perceived by each owner $i$ would be different from the probability measure $P[\sigma, \chi]|v_{i,t},h_{t}$.
Such population change in the future is characterized by owner $i$'s estimation of other owners' current $\mathtt{TM}_{-i,t}=1$ and the planned $\{\mathtt{TM}_{j, \tau_{j}} = 1\}_{j\neq i}$, for $\tau_{j}\in \mathbb{T}_{t+1}$, $j\neq i$.
Let $\mathtt{pm}_{i,t}\equiv \{\mathtt{TM}_{j, \tau_{j}} = 1\}_{j\neq i}$, for $\tau_{j}\in \mathbb{T}_{t}$, $j\neq i$.
Additionally, let $\overline{\mathtt{pm}}\equiv\{\bar{\tau}_{i}\}_{i\in \mathbb{I}}$ denote the population change estimated in the ex-ante stage.
We denote the probability measures given $\mathtt{pm}_{i,t}$ and $\overline{\mathtt{pm}}$ as $P[\sigma, \chi]|v_{i,t},h_{t};\mathtt{pm}_{i,t}$ and $P[\sigma, \chi]|\overline{\mathtt{pm}}$, respectively.

We use the notion of \textit{population prediction model} (PPM) to describe how each owner $i$ estimates $\mathtt{pm}_{i,t}$, for all $i\in\mathbb{I}$, $t\in\mathbb{T}$.
Similar to the public history $h_{t}$, we assume that the buyer publicly discloses who leaves the market at the end of each period $t$ after all current decisions have been made.
As a result, each owner $i$ does not have to form posterior beliefs about the realized $\{\mathtt{TM}_{-i,s}\}_{s=0}^{t-1}$. Thus, his $\mathtt{pm}_{i,t}$ only predicts the unrealized $\{\mathtt{TM}_{-i,\tau}\}$, for $\tau\in\mathtt{T}_{t}$, of the remaining owners (i.e. who participate in period $t$).
The PPM includes an optimal stopping rule, denoted as $\phi^{\chi}\equiv\{\phi^{\chi}_{i}\}_{i\in\mathbb{I}}$, that governs each owner's choice of $\mathtt{TM}_{i,t} \in \{1,0\}$.
%
%
With a slight abuse of notation, let $J^{\Gamma}_{i}(\chi, \tau; \overline{\mathtt{pm}}_{i})$ and  $J^{\Gamma}_{i,t}(v_{i,t}, h_{t}, \tau; \chi, \mathtt{pm}_{i,t})$, respectively, denote owner $i$'s expected payoffs defined in (\ref{eq:owner_ex_ante_payoff}) and (\ref{eq:owner_interim_payoff}) by replacing $T$ by any $\tau\in\mathbb{T}$, in which the expectations are under the probability measures $P[\sigma, \chi]|\overline{\mathtt{pm}}$ and $P[\sigma, \chi]|v_{i,t},h_{t};\mathtt{pm}_{i,t}$.

For any reporting strategy profile $\chi$, owner $i$'s stopping rule is optimal if there exists $\overline{\mathtt{pm}}=\{\bar{\tau}_{i}\}_{i\in \mathbb{I}}$ with $\overline{\mathtt{pm}}_{i}=\{\bar{\tau}_{j}\}_{j\neq i }$ such that, for all $i\in\mathbb{I}$, $\bar{\tau}_{i}\in\overline{\mathtt{pm}}$,
\begin{equation}\tag{$\mathtt{OS}$}\label{eq:optimal_stopping_def}
    \begin{aligned}
    \sup_{\tau\in \mathbb{T}} J^{\Gamma }_{i}(\chi,\tau; \overline{\mathtt{pm}}_{i}) = J^{\Gamma }_{i}(\chi,\bar{\tau}_{i};\overline{\mathtt{pm}}_{i}).
    \end{aligned}
\end{equation}
Basically, a stopping rule is optimal if there exists a time horizon $\bar{\tau}_{i}$ such that owner $i$'s ex-ante expected payoff is maximized under a given mechanism $\Gamma$, owners' reporting strategy profile $\chi$, and other owners' $\bar{\tau}_{-i}$.
Here, the dependence of $J^{\Gamma }_{i}(\chi,\bar{\tau}_{i};\overline{\mathtt{pm}}_{i})$ on $\overline{\mathtt{pm}}_{i}$ is only through the probability measure $P[\sigma, \chi]|\overline{\mathtt{pm}} = P[\sigma, \chi]|\{\overline{\mathtt{pm}}_{i},\overline{\mathtt{pm}}_{-i} \}$.

One special case is when each owner $i$ is \textit{bounded rational} in that his PPM assumes that each $\bar{\tau}_{j} = T$ in $\mathtt{pm}_{i}$ and $\mathtt{pm}_{i,t}=\{\mathtt{TM}_{i,T}=1\}$, for all $i\in\mathbb{I}$, $t\in\mathbb{T}$.
In other words, each bounded rational owner's stopping decision is made by expecting that all other owners who participate in period $t$ would not leave until the final period $T$.
We refer to such PPM as bounded-rational PPM (BPM).


To characterize the optimal stopping problem (\ref{eq:optimal_stopping_def}), we introduce the \textit{value function} $U^{\Gamma}_{i,t}(\cdot;\chi,  \mathtt{pm}_{i,t}): V_{i,t} \mapsto \mathbb{R}$ as follows: for any $v_{i, t} \in V_{i,t}$, $h_{t}\in H_{t}$, $t\in\mathbb{T}$, $i\in \mathbb{I}$,
\begin{equation}\label{eq:value_optimal_stopping}
    \begin{aligned}
    U^{\Gamma}_{i, t}(v_{i,t}, h_{t};\chi,  \mathtt{pm}_{i,t})\equiv&\sup_{\tau\in\mathbb{T}_{t}}J^{\Gamma}_{i,t}(v_{i,t}, h_{t}, \tau; \chi, \mathtt{pm}_{i,t}).
    \end{aligned}
\end{equation}
Here, the value $U^{\Gamma}_{i, t}(v_{i,t}, h_{t};\chi,  \mathtt{pm}_{i,t})$ is owner $i$'s maximum period-$t$ ex-interim expected payoff given any $v_{i,t}$ and $h_{t}$.
Since the utility $z_{i,t}$ is bounded and the time horizon is finite, the value $U^{\Gamma}_{i, t}(\cdot; \chi, \mathtt{pm}_{i,t})$ is also bounded; i.e., $|U^{\Gamma}_{i, t}(v_{i,t}, h_{t};\chi,  \mathtt{pm}_{i,t})|< \infty$, for all $v_{i,t}\in V_{i,t}$ and $h_{t}\in H_{t}$.
%
%

\begin{lemma}[\cite{peskir2006optimal}]\label{lemma:optimal_stopping}
Fix any mechanism $\Gamma$ and any reporting strategy profile $\chi$.
The followings are true.
\begin{itemize}
    \item[(i)] The value function can be represented recursively as follows: for all $v_{i,t}\in V_{i,t}$, $h_{t}\in H_{t}$, $t\in\mathbb{T}$, $i\in\mathbb{I}$,
    \begin{equation}\label{eq:value_function_v2}
        \begin{aligned}
        &U^{\Gamma}_{i, t}(v_{i,t}, h_{t};\chi,  \mathtt{pm}_{i,t}) = \max\Big\{ J^{\Gamma}_{i,t}(v_{i,t}, h_{t}, t; \chi, \mathtt{pm}_{i,t}),\\
        & \mathbb{E}^{\sigma}_{\chi}\Big[ U^{\Gamma}_{i, t+1}(\tilde{v}_{i,t+1}, \tilde{h}_{t+1};\chi,  \mathtt{pm}_{i,t}) \Big]  \Big\}.
        \end{aligned}
    \end{equation}
    \item[(ii)] The optimal stopping rule $\phi^{\chi}_{i}$ is described as follows:
    \begin{equation}\label{eq:eq:optimal_stopping_v1}
        \begin{aligned}
        \phi^{\chi}_{i}:& \mathtt{TM}_{i,t} = 1, \forall t\in\mathbb{T}, \text{ if }\\
        & U^{\Gamma}_{i, t}(v_{i,t}, h_{t};\chi,  \mathtt{pm}_{i,t}) =  J^{\Gamma}_{i,t}(v_{i,t}, h_{t}, t; \chi, \mathtt{pm}_{i,t}).
        \end{aligned}
    \end{equation}
    %
\end{itemize}
\end{lemma}

Lemma \ref{lemma:optimal_stopping} shows the optimal stopping rule $\phi^{\chi}_{i}$ in terms of owner $i$'s value function and his ex-interim expected payoff.
Here, (\ref{eq:value_function_v2}) reformulates the value function $U^{\Gamma}_{i, t}$ as a Bellman equation.
The optimal stopping rule $\phi^{\chi}_{i}$ (\ref{eq:eq:optimal_stopping_v1}) is established based on the Bellman equation (\ref{eq:value_function_v2}) and suggests a stopping decision in period $t$ if owner $i$'s value equals his period-$t$ ex-interim expected payoff if he stops immediately in $t$.
Thus, the optimal stopping rule protects each owner from the risk of unbearable expected economic loss by continuing participating in the dynamic market.

To capture the role of the posted-price rule $\rho_{i}$ in affecting each owner $i$'s stopping decision, we define the following notion, for all $i\in\mathbb{I}$, $t\in\mathbb{T}$, $v_{i, t}\in V_{i,t}$, $h_{t}\in H_{t}$,
\begin{equation}\label{eq:G_term_stopping_1}
    \begin{aligned}
    &G^{\Gamma}_{i,t}(v_{i,t}, h_{t};  \chi, \mathtt{pm}_{i,t}) \\
        &\equiv \mathbb{E}^{\sigma}_{\chi}\Big[
        \sup_{\tau'\in \mathbb{T}_{t+1}} J^{\Gamma }_{i,t+1}(\tilde{v}_{i,t+1 },  \tilde{h}_{t+1},\tau'; \chi, \mathtt{pm}_{i,t})\Big] \\
        &-J^{\Gamma }_{i,t}(v_{i_{t}}, h_{t}, t; \chi, \mathtt{pm}_{i,t}) + \rho_{i}(t),
    \end{aligned}
\end{equation}
with $J^{\chi_{i}}_{i,T+1}(\cdot; \chi, \mathtt{pm}_{i,t})=0$.
Then, we can rewrite the optimal stopping rule $\phi^{\chi}_{i}$ defined in (\ref{eq:eq:optimal_stopping_v1}) as follows:
\begin{equation}\label{eq:eq:optimal_stopping_v2}
    \begin{aligned}
        \phi^{\chi}_{i}: & \mathtt{TM}_{i,t} = 1, \forall t\in\mathbb{T}, \text{ if }  G^{\Gamma}_{i,t}(v_{i,t}, h_{t};  \chi, \mathtt{pm}_{i,t}) \leq \rho_{i}(t).
    \end{aligned}
\end{equation}
%
We write $\phi^{*}_{i} = \phi^{\chi^{*}}_{i}$ when the reporting strategy profile $\chi^{*}$ is truthful.
Since $G^{\Gamma}_{i,t}$ is independent of $\rho_{i}(t)$, it is possible for the buyer to influence each owner's optimal stopping decision by proper design of $\rho_{i}$.
When $G^{\Gamma}_{i,t}(v_{i,t}, h_{t};  \chi, \mathtt{pm}_{i}) = \rho_{i}(t)$, owner $i$ is indifferent between continuing and stopping; we assume that, as is standard in mechanism design problems, the corresponding tie-breaking rule is in the buyer's favor.

\begin{remark}[PPM]
Fix a mechanism $\Gamma=<\sigma, \bm{\beta},\bm{\theta}, \bm{\rho}>$.
Let $N_{t}\subseteq \mathbb{I}$ denote the set of owners who participate in period $t$, for all $t\in\mathbb{T}$.
Given the instrumental kernels $K=\{K_{s}\}_{s\in\mathbb{T}_{t}}$ and history $h_{t}\in H_{t}$, owner $i$'s stopping decision $\mathtt{TM}_{i,s}$, for $s\in\mathbb{T}_{t}$, is made according to $\phi^{\chi}_{i}$ which depends on the correspondence $v_{i,t} \mapsto \mathtt{pm}_{i,t}$ that relays on the stopping rule profile $\{\phi^{\chi}_{i}\}_{i\in \mathbb{I}}$ of all the owners.
Since each $\phi^{\chi}_{j}$, for any $j\neq i$, depends on $v_{j,t}$, $h_{t}$, and $\mathtt{pm}_{j,t}$ in which $\mathtt{pm}_{j,t}$ includes owner $j$'s estimation of owner $i$'s $\mathtt{TM}_{i,s}$, denoted as $\mathtt{TM}^{(j)}_{i,s}$.
Here, owner $j$ knows that $\mathtt{TM}^{(j)}_{i,s}$ depends on owner $i$'s estimation of $\mathtt{TM}^{(i)}_{-i,t}$, and so on.
Let $\{\mathtt{TM}_{i,s_{i,t}}=1\}_{i\in N_{t}, s_{i,t}\in\mathbb{T}_{t}}$ denote the profile of the owners' period-$t$ stopping decisions, in which each $\mathtt{TM}_{i,s_{i,t}}=1$ is owner $i$'s (planned) stopping decision with $s_{i,t}\in \mathbb{T}_{t}$ is the smallest period in which $\mathtt{TM}_{i,s_{i,t}}=1$ is optimal.
Hence, the profile $\{s_{i,t} \}_{i\in N_{t}, s_{i,t}\in\mathbb{T}_{t}}\in \mathbb{T}^{|N_{t}|}_{t}$ (equivalently, $\{\mathtt{TM}_{i,s_{i,t}}=1\}_{i\in N_{t}, s_{i,t}\in\mathbb{T}_{t}}$) satisfies the optimality criterion of a pure strategy Bayesian Nash equilibrium (BNE) of a \textit{static finite game} in which each player $i$'s reward is given by
\begin{equation*}
    \begin{aligned}
    &R_{i,t}( s_{i,t},  \bm{s}_{-i,t}|v_{i,t}, h_{t};\chi) \\
        &\equiv \mathbb{E}^{\sigma}_{\chi}\Big[
         J^{\Gamma }_{i,t+1}(\tilde{v}_{i,t+1 },  \tilde{h}_{t+1},\tau'; \chi, \mathtt{TM}_{i, s_{i,t}},  \mathtt{TM}_{-i, s_{-i}})\Big] \\
        &-J^{\Gamma }_{i,t}(v_{i_{t}}, h_{t}, t; \chi, \mathtt{TM}_{i, s_{i}},  \mathtt{TM}_{-i, s_{-i,t}}) + \rho_{i}(t).
    \end{aligned}
\end{equation*}
Let $\mu_{i}(v_{i}, h_{t}, \chi)\in\Delta( \mathbb{T}^{|N_{t}|-1}_{t})$ represents owner $i$'s beliefs over $\bm{s}_{-i,t}$. Then, the profile $\bm{s}_{t} = \{s_{i,t}, \bm{s}_{-i,t}\}$ with the belief system $\{\mu_{i}(v_{i}, h_{t}, \chi)\}_{i\in N_{t}}$ satisfy the following optimality criterion of a pure strategy BNE: for all $i\in \mathbb{I}_{t}$, $t\in\mathbb{T}$, $s'_{i,t}\in \mathbb{T}_{t}$,
\begin{equation}\label{eq:BNE_stopping}
    \begin{aligned}
    &\mathbb{E}^{\mu_{i}}\Big[ R_{i,t}( s_{i,t},  \tilde{\bm{s}}_{-i,t}|v_{i,t}, h_{t};\chi) \Big| v_{i}, h_{t}, \chi\Big]\\
    &\geq \mathbb{E}^{\mu_{i}}\Big[ R_{i,t}( s'_{i,t},  \tilde{\bm{s}}_{-i,t}|v_{i,t}, h_{t};\chi) \Big| v_{i}, h_{t}, \chi\Big],
    \end{aligned}
\end{equation}
where the expectation operator $\mathbb{E}^{\mu_{i}}\Big[\cdot\Big|v_{i}, h_{t}, \chi \Big]$ takes expectation over $\bm{s}_{-i,t}$.
Unfortunately, pure strategy BNE of finite game (because $\mathbb{T}_{t}$ is finite) does not always exist \cite{fudenberg1991game}.
When there is no profile $\bm{s}_{t}$ satisfies (\ref{eq:BNE_stopping}), one possible tie-breaking rule is to make the belief $\mu_{i}(v_{i}, h_{t}, \chi)$ to set probability $1$ to the event $\mathtt{TM}_{-i,T} = 1$.
In the rest of the paper, we restrict attention to the case when the owners are bounded-rational; i.e., they use BPM. 
However, our design regime can be easily extended to when owners use PPM if there exists a profile $\bm{s}_{t}$ satisfies (\ref{eq:BNE_stopping}), for every $t\in\mathbb{T}$.
\end{remark}

\subsection{Dynamic Incentive Compatibility}\label{sec:DIC}

The incentive compatibility condition (\ref{eq:d_ex_interim_IC}) guarantees the optimality of truthful reporting in PBE when owners do not have option to stop in any period.
In this section, we define the incentive compatibility for the mechanism $\Gamma$ in PBE when each owner makes a coupled decision of reporting and stopping in each period.
For the ease of notation, we suppress $\overline{\mathtt{pm}}$ and $\mathtt{pm}_{i,t}$ in the notations of payoff functions.

In any period $t$, owner $i$'s incentive of how to report by choosing a strategy $\chi_{i,t}$ depends on how much ex-interim expected payoff he can obtain by using $\chi_{i,t}$ to report his true preference.
With the Markovian instrumentalness, the probability measure of the owners' future preferences perceived by each owner $i$ depends on owner $i$'s current preference, $v_{i,t}$, public history $h_{t}$, current report $\hat{v}_{i,t}$, and the reporting strategy profile $\{\chi_{s}\}_{s\in\mathbb{T}_{t}}$; its dependence on past true preferences $v^{t}_{i}$ and past privacy allocations $\epsilon^{t}$ is only through current $(v_{i,t}, \epsilon_{t})$. 
As a result, if owner $i$ is incentivized to report truthfully when he has reported truthfully in all past periods, then he is also incentivized to report truthfully even when he has misreported in the past.
Each owner's incentive to report truthfully in each period is guaranteed by \textit{dynamic incentive compatibility} in PBE (DIC).

\begin{definition}[DIC]\label{def:DIC}
A market model $\Gamma=<\sigma, \bm{\beta}, \bm{\theta}, \bm{\rho}>$ is PDIC if, truthful reporting is best response of each owner $i$ to all other owners' truthful reporting in every period. That is, for all $i\in\mathcal{N}$, $t\in\mathbb{T}$, $v_{i,t}\in V_{i,t}$, any truthful history $h_{t}\in H_{t}$, and any reporting strategy $\chi_{i}=\{\chi^{*}_{i;0:t-1}, \chi_{i;t:T}\}$ with truthful $\chi^{*}_{i;0:t-1}$ and arbitrary $\chi_{i;t:T}$,
\begin{equation}\label{eq:DIC_original}
    \begin{aligned}
   \max\Big\{ J^{\Gamma}_{i,t}(v_{i,t}&, h_{t}, t),\mathbb{E}^{\sigma}\Big[ U^{\Gamma}_{i, t+1}(\tilde{v}_{i,t+1}, \tilde{h}_{t+1}) \Big]  \Big\}\\
   \geq \max\Big\{ J^{\Gamma}_{i,t}(v_{i,t}&, h_{t}, t; \chi_{i}),\\
   &\mathbb{E}^{\sigma}_{\chi_{i}}\Big[ U^{\Gamma}_{i, t+1}(\tilde{v}_{i,t+1}, \tilde{h}_{t+1};\chi_{i}) \Big]  \Big\},
    \end{aligned}
\end{equation}
with $J^{\Gamma}_{i,T}(v_{i,T}, h_{T}, T)\geq J^{\Gamma}_{i,T}(v_{i,T}, h_{T}, T; \chi_{i})$.
\end{definition}


The DIC condition (\ref{eq:DIC_original}) is a PBE in which every owner report truthfully while believing with probability $1$ that all other owners report truthfully.
The equilibrium also implies that the stopping rule $\phi_{i} = \phi^{\chi^{*}}_{i}$ defined in (\ref{eq:eq:optimal_stopping_v1}) is optimal.

Define owner $i$'s period-$t$ one-shot deviation strategy as $\chi^{[t]}_{i}\equiv\{\chi^{*}_{i;0:t-1}, \chi^{[t]}_{i,t},  \chi^{*}_{i;t+1:T} \}$, such that $\chi^{[t]}_{i}$ is truthful except period-$t$ $\chi^{[t]}_{i,t}$.
We let $\hat{v}_{i,t}$ to denote the generic report of owner $i$ in period $t$ using $\chi^{[t]}_{i}$ (i.e., $\hat{v}_{i,t} = \chi^{[t]}_{i,t}(v_{i,t}, h_{t})$).
For the ease of notation, we replace the reporting strategy $\chi^{[t]}_{i}$ by $\hat{v}_{i,t}$ in notations (e.g., $\mathbb{E}^{\sigma}_{\hat{v}_{i,t} } = \mathbb{E}^{\sigma}_{\chi^{[t]}_{i} }$); unless otherwise stated.

\begin{proposition}\label{prop:one_shot_deviation_principle}
The market model $<\sigma,\bm{\beta},$ $\bm{\theta}, \bm{\rho}>$ with exist option is DIC with belief $\bm{\mu}$ if and only if, for any $v^{t}_{i}\in V^{t}_{i}$, $\epsilon^{t-1}\in\mathcal{E}^{t-1}$, $t\in\mathbb{T}$, $i\in\mathbb{I}$,
\begin{equation}\tag{$\mathtt{DIC}_{i,t}$}\label{eq:DIC_1_shot}
    \begin{aligned}
   \max\Big\{ J^{\Gamma}_{i,t}(v_{i,t}&, h_{t}, t),\mathbb{E}^{\sigma}\Big[ U^{\Gamma}_{i, t+1}(\tilde{v}_{i,t+1}, \tilde{h}_{t+1}) \Big]  \Big\}\\
   \geq \max\Big\{ J^{\Gamma}_{i,t}(v_{i,t}&, h_{t}, t; \hat{v}_{i,t}),\\
   &\mathbb{E}^{\sigma}_{\chi_{i} }\Big[ U^{\Gamma}_{i, t+1}(\tilde{v}_{i,t+1}, \tilde{h}_{t+1};\hat{v}_{i,t}) \Big]  \Big\},
    \end{aligned}
\end{equation}
with $J^{\Gamma}_{i,T}(v_{i,T}, h_{T}, T)\geq J^{\Gamma}_{i,T}(v_{i,T}, h_{T}, T; \hat{v}_{i, T})$.
%
\end{proposition}

\begin{proof}
See Appendix \ref{app:prop:one_shot_deviation_principle}.
\end{proof}

Proposition \ref{prop:one_shot_deviation_principle} establishes a \textit{one-shot deviation principle} (see, e.g., \cite{blackwell1965discounted}) for the dynamic market model that implies the subgame perfectness of the PBE.
This enables the buyer to restrict attention on the characterizations of DIC when each owner $i$ may deviate from truthful reporting by using any one-shot deviation strategy $\chi^{[t]}_{i}$ for every $i\in\mathbb{I}$, $t\in\mathbb{T}$, while the optimality stopping decision is maintained. 
%
%
Define
\begin{equation}\label{eq:G_term_rewritten_0}
    \begin{aligned}
    &g^{\Gamma}_{i,t}(v_{i,t}, h_{t}, \tau;  \chi) \equiv \mathbb{E}^{\sigma}_{\chi}\Big[
         J^{\Gamma }_{i,t+1}(\tilde{v}_{i,t+1 },  \tilde{h}_{t+1},\tau; \chi)\Big] \\
        &-J^{\Gamma }_{i,t}(v_{i_{t}}, h_{t}, t; \chi) + \rho_{i}(t).
    \end{aligned}
\end{equation}
Then, $G^{\Gamma}_{i,t}(v_{i,t}, h_{t};  \chi) = \sup\limits_{\tau\in \mathbb{T}_{t+1}}$ $g^{\Gamma}_{i,t}(v_{i,t}, h_{t}, \tau;  \chi)$.
The stopping rule $\phi_{i}$ defined in (\ref{eq:eq:optimal_stopping_v1}) is optimal for any given reporting strategy $\chi_{i}$.
However, the realization of $\mathtt{TM}_{i,t}=1$ according to $\phi^{\chi_{i}}_{i}$ depends on his current preference $v_{i,t}$, public history $h_{t}$, his current reporting strategy $\chi_{i,t}$, and planned $\chi_{i;t+1:T}$.
Let 
\begin{equation*}
    \begin{aligned}
    \hat{\tau}_{i}=\inf\Big\{\tau_{i}\in\arg\sup\limits_{\tau\in\mathbb{T}_{t}} g^{\Gamma}_{i,t}(v_{i,t}, h_{t}, \tau;  \chi)  \Big\}.
    \end{aligned}
\end{equation*}
Then, owner $i$ with $\chi_{i}$ is optimal to stop in $t$ if $\hat{\tau}_{i} = t$.
Let $\tau^{*}_{i}$ be such $\hat{\tau}_{i}$ if $\chi_{i}$ is truthful.
Proposition \ref{prop:one_shot_deviation_principle} implies that \textit{(i)} when $\tau^{*}_{i}=t$, owner $i$ has no incentive to use $\chi^{[t]}_{i}$ and stop in $t$, use $\chi^{[t]}_{i}$ and continue, or use $\chi^{*}_{i}$ to report truthfully and continue; \textit{(ii)} when $\tau^{*}_{i}>t$, owner $i$ has no incentives to use $\chi^{[t]}_{i}$ and stop in $t$ or use $\chi^{[t]}_{i}$ and continue.

\subsection{The Buyer's Mechanism Design Problem}\label{sec:buyer_mechanism_design_problem}

The privacy-utility tradeoff implies that the buyer suffers losses of utility that she can extract from the data by providing differential privacy.
In this work, we do not consider the effects of the reduction of data on the buyer's utility due to owners' stopping decisions and restrict attention to the utility loss caused only by differential privacy protection.
For any $\epsilon_{t}\in \mathcal{E}$, $t\in \mathbb{T}$, define the buyer's \textit{utility loss} as:
\begin{equation}
    a_{t}(\epsilon_{t})\equiv L\exp(-\epsilon_{t}),
\end{equation}
where $L\in \mathbb{R}_{++}$ represents the maximum utility loss when $\epsilon_{t}\rightarrow 0$.
The formulation of the buyer's utility loss and each owner's privacy loss in (\ref{eq:flow_loss_0}) captures the tradeoff between the owners' privacy and the buyer's utility extracted from the data: the larger (resp. smaller) $\epsilon_{t}$ becomes, the less (resp. more) private the owners' data is but the more (resp. less) utility the buyer can get from the data.

As the market designer, the buyer aims to determine and commit a mechanism $\Gamma=<\sigma, \bm{\beta}, \bm{\theta}, \bm{\rho}>$ and then publicly releases it as a take-it-or-leave-it offer to the owners in the ex-ante stage.
The buyer's mechanism design problem takes into account the expected population dynamics due to owners' stopping rule (based on BPM) evaluated in the ex-ante stage.
Let $\bm{\tau}\equiv\{\bar{\tau}_{i}\}_{i\in\mathbb{I}}\in \mathbb{T}^{n}$ in which each $\bar{\tau}_{i}$ is the \textit{expected stopping time} of owner $i$ with BPM; i.e., $\bar{\tau}_{i}$ satisfies (\ref{eq:optimal_stopping_def}) when owners use BPM.
Let $\bm{N}\equiv \{N_{t}\}_{t\in \mathbb{T}}$ denote the sequence of expected population sets given $\bm{\tau}$ such that $N_{t}\subseteq \mathbb{I}$ is the set of owners (expectedly) participating in period $t$ by excluding those with $\bar{\tau}_{j}<t$.
The buyer's \textit{ex-ante expected cost} is given by,
\begin{equation}\label{eq:buyer_objective}
    \begin{aligned}
    &C^{\Gamma}(\bar{\bm{\tau}})\equiv \mathbb{E}^{\sigma}\Big[ \sum\limits_{t\in \mathbb{T}\backslash\bar{\bm{\tau}}  }\Big( a_{t}\big(  \sigma_{t}( \tilde{\bm{v}}_{t}) \big) + \sum_{i_{t}\in N_{t} }\beta_{i_{t}}(\tilde{\bm{v}}_{t}) \Big) \\
    & + \sum\limits_{\tau\in \bar{\bm{\tau}} } \Big(a_{\tau}\big(  \sigma_{\tau}( \tilde{\bm{v}}_{\tau}) \big) + \sum_{i_{\tau} \in N_{\tau}} \big(\theta_{i_{\tau}}(\tilde{\bm{v}}_{i_{\tau}}) + \rho_{i_{\tau}}(\tau)\big) \Big)\Big].
    \end{aligned}
\end{equation}
%
%
%
Here, the right-hand side of (\ref{eq:buyer_objective}) in the first line captures the expected total cost of the buyer when no owner is expected to leave in each of the periods in $\mathbb{T}\backslash \bar{\bm{\tau}}$;
the terms in the second line captures the expected total cost of the buyer induced by the owners who are expected to leave in each of the periods in $\bar{\bm{\tau}}$.
The buyer makes ex-ante commitment by determining a mechanism $\Gamma = <\sigma, \bm{\beta}, \bm{\theta}, \bm{\rho}>$ that solves the following constrained optimization problem:
\begin{equation}\label{eq:principal_original_design}
    \begin{split}
        \min_{\sigma, \bm{\beta}, \bm{\theta}, \bm{\rho} } \;\; C^{\sigma, \bm{\beta}, \bm{\theta}, \bm{\rho}}(\bar{\bm{\tau}};\bm{N}),\;\;
        \text{s.t. DIC, IR.} 
    \end{split}
\end{equation}
Here, the dynamic incentive compatibility (DIC) condition is given by \ref{eq:DIC_1_shot}, for all $i\in\mathbb{I}$, $t\in \mathbb{T}$, which guarantee the truthful reporting.
The individual rationality (IR) constraint is imposed to guarantee that each owner $i$ has incentive to participate in the market. 
In this work, we focus on ex-ante individual rationality.
This is captured by a non-negative ex-ante expected payoff. That is, for all $i\in\mathbb{I}$,
\begin{equation}\tag{$\mathtt{IR}_{i}$}\label{eq:ex_ante_IR}
    \begin{aligned}
    & J^{\Gamma}_{i}(\bar{\tau}_{i}) \equiv \mathbf{E}^{\sigma}_{\chi}\Big[ J^{\Gamma}_{i,0}(\tilde{v}_{i,0}, \tau)  \Big] \geq 0.
    \end{aligned}
\end{equation}

The DIC as a PBE induced by the market model $\Gamma$ requires a strong rationality of owners: each owner $i$ is rational in the sense that \textit{(i)} he maintains correct beliefs about all that is unknown (but payoff-relevant) to him and \textit{(ii)} he can accurately forecast and estimate how other owners will respond to any decisions he would make in each period.
Likewise, the buyer's mechanism design problem also requires her strong rationality to adopt accurate beliefs in regard to the dynamics of the environment (i.e., how the instrumentalness drives the dynamics of the owners' preferences and the expected population dynamics due to owners' stopping decisions based on BPM) and to the decision makings of each owner including the owner's beliefs about others.
The theoretical characterizations in this work are based on these assumptions of strong rationality and the owners use BPM in their stopping decisions.

\section{Characterization of DIC}\label{sec:characterization}

In this section, we characterize the DIC of our dynamic data market by providing formulations of the monetary transfer rules (i.e., the compensation rules and the stopping payment rule) in terms of the assignment rule and the sufficient and the necessary conditions for DIC. 
The following assumption holds for this section.
\begin{assumption}\label{assp:full_support}
The probability density $f_{i,t+1}(v_{i,t}|$ $v_{i, t-1}$ $, \epsilon_{t-1})>0$ with $f_{i,0}(v_{i,0})>0$ for all $v_{i,t}\in V_{i,t}$, $v_{i, t-1}\in V_{i,t-1}$, $\epsilon_{t-1}\in \mathcal{E}$, $t\in\mathbb{T}\backslash\{0\}$.
\end{assumption}
Assumption \ref{assp:full_support} considers a full support environment, in which each of owner $i$'s instrumental preferences has a strictly positive probability to occur at every period.

Given any DIC market model $<\sigma, \bm{\beta}, \bm{\theta}, \bm{\rho}>$, truthful reporting strategy $\chi^{*}_{i}$ is optimal for each owner $i$.
For the simplicity, with a slight abuse of notation, let $J^{\Gamma}_{i,t}(v_{i,t}, \hat{v}_{i,t}, h_{t}, \tau)$ denote owner $i$'s period-$t$ ex-interim payoff function when he uses a one-shot deviation strategy to report $\hat{v}_{i,t}$ of his true preference $v_{i,t}$.
%
%
When owner $i$ reports truthfully, we suppress the report; i.e.,  $J^{\Gamma}_{i,t}(v_{i,t},  h_{t}, \tau) = J^{\Gamma}_{i,t}(v_{i,t}, \hat{v}_{i,t}, h_{t}, \tau)$.
Then, we have the following lemma based on the envelope theorem (see, e.g., \cite{milgrom2002envelope,pavan2014dynamic}).

\begin{lemma}\label{lemma:envelope_necessary_condition}
Suppose Assumption \ref{assp:full_support} holds. 
Then, in any DIC market model $\Gamma=<\sigma, \bm{\beta}, \bm{\theta}, \bm{\rho}>$, we have, for all $i\in\mathbb{I}$, $t\in \mathbb{T}$, $\tau\in \mathbb{T}_{t}$, $v_{i,t}\in V_{i,t}$, $h_{t}\in H_{t}$, 
\begin{equation}\label{eq:lemma_envelope}
    \begin{aligned}
    &\frac{\partial J^{\Gamma}_{i,t}(x,  h_{t}, \tau) }{\partial  x}\Big|_{x = v_{i,t}}  \\
    &\equiv \mathbb{E}^{\sigma}\Big[  \sum\limits_{s=t}^{\tau}\big(1 - \exp(\sigma_{s}(\tilde{\bm{v}}_{s}))\mathcal{G}^{s}_{t}(\tilde{v}^{s}_{i,t}|\sigma) \big)\Big] \text{ where}
\end{aligned} 
\end{equation}
$\mathcal{G}^{s}_{t}(\tilde{v}^{s}_{i,t}|\sigma)\equiv\prod_{k=t}^{s} \frac{\partial}{\partial x} K_{i,t}(x, \tilde{\epsilon}^{k-1}; \tilde{c}_{i,k-1})\Big|_{x =\tilde{v}_{i,k-1}}$.
\end{lemma}
%
%
\begin{proof}
See Appendix \ref{app:lemma:envelope_necessary_condition}.
\end{proof}

The following corollary follows the Kolmogorov's Existence Theorem \cite{billingsley2008probability}.
\begin{corollary}\label{corollary:kolmogorov}
If each owner $i$'s intrinsic preference is uniformly distributed over $(0,1)$, then the term $\mathcal{G}^{s}_{t}(\cdot|\sigma)$ becomes:
\begin{equation*}
    \begin{aligned}
    &\mathcal{G}^{s}_{t}(\tilde{\bm{v}}^{s}_{t}|\sigma) 
    =  \prod_{k=t+1}^{s}\frac{-\partial F_{i,k}(x| \tilde{v}_{i, k-1}, \tilde{\epsilon}^{k-1} ) }{f_{i, k }(\tilde{v}_{i, k}| \tilde{v}_{i, k-1}, \tilde{\epsilon}^{k-1}) \partial x}\\
    & \Big|_{x=\psi_{i, k-1}(v_{i, k-2}, \epsilon^{k-2}|\tilde{c}_{i, k-1})},
    \end{aligned}
\end{equation*}
where $\psi_{i,t}(v_{i, t-1}, \epsilon^{t-1}|c_{i,t}) = $ $\inf\{v_{i, t}: F_{i, t}(v_{i, t}$ $|v_{i, t-1},$ $\epsilon^{t-1})$ $\geq c_{i,t}\}$.
\end{corollary}
\begin{proof}
See Appendix \ref{app:lemma:envelope_necessary_condition}.
\end{proof}

Lemma \ref{lemma:envelope_necessary_condition} provides a first-order necessary condition for the optimality of each owner's truthful reporting strategy.
Since $\epsilon_{t}>0$, the term $1-\exp(\epsilon_{t})<0$, for all $\epsilon_{t}$, $t\in\mathbb{T}$.
Then, the monotonicity of $J^{\Gamma}_{i,t}$ with respect to owner $i$'s preference is determined by the sign of $\mathcal{G}^{s}_{t}(\cdot|\sigma)$.
Corollary \ref{corollary:kolmogorov} shows an alternative representation of the first-order condition in terms of the cumulative distribution (cdf) the probability density (pdf) functions associated with the instrumentalness when the owners' intrinsic preference are drawn independently from a uniform distribution.
Consider the following assumption regarding the cdf.

\begin{assumption}\label{assp:monotone_transistions}
For all $i\in\mathcal{I}$, $t\in \mathbb{T}\backslash{\{T\}}$,  $v'_{i, t}\geq v_{i, t}\in V_{i,t}$, $v_{i, t+1}\in V_{i,t+1}$, $\epsilon^{t}\in\mathcal{E}^{t}$, 
\begin{equation}\label{eq:1st_order_dom}
    F_{i_{t+1}}(v_{i_{t+1}}| v'_{i_{t}}, \epsilon^{t}) \leq  F_{i_{t+1}}(v_{i_{t+1}}| v_{i_{t}}, \epsilon^{t}).
\end{equation}
\end{assumption}
Assumption \ref{assp:monotone_transistions} imposes a monotonicity condition to the probability distribution function of each owner in the sense of first-order stochastic dominance. That is, higher preference in current period $t$ leads to a higher preference in the next period $t+1$ probabilistically, given the same $\epsilon^{t}$.
In other words, Assumption \ref{assp:monotone_transistions} assumes that owners who value their privacy more in current period will most probably continue to value their privacy in the next period more than other owners with a relatively lower valuation of privacy in the current period.

The following lemma formally states the monotonicity of $J^{\Gamma}_{i, t}$.

\begin{lemma}\label{lemma:monotonicity_J_1}
Suppose Assumptions \ref{assp:monotone_transistions} and \ref{eq:1st_order_dom} hold. Then, in any DIC market model $\Gamma=<\sigma, \bm{\beta}, \bm{\theta}, \bm{\rho}>$, $J^{\Gamma}_{i,t}(v_{i, t}, h_{t}, \tau)$ is weakly decreasing in $v_{i, t}$ for all $i\in\mathbb{I}$, $t\in\mathbb{T}$, $\tau\in \mathbb{T}_{t}$, $h_{t}\in H_{t}$.
\end{lemma}
\begin{proof}
See Appendix \ref{app:lemma:monotonicity_J_1}.
\end{proof}

Lemma \ref{lemma:monotonicity_J_1} shows that increasing an owner's preference over privacy in any period decreases his ex-interim expected payoff.
In other words, owners who care more about privacy (i.e. with higher preference) incline to stop than owners who care less about their privacy (i.e., with lower preference).

Define, with a slight abuse of notation, 
\begin{equation*}\label{eq:G_term_rewritten_1}
    \begin{aligned}
    G^{\Gamma}_{i,t}(v_{i,t}, h_{t}, \tau) \equiv \sup\limits_{\tau'\in \mathbb{T}_{t+1}} g^{\Gamma}_{i,t}(v_{i,t}, h_{t}, \tau'),
    \end{aligned}
\end{equation*}
where $g^{\Gamma}_{i,t}$ is given by (\ref{eq:G_term_rewritten_1}) and $\tau = \inf\big\{\arg\sup\limits_{\tau'\in \mathbb{T}_{t+1}}$ $g^{\Gamma}_{i,t}(v_{i,t}, h_{t}, \tau') \big\}$.
By Lemmas \ref{lemma:envelope_necessary_condition} and \ref{lemma:monotonicity_J_1}, we have that in any DIC market, $G^{\Gamma}_{i,t}(v_{i,t}, h_{t}, \tau)$ is non-increasing. 
%
%
Hence, it is straightforward to see that the term $G^{\Gamma}_{i,t}(v_{i,t}, h_{t}, \tau)$ in (\ref{eq:G_term_stopping_1}) is also non-increasing.
Based on the stopping rule $\phi^{\chi}_{i}$ in (\ref{eq:eq:optimal_stopping_v2}), we define the \textit{stopping region} as follows: for all $i\in\mathbb{I}$, $t\in\mathbb{T}$, $h_{t}\in H_{t}$,
\begin{equation}\label{eq:stopping_region_original_1}
    \begin{aligned}
    \mathcal{R}^{\Gamma}_{i, t} \equiv \big\{v_{i_{t}}\in V_{i}: G^{\Gamma}_{i,t}(v_{i,t}, h_{t}) \leq \rho_{i}(t)) \big\}.
    \end{aligned}
\end{equation}
Define the \textit{indifference region} of the stopping region $\mathcal{R}^{\chi_{i}}_{i_{t}}$ is given as,
\begin{equation*}
    \mathcal{S}^{\Gamma}_{i,t}\equiv \big\{ v_{i_{t}}\in V_{i}: G^{\Gamma}_{i,t}(v_{i,t}, h_{t}) = \rho_{i}(t)) \big\}.
\end{equation*}
Specifically, owner $i$ with a preference $v_{i,t}\in \mathcal{S}^{\Gamma}_{i,t}$ is indifferent between stopping and continuing; as is standard, we assume that the tie-breaking rule is in the buyer's favor.

\begin{proposition}\label{prop:threshold_rule_optimal_stopping}
Suppose Assumptions \ref{assp:full_support} and \ref{assp:monotone_transistions} hold.
%
%
In any DIC market $\Gamma=< \sigma, \bm{\beta}, \bm{\theta}, \bm{\rho}>$, the optimal stopping rule $\phi_{i}$ given by (\ref{eq:eq:optimal_stopping_v2}) is a threshold rule. That is, there exists a unique interval $[\kappa^{l}_{i}(t), \kappa^{r}_{i}(t)] = \mathcal{S}^{\Gamma}_{i,t}$ with $\kappa^{l}_{i}(t) \leq  \kappa^{r}_{i}(t)$ and $\kappa^{l}_{i}(T) =  \kappa^{r}_{i}(T) = \underline{v}_{i,t}$, such that, the stopping region $\mathcal{R}^{\Gamma}_{i, t}$ in (\ref{eq:stopping_region_original_1}) is equivalent to 
\begin{equation*}
    \mathcal{R}^{\Gamma}_{i, t} = \big\{v_{i,t}\in V_{i,t}: v_{i,t} \geq \kappa^{\ell}_{i}(t) \big\}.
\end{equation*}
We refer to $\kappa^{\ell}_{i}:\mathbb{T}\mapsto V_{i,t}$ as the threshold function of owner $i$, for all $i\in\mathbb{I}$.
\end{proposition}
\begin{proof}
See Appendix \ref{app:prop:threshold_rule_optimal_stopping}.
\end{proof}

With the threshold optimal stopping rule (threshold rule), it is optimal for each owner $i$ to stop in $t$ when his preference $v_{i,t}\geq \kappa^{\ell}_{i}(t)$, for all $i\in\mathbb{I}$, $t\in\mathbb{T}$.

Define, 
\begin{equation}\label{eq:J_without_beta}
    \begin{aligned}
    &\bar{J}^{\Gamma}_{i, t}(v_{i,t}, \hat{v}_{i,t},h_{t}, \tau)\\
    &\equiv  \mathbb{E}^{ \bm{\kappa}_{-i,t} }\Big[-\ell(v_{i,t}, \tilde{\epsilon}_{t})\Big] + \mathbb{E}^{\sigma}_{\hat{v}_{i,t}}\Big[ \sum\limits_{s=t+1}^{\tau} z_{i,s}(\tilde{v}_{i,t}, \tilde{\epsilon}_{s}, \tilde{p}_{i,s} ) \Big],
    \end{aligned}
\end{equation}
%
with $\sum^{\tau}_{s=t+1}z_{i,s}(\cdot) = 0$ when $\tau=t$, where the first expectation is taken over other owners' contemporaneous preferences.
Hence, $\bar{J}^{\Gamma}_{i, t}$ $(v_{i,t}, \hat{v}_{i,t},h_{t}, \tau)$ is owner $i$'s period-$t$ ex-interim expected payoff-to-go without period-$t$ payment when owner $i$ uses one-shot deviation reporting strategy to report $\hat{v}_{i,t}$ of $v_{i,t}$.
Next, we introduce the notion of \textit{distance}, denoted by $d^{S}$ and $d^{-S}$ where the superscripts $S$ and $-S$ refer to \textit{stop} and \textit{non-stop}, respectively.
Specifically, for all $i\in\mathbb{I}$, $t\in\mathbb{T}$, $v_{i,t}, \hat{v}_{i,t}\in V_{i,t}$,
\begin{equation}\label{eq:distance_S_1}
    \begin{aligned}
    &d^{S}_{i,t}(\hat{v}, v_{i,t})\equiv\bar{J}^{\Gamma}_{i, t}(\hat{v}_{i,t}, \hat{v}_{i,t},h_{t}, t) -\bar{J}^{\Gamma}_{i, t}(v_{i,t}, \hat{v}_{i,t},h_{t}, t).
    \end{aligned}
\end{equation}
and, for any $\tau\in\mathbb{T}_{t}$,
\begin{equation}\label{eq:distance_S_2}
    \begin{aligned}
    &d^{-S}_{i,t}(\hat{v}, v_{i,t}; \tau) \equiv \bar{J}^{\Gamma}_{i, t}(\hat{v}_{i,t}, \hat{v}_{i,t},h_{t}, \tau) -\bar{J}^{\Gamma}_{i, t}(v_{i,t}, \hat{v}_{i,t},h_{t}, \tau).
    \end{aligned}
\end{equation}
Let 
\begin{equation}\label{eq:integral_envelope}
    \begin{aligned}
    &\Lambda^{\sigma}_{i,t}(v_{i,t}, v'_{i,t};\tau)\\
    &\equiv \int_{v'_{i,t}}^{v_{i,t}} \mathbb{E}^{\sigma}\Big[ \sum\limits_{s=t}^{\tau}\big(1 - \exp(\sigma_{s}(\tilde{\bm{v}}_{s}))\mathcal{G}^{s}_{t}(\tilde{v}^{s}_{i,t}|\sigma) \big)\Big| x, h_{t}\Big] dx.
    \end{aligned}
\end{equation}
%
%
    
%
The integrand of (\ref{eq:integral_envelope}) is from the envelope condition in (\ref{eq:lemma_envelope}).
Then, we have the following theorem.

\begin{theorem}\label{theorem:sufficient_condition}
Suppose Assumptions \ref{assp:full_support} and \ref{assp:monotone_transistions} hold.
In any DIC market, the following statements hold.
\begin{itemize}
    \item[(i)] The compensation rule $\bm{\beta}$ can be represented in terms of the assignment rule $\sigma_{t}$, i.e.,
    \begin{equation}\label{eq:thm_beta}
        \begin{aligned}
            &\beta_{i, t}(\bm{v}_{t}) = \sup_{\tau\in\mathbb{T}_{t}} \Lambda^{\sigma}_{i,t}(v_{i,t}, \bar{v}_{i,t};\tau )
            - \mathbb{E}_{t}^{\sigma}\Big[\\
            &\sup_{\tau\in\mathbb{T}_{t+1}} \Lambda^{\sigma}_{i,t}(\tilde{v}_{i,t+1}, \bar{v}_{i,t+1};\tau )\Big]+\ell(v_{i,t}, \sigma_{t}(v_{i,t}, \bm{v}_{-i, t})),
        \end{aligned}
    \end{equation}
    where $\bar{v}_{i,k}\in V_{i,k}$ is the maximum preference in period $k$, for all $k\in\mathbb{T}$.
    \item[(ii)] When owner $i$ decides to stop at $t$, the rules $\theta_{i,t}$ and $\rho_{i}$, respectively, are given in terms of $\sigma$ as follows:
    \begin{equation}\label{eq:thm_theta}
        \theta_{i, t}(\bm{v}_{t}) = \Lambda^{\sigma}_{i,t}(v_{i, t}, \bar{v}_{i,t},\tau) +\ell(v_{i, t}, \sigma_{t}(v_{i, t}, \bm{v}_{-i, t})),
    \end{equation}
    \begin{equation}\label{eq:thm_rho}
        \begin{aligned}
        &\rho_{i}(t) = \mathbb{E}^{\sigma}\Big[\sum\limits_{s=t}^{T-1}\Big(- \Lambda^{\sigma}_{i,s}(\tilde{v}_{i, s}\wedge \kappa^{\ell}_{i}(s), s;v'_{i,s})       \\
        & +\Lambda^{\sigma}_{i,s+1}(\tilde{v}_{i, s+1}\wedge \kappa^{\ell}_{i}(s+1), \bar{v}_{i,s+1};s+1)     \Big) \\
        &- \Big( \sum\limits_{\tau'\in \mathbb{T}_{s+1}}\Lambda^{\sigma}_{i,s+1}(\tilde{v}_{i, s+1}\wedge \kappa^{\ell}_{i}(s+1), \bar{v}_{i,s+1};\tau') \\
        &-\sum\limits_{\tau'\in \mathbb{T}_{s}}\Lambda^{\sigma}_{i,s}(\tilde{v}_{i, s}\wedge \kappa^{\ell}_{i}(s), \bar{v}_{i,s};\tau')\Big)   \Big|\kappa^{\ell}_{i}(t), \emptyset\Big].
        \end{aligned}
    \end{equation}
    \item[(iii)] The privacy allocation rule $\sigma$ satisfies the following conditions:
    \begin{equation}\label{eq:thm_sigma_C1}
        \Lambda^{\sigma}_{i,t}(\hat{v}_{i, t}, v_{i, t}; t)  \leq d^{S}_{i, t}(\hat{v}_{i, t}, v_{i, t}),
    \end{equation}
    \begin{equation}\label{eq:thm_sigma_C2}
        \begin{aligned}
        \sup&{}_{\tau\in\mathbb{T}_{t}} \Lambda^{\sigma}_{i,t}(\hat{v}_{i, t}, \bar{v}_{i,t}; \tau) - \sup_{\tau\in\mathbb{T}_{t}} \Lambda^{\sigma}_{i,t}(v_{i, t}, \bar{v}_{i,t}; \tau)\\
        \leq & \inf_{\tau\in \mathbb{T}_{t}}\Big\{d^{-S}_{i, t}(\hat{v}_{i, t}, v_{i, t};\tau) \Big\} - \sup_{\tau\in\mathbb{T}_{t}}\rho_{i}(\tau).
        \end{aligned}
    \end{equation}
\end{itemize}
\end{theorem}
\begin{proof}
See Appendix \ref{app:theorem:sufficient_condition}.
\end{proof}

Theorem \ref{theorem:sufficient_condition} establishes a design regime for DIC market model.
Specifically, (\ref{eq:thm_beta}) and (\ref{eq:thm_theta}) give the designs of preference-related payment rules $\bm{\beta}$ and $\bm{\theta}$ in terms of the privacy allocation rule $\sigma$, respectively, while (\ref{eq:thm_rho}) constructs the preference-independent posted-price payment rule $\bm{\rho}$ in terms of $\sigma$.
Given the constructions (\ref{eq:thm_beta})-(\ref{eq:thm_rho}), the conditions (\ref{eq:thm_sigma_C1}) and (\ref{eq:thm_sigma_C2}) constitute a sufficient condition for DIC.
Here, $\Lambda^{\sigma}_{i,t}(v_{i, t}, \bar{v}_{i,t};\tau)$ is the \textit{information rent} for any $\tau\in\mathbb{T}_{t}$ of owner $i$ with current preference $v_{i,t}$ that captures the payoff he can expect by pretending to have the highest preference $\bar{v}_{i,t}$ due to the buyer's not knowing the true preference $v_{i,t}$ while assuming other owners are truthful.
From Lemma \ref{lemma:monotonicity_J_1}, we have that owner $i$'s information rent $\Lambda^{\sigma}_{i,t}(v_{i, t}, \bar{v}_{i,t}; \tau)$ is non-negative, for all $i\in\mathbb{I}$, $t\in \mathbb{T}$, $\tau\in\mathbb{T}_{t}$, $v_{i,t}\in V_{i,t}$.
Hence, each owner $i$ with $v_{i, t} = \bar{v}_{i,t}$ has no information rent for privacy protection, which coincides with the setting that owners have tendency for more privacy protection.
Given the information rents, we can interpret each payment rule as follows.
The rule $\beta_{i,t}$ in (\ref{eq:thm_beta}) in constructed by the maximum information rent given the optimal stopping rule, the expected future information rent, and the current-period immediate privacy loss.
The rule $\theta_{i, t}$ is (\ref{eq:thm_theta}) is constructed by the current-period one-stage information rent and the immediate privacy loss.
The rule $\bm{\rho}$ is independent of any realizations of owners' preferences.
For a typical owner $i$, $\rho_{i}(t)$ in (\ref{eq:thm_rho}) is formulated as an expected combination of information rent, in which the period-$t$ ex-interim expectation is taken by letting current preference be the threshold $\kappa^{\ell}_{i}(t)$ with empty history, and the stochastic process from $t+1$ onward is constrained; i.e., forcing the realization of $\tilde{v}_{i, s}$ to be the threshold value $\kappa^{l}_{i}(s)$ if it is above $\kappa^{l}_{i}(s)$, for all $s\in\mathbb{T}_{t+1}$.
The formulation (\ref{eq:thm_rho}) obtains a relationship between the rule $\rho_{i}$ and the threshold function $\kappa^{l}_{i}$, given the privacy allocation rule $\sigma$, such that the design of $\rho_{i}$ can be equivalent to the design of $\kappa^{\ell}_{i}$, for $i\in\mathbb{I}$.

The following corollary directly follows Theorem \ref{theorem:sufficient_condition}.
\begin{corollary}
Suppose Assumptions \ref{assp:full_support} and \ref{assp:monotone_transistions} hold.
In any DIC market without the posted-price payment rule $\bm{\rho}$ (i.e., $\rho_{i}(t)=0$, for all $t\in\mathbb{T}$, $i\in\mathbb{I}$), the stopping rule is optimal if and only if there exists a threshold function $\kappa^{\ell}_{i}$ that solves the following equation:
\begin{equation}
        \begin{aligned}
        &0 = \mathbb{E}^{\sigma}\Big[\sum\limits_{s=t}^{T-1}\Big(- \Lambda^{\sigma}_{i,s}(\tilde{v}_{i, s}\wedge \kappa^{\ell}_{i}(s), s;v'_{i,s})       \\
        & +\Lambda^{\sigma}_{i,s+1}(\tilde{v}_{i, s+1}\wedge \kappa^{\ell}_{i}(s+1), \bar{v}_{i,s+1};s+1)     \Big) \\
        &- \Big( \sum\limits_{\tau'\in \mathbb{T}_{s+1}}\Lambda^{\sigma}_{i,s+1}(\tilde{v}_{i, s+1}\wedge \kappa^{\ell}_{i}(s+1), \bar{v}_{i,s+1};\tau') \\
        &-\sum\limits_{\tau'\in \mathbb{T}_{s}}\Lambda^{\sigma}_{i,s}(\tilde{v}_{i, s}\wedge \kappa^{\ell}_{i}(s), \bar{v}_{i,s};\tau')\Big)   \Big|\kappa^{\ell}_{i}(t), \emptyset\Big].
        \end{aligned}
    \end{equation}
%
\end{corollary}

Let $e_{i}=\{e_{i, 0},e_{i, 1},\dots, e_{i, T}\}$ denote a sequence of threshold values generated by a threshold function $\kappa^{l}_{i}$, where each $e_{i, t} = \kappa^{l}_{i}(t)$.
From the stopping region $\mathcal{R}^{\Gamma}_{i, t}$, if the buyer can freely choose $\kappa^{\ell}_{i}(t)\in V_{i,t}$, for all $t\in\mathbb{T}$, then, we say that the buyer can control owner $i$'s stopping decision; i.e., she can make owner $i$ to stop or to continue in any period.
We write $\mathtt{DIC}[\bm{\kappa}^{\ell}]$ as a set of all privacy allocation rules that satisfy (\ref{eq:thm_sigma_C1}) and (\ref{eq:thm_sigma_C1}), when $\bm{\rho}$ is constructed in (\ref{eq:thm_rho}) given threshold functions $\bm{\kappa}^{\ell}$.

\begin{corollary}\label{corollary:limited_stopping_effect}
Suppose Assumptions \ref{assp:full_support} and \ref{assp:monotone_transistions} hold.
Suppose additionally that when each owner is indifferent between stopping or continuing, he chooses to stay in the market. The followings are true.
\begin{itemize}
    \item[(i)] The buyer is able to prevent owner $i$ to leave the market before $t=T$ if and only if there exists $\sigma\in \mathtt{DIC}[\{\bar{v}^{T}_{i}, \bm{\kappa}^{l}_{-i}\}]$.
    \item[(ii)] The buyer is able to make owner $i$ to leave the market at any specific period $t\in\mathbb{T}$ (not before $t$) if and only if there exists $\sigma\in \mathtt{DIC}[\{\{\bar{v}^{t-1}_{i}, \underline{v}_{i,t}, v^{T}_{i, t+1}\}, \bm{\kappa}^{l}_{-i}\}]$, where $v^{T}_{i, t+1}=\{v_{i, s }\}_{s=t+1}^{T}$.
\end{itemize}
\end{corollary}

Corollary \ref{corollary:limited_stopping_effect} shows the restrictions of the buyer's ability to control the owners' stopping decisions.
These restrictions are specified by $\mathtt{DIC}[\{\bar{v}^{T}_{i}, \bm{\kappa}^{l}_{-i}\}]$ and $\mathtt{DIC}[\{\{\bar{v}^{t-1}_{i}, \underline{v}_{i,t}, v^{T}_{i, t+1}\}, \bm{\kappa}^{l}_{-i}\}]$, which requires the design of $\sigma$ and the choices of the thresholds to satisfy the sufficient conditions in Theorem \ref{theorem:sufficient_condition}.

We establish a necessary condition of DIC based on the result obtained in Lemma \ref{lemma:envelope_necessary_condition}.

\begin{proposition}\label{prop:necessary_condition}
Suppose Assumptions \ref{assp:full_support} and \ref{assp:monotone_transistions} hold.
Let the payment rule $\bm{\beta}$ and $\bm{\theta}$ be formulated in (\ref{eq:thm_beta}) and (\ref{eq:thm_theta}), respectively, and let the posted-price payment rule $\bm{\rho}$ be formulated in (\ref{eq:thm_rho}).
In any DIC market $\Gamma=<\sigma, \bm{\beta}, \bm{\theta}, \bm{\rho}>$, the privacy allocation rule $\sigma$ satisfies the followings:
\begin{equation}\label{eq:necessary_C1}
        \Lambda^{\sigma}_{i,t}(\hat{v}_{i, t}, \bar{v}_{i,t}; t) - \Lambda^{\sigma}_{i}(v_{i, t}, \bar{v}_{i,t}; t) \leq d^{S}_{i,t}(\hat{v}_{i, t}, v_{i, t}),
    \end{equation}
    \begin{equation}\label{eq:necessary_C2}
        \begin{aligned}
            \sup_{\tau\in\mathbb{T}_{t}} \Lambda^{\sigma}_{i,t}(\hat{v}_{i, t}, \bar{v}_{i,t}; \tau) - \sup_{\tau\in\mathbb{T}_{t}} \Lambda^{\sigma}_{i,t}(v_{i, t}, \bar{v}_{i,t}; \tau)&\\
            \leq \sup_{\tau\in\mathbb{T}_{t}} d^{-S}_{i, t}(\hat{v}_{i, t},  v_{i, t};\tau)&.
        \end{aligned}
    \end{equation}
\end{proposition}
\begin{proof}
See Appendix \ref{app:prop:necessary_condition}.
\end{proof}

\section{Optimal Market Design Problem and Its Relaxation}\label{sec:relaxed_market_design}

From the formulations of $<\bm{\beta}, \bm{\theta}, \bm{\rho}>$ in Theorem \ref{theorem:sufficient_condition}, $\bar{\bm{\tau}}=\{\bar{\tau}_{i}\}_{i\in \mathbb{I}}$ can be characterized by the assignment rule $\sigma$, the threshold rule $\bm{\kappa}^{l}$, and the endogenous dynamics; i.e., we can write (with a slight abuse of notation) $\bar{\tau}_{i} = \bar{\tau}_{i}(\sigma, \kappa^{l}_{i}) = \sum_{t=0}^{T}t\cdot P\big(v_{i, t}\geq \kappa^{l}_{i}(t)\big)=\mathbb{E}^{\sigma}\Big[\sum^{T}_{t=0 }t\cdot F_{i, t}(\kappa^{l}_{i}(t)| v_{i, t-1}, $ $\sigma^{t-1}(\bm{v}^{t-1} )) \Big]$.
Since our market model is finite-horizon, $\bar{\tau}_{i}$ exists for all $i\in\mathbb{I}$.
Based on Theorem \ref{theorem:sufficient_condition}, we apply a first-order approach \cite{rogerson1985first,tadelis2005lectures} to rewrite the buyer's objective function (\ref{eq:buyer_objective}) as follows (by integration by parts):
\begin{equation}\label{eq:principal_objective_rewrittern}
    \begin{split}
        &C^{\Gamma}(\bar{\bm{\tau}};\bm{N}) = \sum\limits_{i\in \mathbb{I}}  J^{\Gamma}_{i, 0}(\underline{v}_{i,0}, \bar{\tau}_{i})+ \mathbb{E}^{\sigma}\Big[ \sum\limits_{t=0}^{\bar{\bm{\tau}}}L\exp(-\sigma_{t}( \tilde{\bm{v}}_{t}))\\
        &+ \sum\limits_{i\in\mathbb{I}, \tau'_{i}\in \bar{\bm{\tau}} }\sum\limits_{t=0 }^{ \tau'_{i}} \tilde{v}_{i, t}[\exp( \sigma_{t}( \tilde{\bm{v}}_{t}))-1]\\
        & +\sum\limits_{i\in\mathbb{I}, \tau'_{i}\in \bar{\bm{\tau}} }\sum\limits_{t=0 }^{ \tau'_{i}}\frac{[\exp(\sigma_{t}( \tilde{\bm{v}}_{t}))  )-1] (1-F_{i, 0}(\tilde{v}_{i, 0}))  }{ f_{i}(\tilde{v}_{i, 0})/\mathcal{G}^{t}_{0}(\tilde{\bm{v}}^{t}_{0}|\sigma)}  \Big].
    \end{split}
\end{equation}

Here, (\ref{eq:principal_objective_rewrittern}) is a relaxed objective function by letting owners to make decisions at stationary points and substituting $<\bm{\beta}, \bm{\theta}, \bm{\rho}>$ given in (\ref{eq:thm_beta})-(\ref{eq:thm_rho}), respectively. 
From Lemma \ref{lemma:monotonicity_J_1}, we have that $J_{i, t}$ is weakly decreasing.
Hence, if the mechanism $\Gamma$ induces $J^{\Gamma}_{i, 0}(\bar{v}_{i,0}, \tau'_{i})\geq 0$, for all $i\in\mathbb{I}$, then the IR constraint is satisfied; i.e., $J^{\Gamma}_{i}(\tau'_{i})\geq 0$ for all $i\in\mathbb{I}$, $\tau'_{i}\in \mathbb{T}$.
With a slight abuse of notation, let $J^{\sigma, \bm{\kappa}^{\ell}}_{i,t}(\cdot) = J^{\Gamma}_{i,t}(\cdot)$ when $<\bm{\beta}, \bm{\theta}, \bm{\rho}>$ satisfies (\ref{eq:thm_beta})-(\ref{eq:thm_rho}), respectively.
Let $\bar{C}^{\sigma, \bm{\kappa}^{\ell}}(\bar{\tau};\bm{N}) = C^{\Gamma}(\bar{\bm{\tau}};\bm{N}) -\sum\limits_{i\in \mathbb{I}}  J^{\sigma, \bm{\kappa}^{\ell}}_{i,0}(\underline{v}_{i,0},\bar{\tau}_{i})$.
Hence, based on (\ref{eq:principal_objective_rewrittern}) we can relax the buyer's mechanism design problem (\ref{eq:principal_original_design}) as follows:
Hence, based on (\ref{eq:principal_objective_rewrittern}) we can relax the buyer's mechanism design problem (\ref{eq:principal_original_design}) as follows:
\begin{equation}\label{eq:buyer_relaxed}
    \begin{aligned}
    \min_{\sigma, \bm{\kappa}^{\ell}} \;\; \bar{C}^{\sigma, \bm{\kappa}^{\ell} }(\bar{\bm{\tau}};\bm{N}), \text{ s.t., } J^{\sigma, \bm{\kappa}^{\ell}}_{i,0}(\underline{v}_{i,0},\bar{\tau}_{i})\geq 0, \forall i\in\mathbb{I}.
    \end{aligned}
\end{equation}
Therefore, the buyer's mechanism design problem of finding optimal $\Gamma=<\sigma, \bm{\beta}, \bm{\theta}, \bm{\rho}>$ by satisfying the DIC and IR constraints is relaxed to (\ref{eq:buyer_relaxed}), which requires $J^{\sigma, \bm{\kappa}^{\ell}}_{i,0}(\underline{v}_{i,0},\bar{\tau}_{i})$, for all $i\in\mathbb{I}$.

If the market is complete-information (i.e., the realizations of preferences $\{v_{i, t}\}_{i\in\mathbb{I}, t\in\mathbb{T}}$ are common knowledge), then the buyer design the market by solving the optimization problem (\ref{eq:principal_original_design}) constrained only by the individual rationality constraint $\text{IR}_{i}$, for all $i\in\mathbb{I}$.
Thus, each owner $i$ only makes stopping decision in each period.
Since we restrict attention to the BPM, the buyer can treat each owner $i$ separately and choose the payment rules by making $\text{IR}_{i}$ binding.
Let $\vec{\Gamma} = <\vec{\sigma}, \bm{\vec{\beta}}, \bm{\vec{\theta}}, \bm{\rho}>$ denote the resulting optimal market model (\textit{first-best} mechanism) in the complete-information environment.
It is straightforward to see that the optimal payment rules are constructed as $\vec{\beta}_{i, t}(v_{i, t}, \bm{v}_{-i, t}) = \ell(v_{i, t}, \sigma_{t}(v_{i, t}, \bm{v}_{-i, t}))$ and $\vec{\theta}_{i, t}(v_{i, t}, \bm{v}_{-i, t}) = \ell(v_{i, t}, \sigma_{t}(v_{i, t}, \bm{v}_{-i, t}))$.
The posted-price rule $\vec{\rho}_{i}$ is the same as (\ref{eq:thm_rho}) since it is independent of owners' preference.
Suppose that $\sigma^{*}$ and $\bm{\kappa}^{\ell*}$ solves (\ref{eq:buyer_relaxed}).
Let $\bm{\beta}^{*}, \bm{\theta}^{*}, \bm{\rho}^{*}$ be formulated according to (\ref{eq:thm_beta})-(\ref{eq:thm_rho}), respectively.
From its definition in (\ref{eq:integral_envelope}), $\Lambda^{\sigma}_{i,t}(v_{i,t}, \bar{v}_{i,t};\tau)\geq 0$, for all $i\in\mathbb{I}$, $t\in\mathbb{T}$, $\tau\in\mathbb{T}_{t}$, $v_{i,t}\in V_{i,t}$.
According to (\ref{eq:thm_beta}) and (\ref{eq:thm_theta}), for each owner $i$ with the highest preference $v_{i,t}=\bar{v}_{i,t}$, the payment satisfies $\vec{\beta}_{i, t}(\bar{v}_{i,t}, \bm{v}_{-i, t}) \geq \beta^{*}_{i, t}(\bar{v}_{i,t}, \bm{v}_{-i,t})$ and $\vec{\theta}_{i, t}(\bar{v}_{i,t}, \bm{v}_{-i, t}) = \theta^{*}_{i, t}(\bar{v}_{i,t}, \bm{v}_{-i, t})$.
Hence, compared with $<\bm{\vec{\beta}}, \bm{\vec{\theta}}>$, the payment rules $<\bm{\beta}^{*}, \bm{\theta}^{*}>$ takes into account owners' net expected information rent (net rent), given as
\begin{equation}\label{eq:net_expected_rent}
    \mathbb{E}^{\sigma; \mu_{i}}\Big[\sum_{i\in\mathbb{I}, \tau'_{i}\in \bar{\bm{\tau}} }\sum_{t=0 }^{ \tau'_{i}}\frac{[\exp(\sigma_{t}( \tilde{\bm{v}}_{t}))  )-1] (1-F_{i, 0}(\tilde{v}_{i, 0}))  }{ f_{i,0}(\tilde{v}_{i, 0})/\mathcal{G}^{t}_{0}(\tilde{\bm{v}}^{t}_{0}|\sigma)}  \Big].
\end{equation}
Hence, the buyer's optimal $\sigma^{*}$ is chosen to reduce this net rent. 
For each owner $i$ with the hightest preference $\bar{v}_{i,t}$, it is straightforward to see that $\vec{\beta}_{i, t}(\bar{v}_{i}, \bm{v}_{-i, t}) \geq \beta^{*}_{i, t}(\bar{v}_{i,t}, \bm{v}_{-i, t})$ and $\vec{\theta}_{i, t}(\bar{v}_{i}, \bm{v}_{-i, t}) = \theta^{*}_{i, t}(\bar{v}_{i,t}, \bm{v}_{-i, t})$.
However, due to the Markovian instrumentalness, the net rent (\ref{eq:net_expected_rent}) can take various forms and the relationship between differential privacy and the reduction of the net rent is in general unclear.
Therefore, the tradeoff of privacy and the buyer's payoff in dynamic market in general does not coincide with the fundamental privacy-utility tradeoff of differential privacy (described in \ref{sec:background_differential_privacy}).

Due to the dynamics of the market, the constrained optimization problem (\ref{eq:buyer_relaxed}) is in general analytically intractable.
Computationally solving (\ref{eq:buyer_relaxed}) in general involves approximations, which may inevitably violate the DIC conditions we obtained in Theorem \ref{theorem:sufficient_condition} and Proposition \ref{prop:necessary_condition}.
It is beyond the scope of this paper to conduct algorithmic analysis of such computational approximations and to design efficient algorithms to solve (\ref{eq:buyer_relaxed}) numerically and we put them to our future work.
To address the intractability in practical mechanism design, one way is to use a weaker version of incentive compatibility which is known as $\delta$-incentive compatibility or $\delta$ approximate incentive compatibility.
For our dynamic environment, we define the notion of $<\delta^{S}_{i, t}, \delta^{-S}_{i, t}>$-DIC: the market model is $\delta^{S}_{i, t}$-DIC when it is optimal for owner $i$ to stop; the market model is $\delta^{-S}_{i, t}$-DIC when it is optimal for owner $i$ to continue.

Define, 
\begin{equation}\label{eq:h_S}
    \begin{aligned}
&\Xi^{S}_{i, t}\equiv \sup_{v_{i, t}, \hat{v}_{i, t}} \Big\{\ell\big(\hat{v}_{i, t}, \sigma_{t}(\hat{v}_{i, t},  \bm{v}_{-i, t})\big) \\
&-  \ell\big(v_{i, t}, \sigma_{t}(\hat{v}_{i, t},  \bm{v}_{-i, t})\big) +  \Lambda^{\sigma}_{i}(\hat{v}_{i, t}, v_{i, t}; t) \Big\},
\end{aligned}
\end{equation}
\begin{equation}\label{eq:H_bar_S}
\begin{aligned}
&\Xi^{-S}_{i, t}\equiv \sup_{v_{i, t}, \hat{v}_{i, t}} \Big\{\sup_{\tau\in\mathbb{T}_{t}} \bar{J}^{\Gamma}_{i, t}(v_{i,t}, \hat{v}_{i,t},h_{t}, \tau) \\
&- \sup_{\tau\in\mathbb{T}_{t}} \bar{J}^{\Gamma}_{i, t}(\hat{v}_{i,t}, \hat{v}_{i,t},h_{t}, \tau)\\
& + \sup_{\tau\in\mathbb{T}_{t}} \Lambda^{\sigma}_{i,t}(v_{i,t}, \bar{v}_{i,t}; \tau) - \sup\limits_{\tau\in \mathbb{T}_{t}} \Lambda^{\sigma}_{i,t}(v_{i,t}, \bar{v}_{i,t}; \tau)
\end{aligned}
\end{equation}
where $\bar{J}^{\Gamma}_{i, t}$ is defined in (\ref{eq:J_without_beta}).
We have the following proposition.
\begin{proposition}\label{prop:delta_DIC}
Suppose Assumptions \ref{assp:full_support} and \ref{assp:monotone_transistions} hold.
Let $<\bm{\beta}, \bm{\theta}, \bm{\rho}>$ be constructed in (\ref{eq:thm_beta})-(\ref{eq:thm_rho}), respectively.
Then, the market model is $<\delta^{S}_{i, t}, \delta^{-S}_{i, t}>$-DIC with $\delta^{S}_{i, t} = \Xi^{S}_{i, t}$ and $\delta^{-S}_{i, t} = \Xi^{-S}_{i, t}+\sup_{\tau\in\mathbb{T}_{t}}\rho_{i}(\tau)$, when $\delta^{S}_{i, t}> 0$ and $\Xi^{-S}_{i, t}+\sup_{\tau\in\mathbb{T}_{t}}\rho_{i}(\tau)> 0$; the market model is DIC, when $\delta^{S}_{i, t}\leq 0$ and $\Xi^{-S}_{i, t}+\sup_{\tau\in\mathbb{T}_{t}}\rho_{i}(\tau)\leq 0$.
\end{proposition}

\begin{proof}

See Appendix \ref{app:prop:delta_DIC}.


\end{proof}

Proposition \ref{prop:delta_DIC} establishes a sufficient condition for a relaxed DIC criterion for our dynamic market model. 
The result can be used as a worst-case analysis of the dynamic market model when there are opportunities for the owners to misreport their true preferences over privacy protection.

\section{Conclusion}\label{sec:conclusion}

This work has proposed a dynamic market model for trading data privacy when the data owners have a dual preference for privacy protection offered by the data buyer.
The market model consists of one privacy allocation rule profile, two preference-dependent payment rule profiles, and one preference-independent posted-price payment rule profile.
The preference-dependent payment rules are used to determine monetary values to compensate the owners for their privacy loss while the posted-price payment rule is used by the buyer to influence the owners' stopping decisions.
We have studied a mechanism design problem in a dynamic environment where owners' privacy preferences evolve over time due to \textit{(i)} the time-varying of the owners' intrinsic component of preference and \textit{(ii)} the dynamic instrumentalness due to the buyer's sequential multiple usages of data.
The owners are allowed to leave the market at the end of each period of data usage if his expected loss of continuing is beyond his tolerance.
An optimal stopping problem has been modeled for the owners in a relaxed manner when each owner expects the future payoffs, he assumes that all other owners do not plan to leave the market.
Under a monotonicity assumption about the instrumentalness of the owners' preferences, the optimal stopping rule has been transformed into a threshold-based stopping rule with a profile of threshold functions.
By taking into consideration the owners' coupled deviations from truthful reporting and optimal stopping, a new notion of dynamic incentive compatibility based on the Bellman equation has been defined as an essential design restriction of the buyer's optimal market model.

We have provided a solid theoretic design regime for the dynamic incentive-compatible market model by characterizing the preference-dependent payment rules in terms of the privacy allocation rule.
The posted-price payment rule has been characterized in terms of the privacy allocation rule and the threshold functions to maintain the optimality of the owners' stopping decisions and to support the guarantee of the dynamic incentive compatibility.
A restriction of the buyer's ability to control the owners' stopping decisions has been captured by establishing the relationships between the privacy allocation rule and the threshold function.
The buyer's optimal market design problem by determining four decision rule profiles with the individual rationality and the dynamic incentive compatibility constraints has been relaxed to an optimization problem of determining the privacy allocation rule profile and the threshold function profile with a modified individual rationality constraint set.
An approximated dynamic incentive-compatible mechanism design principle has been provided to address the inevitable violation of incentive compatibility when optimal mechanism design is solved approximately.
Designing efficient algorithms to computationally solve the buyer's optimal market design with an analysis of the violation of the theoretical implementability is our natural future work.


\renewcommand\arraystretch{1}  %


\begin{table*}[htbp]
	\centering
	\caption{Summary of Notations}
	\begin{tabularx}{\textwidth}{c>{\raggedright}X}
		\toprule  
		Symbol & Meaning\tabularnewline
		\toprule  
		$\tilde{y}_{i,t}, y_{i,t}$ & random variable, realization of random variable (of owner $i$ in period $t$).
		\tabularnewline
		\midrule 
		$\mathbb{I}$, $\mathbb{T}$, $\mathbb{T}_{t}$ ,$\mathcal{C}_{i}$,
		$V_{i,t}$, $\mathcal{E}$ & set of owners, set of periods, set of periods starting from period $t$, set of owner $i$'s intrinsic preferences, set of owner $i$'s period-$t$ preferences, set of privacy allocations\tabularnewline \midrule
		$h_{t}$, $H_{t}$ & public history, set of public histories
		\tabularnewline
		\midrule 
	    $K^{\mathcal{C}}_{i}$, $K_{t}$& prior distribution of owner $i$'s intrinsic preference, instrumental kernel function
		\tabularnewline
		\midrule 
		$\chi=\{\chi_{i,t}\}_{i\in\mathbb{I},t\in\mathbb{T}}$ & reporting strategy profile, each $\chi_{i,t}$ is owner $i$'s period-$t$ strategy\tabularnewline
		\midrule 
		$\mathtt{TM}_{i,t}\in\{0,1\}$ & owner $i$'s period-$t$ stopping decision: $\mathtt{TM}_{i,t}=1$ means stop and $\mathtt{TM}_{i,t}=0$ means continue\tabularnewline \midrule
		$\sigma=\{\sigma_{t}\}_{t\in\mathbb{T}}$ & privacy allocation rule profile, each $\sigma_{t}$ is period-$t$ rule\tabularnewline
		 \midrule
		 $\bm{\beta}={\beta_{i,t}}_{i\in\mathbb{I}, t\in\mathbb{T}}$ & preference-dependent payment rule profile, each $\beta_{i,t}$ is the rule for owner $i$'s period-$t$ (non-stopping) payment\tabularnewline
		 \midrule
		 $\bm{\theta}={\theta_{i,t}}_{i\in\mathbb{I}, t\in\mathbb{T}}$ & preference-dependent payment rule profile, each $\theta_{i,t}$ is the rule for owner $i$'s period-$t$ (stopping) payment \tabularnewline
		 \midrule
		 $\bm{\rho}=\{\rho_{i}\}_{i\in\mathbb{I}}$ & preference-independent posted-price payment rule, each $\rho_{i}$ is the rule for owner $i$ if he stops in $t$\tabularnewline \midrule
		 $p_{i,t}$ & realized payment for owner $i$ in period $t$
		 \tabularnewline \midrule
		 $\ell(v_{i,t}, \epsilon_{t})$ & owner $i$'s loss function in period $t$
		 \tabularnewline \midrule
		 $z_{i,t}(v_{i,t}, \epsilon_{t}, p_{i,t})$ & owner $i$'s one-period payoff function\tabularnewline \midrule
		 $J^{\Gamma}_{i,t}(v_{i,t}, h_{t}, \tau; \chi)$  & owner $i$'s period-$t$ ex-interim expected payoff, $\tau\in\mathbb{T}_{t}$ \tabularnewline \midrule
		 $J^{\Gamma}_{i}(\chi, \tau)$ & owner $i$'s ex-ante expected payoff\tabularnewline \midrule
		 $\mathtt{pm}_{i,t}=\{\mathtt{TM}_{j,\tau_{j}}=1\}_{j\neq i}$ & owner $i$'s estimation of other owners' stopping decision (including the planned ones) in period $t$\tabularnewline \midrule
		 $\overline{\mathtt{pm}}=\{\bar{\tau}_{i}\}_{i\in\mathbb{I}}$ & collection of expected stopping times (i.e., population dynamics) of the owners evaluated by each owner in the ex-ante stage\tabularnewline \midrule
		 $\phi^{\chi}_{i}$ & owner $i$'s stopping rule when the owners' reporting strategy profile is $\chi$\tabularnewline \midrule
		 $C^{\Gamma}(\bar{\bm{\tau}})$ & the buyer's ex-ante expected cost, where $\bar{\bm{\tau}}$ is a collection of expected stopping times evaluated in the ex-ante stage 
		 \tabularnewline
		\bottomrule  
	\end{tabularx}
 \label{table:notations}
\end{table*}

\appendix
\section{Proof of Proposition \ref{prop:one_shot_deviation_principle}}\label{app:prop:one_shot_deviation_principle}

The \textit{only if} part is straightforward due to the optimality of truthful reporting.
Hence, we omit it here and focus on the \textit{if} part.
For the ease of notation, we suppress the public history in the notations of the rules.
The proof is constructed by establishing contradictions.
Fix a profile $\Gamma = <\sigma, \bm{\beta},\bm{\theta}, \bm{\rho}>$.
Suppose that the truthful reporting strategy $\chi^{*}_{i}$ satisfies (\ref{eq:DIC_1_shot}) for any period-$t$ one-shot deviation strategy $\chi^{[t]}_{i}$ for any $t\in\mathbb{T}$, but it violates the DIC defined in (\ref{eq:DIC_original}).
In other words, there exists another reporting strategy $\chi^{(1)}_{i}\equiv\{\chi^{(1)}_{i, t}\}_{t\in\mathbb{T}}$ and some instrumental preference $v_{i, t}$ such that $U^{\Gamma}_{i, t}(v_{i, t},h_{t};\chi^{(1)})> U^{\Gamma}_{i, t}(v_{i, t},h_{t}:\chi^{*})$.
Let $\phi^{\chi^{\star}_{i}}_{i}$ and $\phi^{\chi^{(1)}_{i}}_{i}$ denote the optimal stopping rules given $\chi^{\star}_{i}$ and $\chi^{(1)}_{i}$, respectively.
Suppose that at period $t$, $\phi^{\chi^{\star}_{i}}_{i}$ calls for stopping but $\phi^{\chi^{(1)}_{i}}_{i}$ calls for continuing, i.e.,
\begin{equation*}
    J^{\Gamma}_{i,t}(v_{i,t}, h_{t}, t; \chi^{*}_{i}) < \mathbb{E}^{\sigma}_{\chi^{(1)}_{i}}\Big[U^{\Gamma}_{i,t+1}(\tilde{v}_{i,t+1}, \tilde{h}_{t+1};\chi^{(1)}_{i} )\Big| v_{i,t}, h_{t}\Big].
\end{equation*}

Equivalently, there exists some constant $\eta>0$ such that
\begin{equation}\label{eq:appendix_A_1}
    \begin{aligned}
    J^{\Gamma}_{i,t}(v_{i,t}, h_{t}; \chi^{*}_{i}) + 2\eta\leq \mathbb{E}^{\sigma}_{\chi^{(1)}_{i}}\Big[U^{\Gamma}_{i,t+1}(\tilde{v}_{i,t+1}, \tilde{h}_{t+1};\chi^{(1)}_{i} )\Big| v_{i,t}, h_{t}\Big].
    \end{aligned}
\end{equation}
Consider another reporting strategy $\chi^{(2)}_{i}\equiv\{\chi^{(2)}_{i, t}\}_{t\in\mathbb{T}}$, such that $\chi^{(2)}_{i, s} = \chi^{(1)}_{i, s}$, for all $s\in \mathbb{T}_{t,t+k}$, for some $k>0$, and 
\begin{equation}\label{eq:appendix_A_2}
    \begin{aligned}
    \mathbb{E}^{\sigma}_{\chi^{(1)}_{i}}\Big[U^{\Gamma}_{i,t+1}(\tilde{v}_{i,t+1}, \tilde{h}_{t+1};\chi^{(1)}_{i} )\Big| v_{i,t}, h_{t}\Big] -\eta \leq \mathbb{E}^{\sigma}_{\chi^{(2)}_{i}}\Big[U^{\Gamma}_{i,t+1}(\tilde{v}_{i,t+1}, \tilde{h}_{t+1};\chi^{(2)}_{i} )\Big| v_{i,t}, h_{t}\Big].
    \end{aligned}
\end{equation}
Hence, (\ref{eq:appendix_A_1}) and (\ref{eq:appendix_A_2}) yield:
\begin{equation}\label{eq:appendix_A_3}
    \begin{aligned}
    J^{\Gamma}_{i,t}(v_{i,t}, h_{t}, t; \chi^{*}_{i}) + \eta \leq \mathbb{E}^{\sigma}_{\chi^{(2)}_{i}}\Big[U^{\Gamma}_{i,t+1}(\tilde{v}_{i,t+1}, \tilde{h}_{t+1};\chi^{(2)}_{i} )\Big| v_{i,t}, h_{t}\Big].
    \end{aligned}
\end{equation}
Let $\chi'_{i} = \{\chi'_{i,t}\}_{t\in\mathbb{T}}$ denote any reporting strategy, such that $\chi'_{i,s} = \chi^{(2)}_{i,s}$, for all $s\in\mathbb{T}_{t,t+k}$, and $\chi'_{i, s'}$ is truthful for all $s'\in \mathbb{T}\backslash \mathbb{T}_{t,t+k}$, for some $k>0$.
Hence, (\ref{eq:appendix_A_3}) tells us that a deviation using any such $\chi'_{i}$ is enough to obtain a non-negative profit.
Next, consider a one-shot deviation reporting strategy $\chi^{[s]}_{i}$ for some $s\in \mathbb{T}_{t,t+k}$, for $k>0$, such that $\chi^{[s]}_{i, s} = \chi^{(2)}_{i, s}$.
Then, (\ref{eq:appendix_A_3}) gives
\begin{equation}\label{eq:appendix_A_4}
    \begin{aligned}
    J^{\Gamma}_{i,t}(v_{i,t}, h_{t}, t; \chi^{*}_{i}) < \mathbb{E}^{\sigma}_{\chi^{[t+k-1]}_{i}}\Big[U^{\Gamma}_{i,t+1}(\tilde{v}_{i,t+1}, \tilde{h}_{t+1};\chi^{[t+k-1]}_{i} )\Big| v_{i,t}, h_{t} \Big].
    \end{aligned}
\end{equation}
Let $t' = t + k$. From (\ref{eq:value_function_v2}) in Lemma \ref{lemma:optimal_stopping}, we have, for all $v_{i, t'}\in V_{i,t}$,
\begin{equation}\label{eq:appendix_A_5}
    \begin{aligned}
    &U^{\Gamma}_{i, t'-1}(v_{i,t'-1}, h_{t'-1}; \chi^{[t'-1]}_{i}) = \max\Big\{ J^{\Gamma}_{i,t'-1}(v_{i,t-1}, h_{t'-1}, t'-1; \chi^{[t'-1]}_{i}), \\
    & \mathbb{E}^{\sigma}_{\chi^{[t'-1]}_{i}}\Big[U^{\Gamma}_{i,t'}(\tilde{v}_{i,t'}, \tilde{h}_{t'}; \chi^{[t'-1]}_{i} )   \Big| v_{i,t'-1}, h_{t'-1}\Big] \Big\}.
    \end{aligned}
\end{equation}
Since the truthful reporting strategy $\chi^{*}$ satisfies (\ref{eq:DIC_1_shot}), we have
\begin{equation*}
    \begin{aligned}
    U^{\Gamma}_{i, t'-1}(v_{i,t'-1}, h_{t'-1}; \chi^{[t'-2]}_{i})  \geq U^{\Gamma}_{i, t'-1}(v_{i,t'-1}, h_{t'-1}; \chi^{[t'-1]}_{i}).
    \end{aligned}
\end{equation*}
Hence, 
\begin{equation*}
    \begin{aligned}
    \mathbb{E}^{\sigma}_{\chi^{[t'-2]}_{i}}\Big[U^{\Gamma}_{i,t+1}(\tilde{v}_{i,t'}, \tilde{h}_{t'}; \chi^{[t'-2]}_{i} )   \Big| v_{i,t}, h_{t}\Big] \geq \mathbb{E}^{\sigma}_{\chi^{[t'-1]}_{i}}\Big[U^{\Gamma}_{i,t+1}(\tilde{v}_{i,t'}, \tilde{h}_{t'}; \chi^{[t'-1]}_{i} )   \Big| v_{i,t}, h_{t}\Big].
    \end{aligned}
\end{equation*}
From (\ref{eq:appendix_A_4}), we have
\begin{equation*}
    \begin{aligned}
    \mathbb{E}^{\sigma}_{\chi^{[t'-2]}_{i}}\Big[U^{\Gamma}_{i,t+1}(\tilde{v}_{i,t'}, \tilde{h}_{t'}; \chi^{[t'-2]}_{i} )   \Big| v_{i,t}, h_{t}\Big] > J^{\Gamma}_{i,t}(v_{i,t}, h_{t}, t; \chi^{*}_{i}).
    \end{aligned}
\end{equation*}
Backward induction yields:
\begin{equation*}
    \begin{aligned}
    \mathbb{E}^{\sigma}_{\chi^{[t]}_{i}}\Big[U^{\Gamma}_{i,t+1}(\tilde{v}_{i,t'}, \tilde{h}_{t'}; \chi^{[t]}_{i} )   \Big| v_{i,t}, h_{t}\Big]J^{\Gamma}_{i,t}(v_{i,t}, h_{t}, t; \chi^{*}_{i}),
    \end{aligned}
\end{equation*}
which contradicts the setting that $\chi^{*}$ satisfies (\ref{eq:DIC_1_shot}).
Similar procedures can be used for the other cases: \textit{(i)} $\phi^{\chi^{\star}_{i}}_{i}$ calls for stopping and $\phi^{\chi^{(1)}_{i}}_{i}$ calls for stopping, \textit{(ii)} $\phi^{\chi^{\star}_{i}}_{i}$ calls for continuing but $\phi^{\chi^{(1)}_{i}}_{i}$ calls for stopping, \textit{(iii)} $\phi^{\chi^{\star}_{i}}_{i}$ calls for continuing and $\phi^{\chi^{(1)}_{i}}_{i}$ calls for continuing.

\hfill$\square$

\section{Proofs of Lemma \ref{lemma:envelope_necessary_condition} and Corollary \ref{corollary:kolmogorov} }\label{app:lemma:envelope_necessary_condition}

Here, we prove Lemma \ref{lemma:envelope_necessary_condition} and Corollary \ref{corollary:kolmogorov} together.
For the ease of notation, we suppress the public history in the notations of the rules.
In DIC, truthful reporting is optimal for all owners.
Hence, Lemma is directly from the envelope theorem. That is, for all $i\in\mathbb{I}$, $t\in \mathbb{T}$, $\tau\in \mathbb{T}_{t}$, $v_{i,t}\in V_{i,t}$, $h_{t}\in H_{t}$, 
\begin{equation*}
    \begin{aligned}
    &\frac{\partial J^{\Gamma}_{i,t}(x,  h_{t}, \tau) }{\partial  x}\Big|_{x = v_{i,t}}  \equiv \mathbb{E}^{\sigma}\Big[  \sum\limits_{s=t}^{\tau}\big(1 - \exp(\sigma_{s}(\tilde{\bm{v}}_{s}))\mathcal{G}^{s}_{t}(\tilde{v}^{s}_{i,t}|\sigma) \big)\Big] 
\end{aligned} 
\end{equation*}
where $\mathcal{G}^{s}_{t}(\tilde{v}^{s}_{i,t}|\sigma)\equiv\prod_{k=t}^{s} \frac{\partial}{\partial x} K_{i,t}(x, \tilde{\epsilon}^{k-1}; \tilde{c}_{i,k-1})\Big|_{x =\tilde{v}_{i,k-1}}$.

From Kolmogorov's Existence Theorem \cite{billingsley2008probability}, we have, for any $v_{i,t}\in V_{i,t}$, any $t\in \mathbb{T}\backslash\{T\}$,
\begin{equation*}
    v_{i, t+1} = \inf\{v'_{i, t+1}\in V_{i,t}: F_{i, t+1}(v'_{i, t+1}|v_{i, t},\epsilon^{t})\geq c_{i, t+1}\},
\end{equation*}
where the intrinsic preference $c_{i,t+1}$ is uniformly drawn from $(0,1)$. From Assumption \ref{assp:full_support}, we have
\begin{equation*}
    \begin{aligned}
    &\frac{\partial \tilde{v}_{i, s}}{\partial x}\Big|_{x=v_{i, t}} = \frac{\partial \tilde{\bm{v}}_{s}}{\partial x}\Big|_{x=v_{i, t}}  
    =  \prod_{k=t+1}^{s}\frac{-\partial F_{i,k}(x| \tilde{v}_{i, k-1}, \tilde{\epsilon}^{k-1} ) }{f_{i, k }(\tilde{v}_{i, k}| \tilde{v}_{i, k-1}, \tilde{\epsilon}^{k-1}) \partial x}\\
    & \Big|_{x=\psi_{i, k-1}(v_{i, k-2}, \epsilon^{k-2}|\tilde{c}_{i, k-1})}\\
    &= \mathcal{G}^{s}_{t}(\tilde{v}^{s}_{i,t}|\sigma)
    \end{aligned}
\end{equation*}
where $\psi_{i,t}(v_{i, t-1}, \epsilon^{t-1}|c_{i,t}) = $ $\inf\{v_{i, t}: F_{i, t}(v_{i, t}$ $|v_{i, t-1},$ $\epsilon^{t-1})$ $\geq c_{i,t}\}$.

\hfill$\square$

\section{Proof of Lemma \ref{lemma:monotonicity_J_1} }\label{app:lemma:monotonicity_J_1}

From Lemma \ref{lemma:envelope_necessary_condition}, we have
\begin{equation*}
    \begin{aligned}
    &\frac{\partial J^{\Gamma}_{i,t}(x,  h_{t}, \tau) }{\partial  x}\Big|_{x = v_{i,t}}  \equiv \mathbb{E}^{\sigma}\Big[  \sum\limits_{s=t}^{\tau}\big(1 - \exp(\sigma_{s}(\tilde{\bm{v}}_{s}))\mathcal{G}^{s}_{t}(\tilde{v}^{s}_{i,t}|\sigma) \big)\Big].
\end{aligned} 
\end{equation*}
From Assumption \ref{assp:monotone_transistions}, the term (according to Corollary \ref{corollary:kolmogorov}) $\mathcal{G}^{s}_{t}(v^{s}_{i,t}|\sigma)>0$, for all $v^{s}_{i,t}\in V^{t,s}_{i}$, $s\in \mathbb{T}_{t}$.
Since $\exp(x)\geq 1$, for all $x\geq 0$, then $\frac{\partial J^{\Gamma}_{i,t}(x,  h_{t}, \tau) }{\partial  x}\Big|_{x = v_{i,t}} \leq 0$. Hence, $J^{\Gamma}_{i,t}(v_{i,t},  h_{t}, \tau)$ is weakly decreasing in $v_{i,t}$, for all $i\in\mathbb{I}$, $t\in\mathbb{T}$, $\tau\in \mathbb{T}_{t}$, $h_{t}\in H_{t}$.

\hfill$\square$

\section{Proof of Proposition \ref{prop:threshold_rule_optimal_stopping}}\label{app:prop:threshold_rule_optimal_stopping}
%

%
Let $\chi^{*}_{i}$ and $\chi^{[t]}_{i}$, respectively, denote owner $i$'s truthful reporting strategy and period-$t$ one-shot deviation strategy.
Recall the term $g^{\Gamma}_{i,t}(v_{i,t}, h_{t}, \tau;  \chi)$ given in (\ref{eq:G_term_rewritten_1}).
With a slight abuse of notation, let $\tau_{i}[\chi_{i}, v_{i, t}]$ denote the minimum time horizon, given the reporting strategy $\chi_{i}$ and owner $i$'s current preference $v_{i,t}$, such that
\begin{equation}
    g^{\Gamma}_{i,t}(v_{i,t}, h_{t}, \tau_{i}[\chi_{i}, v_{i, t}];  \chi_{i}) = G^{\Gamma}_{i,t}(v_{i,t}, h_{t};  \chi).
\end{equation}
By Lemma \ref{lemma:monotonicity_J_1}, $G^{\Gamma}_{i,t}(v_{i,t}, h_{t};  \chi)$ is weakly decreasing in any DIC market model.

Suppose that the indifference region contains two different intervals, $[\kappa^{l}_{i}(t), \kappa^{r}_{i}(t)]$ and $[\bar{\kappa}^{l}_{i}(t), \bar{\kappa}^{r}_{i}(t)]$ with no intersections ($\kappa^{l}_{i}(t)\neq \bar{\kappa}^{l}_{i}(t)$).
With a slight abuse of notation, let $\phi^{\chi^{*}_{i}}_{i}[\kappa^{l}_{i}]$ denote the threshold rule in the DIC market and let $\tau_{i}[\chi_{i}, v_{i, t}; \kappa^{l}_{i}]$ denote the term $\tau_{i}[\chi_{i}, v_{i, t}]$ define above, when the threshold function is $\kappa^{l}_{i}$.
Suppose $\tau_{i}[\chi^{*}_{i}, v_{i, t}; \kappa^{l}_{i}] = \tau_{i}[\chi^{*}_{i}, v_{i, t}; \bar{\kappa}^{l}_{i}]$, for some $v_{i, t}\in V_{i,t}$, $t\in \mathbb{T}$.
Assume without loss of generality $\kappa^{l}_{i}(t)> \bar{\kappa}^{l}_{i}(t)$, for some $t\in \mathbb{T}$.
Then, 
\begin{equation*}
    \begin{aligned}
    P(\tau_{i}[\chi^{*}_{i}, v_{i,t}; \kappa^{l}_{i}] = t ) =& P(v_{i,t}\geq \kappa^{l}_{i}(t),  \tau_{i}[\chi^{*}_{i}, v_{i,t-1}; \kappa^{l}_{i}]>t-1)\\
    =& \mathbb{E}\Big[\mathbb{E}\Big[\mathbf{1}_{\{v_{i,t}\geq \kappa^{l}_{i}(t)\}}\Big] \mathbf{1}_{\{v_{i,t-1}< \kappa^{l}_{i}(t-1)\}}   \Big]. 
\end{aligned}
\end{equation*}
Hence, we have
\small{
\begin{equation}\label{eq:Appendix_D_1}
    \begin{split}
        P( & \tau_{i}[\chi^{*}_{i}, v_{i, t}; \bar{\kappa}^{l}_{i}] = t ) - P(\tau_{i}[\chi^{*}_{i}, v_{i, t}; \kappa^{l}_{i}] = t )\\
        =& \mathbb{E}\Big[\mathbb{E}\Big[\mathbf{1}_{\{v_{i, t}\geq \bar{\kappa}^{l}_{i}(t)\}}\Big] \mathbf{1}_{\{v_{i, t-1}< \bar{\kappa}^{l}_{i}(t-1)\}}   \Big]-\mathbb{E}\Big[\mathbb{E}\Big[\mathbf{1}_{\{v_{i, t}\geq \kappa^{l}_{i}(t)\}}\Big] \mathbf{1}_{\{v_{i, t-1}< \kappa^{l}_{i}(t-1)\}}   \Big]\\
        =& \mathbb{E}\Big[\mathbb{E}\Big[\mathbf{1}_{\{\bar{\kappa}^{l}_{i}(t)\leq v_{i, t} \leq \kappa^{l}_{i}(t)\}}\Big] \mathbf{1}_{\{v_{i, t-1}< \bar{\kappa}^{l}_{i}(t-1)\}}   \Big].
    \end{split}
\end{equation}
}
Due to Assumption \ref{assp:full_support} and the setting $\kappa^{l}_{i}(t)> \bar{\kappa}^{l}_{i}(t)$, the right-hand side of (\ref{eq:Appendix_D_1}) is strictly positive. However, since $\tau_{i}[\chi^{*}_{i}, v_{i, t}; \kappa^{l}_{i}] = \tau_{i}[\chi^{*}_{i}, v_{i, t}; \bar{\kappa}^{l}_{i}]$, $P( \tau_{i}[\chi^{*}_{i}, v_{i, t}; \bar{\kappa}^{l}_{i}] = t ) - P(\tau_{i}[\chi^{*}_{i}, v_{i, t}; \kappa^{l}_{i}] = t )=0$, which gives a contradiction.
Therefore, the threshold function is unique.

\hfill$\square$

\section{Proof of Theorem \ref{theorem:sufficient_condition} }\label{app:theorem:sufficient_condition}

We first prove that, given any $\Lambda^{\sigma}_{i}$ that satisfies the conditions (\ref{eq:thm_sigma_C1}) and (\ref{eq:thm_sigma_C2}), the market model with $\beta_{i, t}$, $\theta_{i, t}$, and $\rho_{i}$ constructed in (\ref{eq:thm_beta})-(\ref{eq:thm_rho}), respectively, is DIC. After that, we prove that the formulation of $\Lambda^{\sigma}_{i}$ in (\ref{eq:integral_envelope}) is valid.
For the ease of notation, we suppress the public history in the notations; unless otherwise stated.

We fix other owners' period-$t$ instrumental preference as $\bm{v}_{-i, t}$, for any $t\in\mathbb{T}$.
Let $v_{i, t}\in V_{i}$ and $\hat{v}_{i, t}\in V_{i}$ be any two instrumental preferences at any period $t\in \mathbb{T}$.
The formulation of $\theta_{i, t}$ in (\ref{eq:thm_theta}) yields 
\begin{equation}\label{eq:Appendix_E_1}
    \begin{aligned}
    &\theta_{i, t}(\hat{v}_{i, t},\bm{v}_{-i, t})-\theta_{i, t}(v_{i, t},\bm{v}_{-i, t}) \\
    =& \Lambda^{\sigma}_{i}(\hat{v}_{i, t}, \bar{v}_{i}; t) +\ell(\hat{v}_{i, t}, \sigma_{t}(\hat{v}_{i, t}, \bm{v}_{-i, t})) -\Lambda^{\sigma}_{i}(v_{i, t}, \bar{v}_{i}; t) -\ell(v_{i, t}, \sigma_{t}(v_{i, t}, \bm{v}_{-i, t}))\\
    =& \Lambda^{\sigma}_{i}(\hat{v}_{i, t}, \bar{v}_{i}; t) - \Lambda^{\sigma}_{i}(v_{i, t}, \bar{v}_{i}; t) - \big(-\ell(\hat{v}_{i, t}, \sigma_{t}(\hat{v}_{i, t}, \bm{v}_{-i, t})) + \ell(v_{i, t}, \sigma_{t}(\hat{v}_{i, t}, \bm{v}_{-i, t}))   \big) \\
    &-\big( \ell(v_{i, t}, \sigma_{t}(v_{i, t}, \bm{v}_{-i, t})) -\ell(v_{i, t}, \sigma_{t}(\hat{v}_{i, t}, \bm{v}_{-i, t}))   \big).
    \end{aligned}
\end{equation}
From the definition of $d^{S}_{i, t}$ in (\ref{eq:distance_S_1}) and  condition (\ref{eq:thm_sigma_C1}), the right-hand side (RHS) of (\ref{eq:Appendix_E_1}) becomes:
\begin{equation}\label{eq:Appendix_E_2}
    \begin{aligned}
    \theta_{i, t}(\hat{v}_{i, t},\bm{v}_{-i, t})-\theta_{i, t}(v_{i,t},\bm{v}_{-i, t})
    = &\Lambda^{\sigma}_{i}(\hat{v}_{i, t}, \bar{v}_{i}; t) - \Lambda^{\sigma}_{i}(v_{i, t}, \bar{v}_{i}; t) - d^{S}_{i,t}(\hat{v}_{i,t}, v_{i, t})\\
    & - \big( \ell(v_{i, t}, \sigma_{t}(v_{i, t}, \bm{v}_{-i, t})) -\ell(v_{i, t}, \sigma_{t}(\hat{v}_{i, t}, \bm{v}_{-i, t}))   \big)\\
    \leq& -\ell(v_{i, t}, \sigma_{t}(v_{i, t}, \bm{v}_{-i, t})) +\ell(v_{i, t}, \sigma_{t}(\hat{v}_{i, t}, \bm{v}_{-i, t})).
    \end{aligned}
\end{equation}
Rearranging (\ref{eq:Appendix_E_2}) gives
\begin{equation}\label{eq:Appendix_E_2_2}
    -\ell(v_{i, t}, \sigma_{t}(v_{i, t}, \bm{v}_{-i, t})) + \theta_{i, t}(v_{i, t},\bm{v}_{-i, t}) \geq -\ell(v_{i, t}, \sigma_{t}(\hat{v}_{i, t}, \bm{v}_{-i, t})) + \theta_{i, t}(\hat{v}_{i, t},\bm{v}_{-i, t}).
\end{equation}

Next, we apply similar procedures to $\beta_{i, t}$.
From the formulation of $\beta_{i, t}$ in (\ref{eq:thm_beta}), we have
\begin{equation}\label{eq:Appendix_E_3}
    \begin{aligned}
    &\beta_{i, t}(\hat{v}_{i, t}, \bm{v}_{-i, t}) - \beta_{i, t}(v_{i, t}, \bm{v}_{-i, t})\\
    =&\sup_{\tau\in\mathbb{T}_{t}} \Lambda^{\sigma}_{i}(\hat{v}_{i, t}, \bar{v}_{i}; \tau)- \sup_{\tau\in\mathbb{T}_{t}} \Lambda^{\sigma}_{i}(v_{i, t}, \bar{v}_{i}; \tau) +\ell(\hat{v}_{i, t}, \sigma_{t}(\hat{v}_{i, t}, \bm{v}_{-i, t})) - \ell(v_{i, t}, \sigma_{t}(v_{i,t}, \bm{v}_{-i, t}))  \\
    &+ \mathbb{E}^{\sigma}\Big[ \sup_{\tau\in\mathbb{T}_{t+1}} \Lambda^{\sigma}_{i}(\tilde{v}_{i, t+1}, \bar{v}_{i}; \tau)\Big| v_{i,t}\Big]  -\mathbb{E}^{\sigma}\Big[ \sup_{\tau\in\mathbb{T}_{t+1}} \Lambda^{\sigma}_{i}(\tilde{v}_{i, t+1}, \bar{v}_{i}; \tau)\Big| \hat{v}_{i,t}\Big].
    \end{aligned}
\end{equation}
We apply the formulations of $\beta_{i, t}$ and $\theta_{i, t}$, respectively, in (\ref{eq:thm_beta}) and (\ref{eq:thm_theta}) to (\ref{eq:Appendix_E_3}) and obtain the following, for any $\tau\in\mathbb{T}_{t}$:
\begin{equation}\label{eq:Appendix_E_4}
    \begin{aligned}
    &\beta_{i, t}(\hat{v}_{i, t}, \bm{v}_{-i, t}) - \beta_{i, t}(v_{i, t}, \bm{v}_{-i, t})
    = \sup_{\tau\in\mathbb{T}_{t}} \Lambda^{\sigma}_{i}(\hat{v}_{i, t}, \bar{v}_{i}; \tau)- \sup_{\tau\in\mathbb{T}_{t}} \Lambda^{\sigma}_{i}(v_{i, t}, \bar{v}_{i}; \tau)\\
    &+ \mathbb{E}^{\sigma}\Big[\sum_{s=t}^{T} -\ell(\tilde{v}_{i, t}, \sigma_{s}(\bm{\tilde{v}}_{s })) + \sum_{s=t+1}^{T-1} \beta_{i_{s}}( \bm{\tilde{v}}_{s}) + \theta_{i_{T}}( \bm{\tilde{v}}_{T})\Big| v_{i,t}\Big]\\
    &-\mathbb{E}^{\sigma}\Big[\sum_{s=t}^{\tau} -\ell(\tilde{v}_{i, t}, \sigma_{s}(\bm{\tilde{v}}_{s })) + \sum_{s=t+1}^{\tau-1} \beta_{i, s}( \bm{\tilde{v}}_{s}) +  \Lambda^{\sigma}_{i}(\tilde{v}_{i,\tau}, \bar{v}_{i};\tau)\Big| \hat{v}_{i,t}\Big].
    \end{aligned}
\end{equation}
Applying the condition (\ref{eq:thm_sigma_C2}) to (\ref{eq:Appendix_E_4}) obtains
\begin{equation}\label{eq:Appendix_E_5}
    \begin{aligned}
    &\beta_{i, t}(\hat{v}_{i, t}, \bm{v}_{-i, t}) - \beta_{i, t}(v_{i, t}, \bm{v}_{-i, t})
    \leq\inf_{\tau\in \mathbb{T}_{t}}\Big\{d^{-S}_{i,t}(\hat{v}_{i, t}, v_{i, t};\tau) \Big\} - \sup_{\tau\in\mathbb{T}_{t}}\rho_{i}(\tau)\\
    &+ \mathbb{E}^{\sigma}\Big[\sum_{s=t}^{T} -\ell(\tilde{v}_{i, t}, \sigma_{s}(\bm{\tilde{v}}_{s })) + \sum_{s=t+1}^{T-1} \beta_{i_{s}}( \bm{\tilde{v}}_{s}) + \theta_{i_{T}}( \bm{\tilde{v}}_{T})\Big| v_{i,t}\Big]\\
    &-\mathbb{E}^{\sigma}\Big[\sum_{s=t}^{\tau} -\ell(\tilde{v}_{i, t}, \sigma_{s}(\bm{\tilde{v}}_{s })) + \sum_{s=t+1}^{\tau-1} \beta_{i, s}( \bm{\tilde{v}}_{s}) + \Lambda^{\sigma}_{i}(\tilde{v}_{i, \tau}, \bar{v}_{i};\tau)\Big| \hat{v}_{i,t}\Big].
    \end{aligned}
\end{equation}

From the definition of $d^{-S}_{i_{t}}$ and $\bar{J}^{\chi_{i}}_{i_{t}}$, respectively, in (\ref{eq:distance_S_2}) and (\ref{eq:J_without_beta}), we have
\begin{equation*}
   \begin{aligned}
    &\inf_{\tau\in \mathbb{T}_{t}}\Big\{d^{-S}_{i, t}(\hat{v}_{i, t}, v_{i, t};\tau) \Big\}-\mathbb{E}^{\sigma}\Big[\sum_{s=t}^{\tau} -\ell(\tilde{v}_{i, t}, \sigma_{s}(\bm{\tilde{v}}_{s })) + \sum_{s=t+1}^{\tau-1} \beta_{i_{s}}( \bm{\tilde{v}}_{s}) + \theta_{i,\tau}( \bm{\tilde{v}}_{\tau})\Big| \hat{v}_{i,t}\Big]\\
=&\inf_{\tau\in \mathbb{T}_{t}}\Big\{ \mathbb{E}_{t}^{\sigma}\Big[\sum_{s=t}^{\tau}-\ell_{i,s}(\tilde{v}_{i,s}, \sigma_{s}(\bm{\tilde{v}}_{s})) +\sum_{s=t+1}^{\tau-1}\beta_{i, s}(\bm{\tilde{v}}_{s}) + \theta_{i, T}(\bm{\tilde{v}}_{T})\Big| \hat{v}_{i,t} \Big] \\
&-\mathbb{E}^{\sigma}_{\hat{v}_{i,t}}\Big[\sum_{s=t}^{\tau}-\ell_{i, s}(\tilde{v}_{i, s}, \sigma_{s}(\bm{\tilde{v}}_{s})) +\sum_{s=t+1}^{\tau-1}\beta_{i, s}(\bm{\tilde{v}}_{s}) + \theta_{i_{T}}(\bm{\tilde{v}}_{T})\Big| v_{i,t} \Big]\\
&-\mathbb{E}^{\sigma}\Big[\sum_{s=t}^{\tau} -\ell(\tilde{v}_{i, t}, \sigma_{s}(\bm{\tilde{v}}_{s })) + \sum_{s=t+1}^{\tau-1} \beta_{i, s}( \bm{\tilde{v}}_{s}) +  \Lambda^{\sigma}_{i}(\tilde{v}_{i, \tau}, \bar{v}_{i};\tau)\Big| \hat{v}_{i,t}\Big]\Big\}\\
\leq& \inf_{\tau\in \mathbb{T}_{t}}\Big\{ -\mathbb{E}_{\hat{v}_{i,t}}^{\sigma}\Big[\sum_{s=t}^{\tau}-\ell_{i, s}(\tilde{v}_{i, s}, \sigma_{s}(\bm{\tilde{v}}_{s})) +\sum_{s=t+1}^{\tau-1}\beta_{i, s}(\bm{\tilde{v}}_{s}) + \theta_{i, T}(\bm{\tilde{v}}_{T})\Big| v_{i,t} \Big] \Big\}.
\end{aligned}
\end{equation*}

Hence, (\ref{eq:Appendix_E_5}) becomes:
\begin{equation*}
    \begin{aligned}
    &\beta_{i, t}(\hat{v}_{i, t}, \bm{v}_{-i, t}) - \beta_{i, t}(v_{i, t}, \bm{v}_{-i, t}) \leq - \sup_{\tau\in\mathbb{T}_{t}}\rho_{i}(\tau)\\
    &+\inf_{\tau\in \mathbb{T}_{t}}\Big\{ -\mathbb{E}_{\hat{v}_{i,t}}^{\sigma}\Big[\sum_{s=t}^{\tau}-\ell_{i, s}(\tilde{v}_{i, s}, \sigma_{s}(\bm{\tilde{v}}_{s})) +\sum_{s=t+1}^{\tau-1}\beta_{i, s}(\bm{\tilde{v}}_{s}) + \theta_{i_{T}}(\bm{\tilde{v}}_{T})\Big| v_{i,t} \Big] \Big\} \\
    &+ \mathbb{E}^{\sigma}\Big[\sum_{s=t}^{T} -\ell(\tilde{v}_{i, t}, \sigma_{s}(\bm{\tilde{v}}_{s })) + \sum_{s=t+1}^{T-1} \beta_{i, s}( \bm{\tilde{v}}_{s}) + \theta_{i, T}( \bm{\tilde{v}}_{T})\Big| v_{i,t}\Big].
    \end{aligned}
\end{equation*}

From the monotonicity of $J^{\Gamma}_{i, t}$ in Lemma \ref{lemma:monotonicity_J_1} and the formulation of $\rho_{i}$ in (\ref{eq:thm_rho}), we have, for any $\tau'\in\mathbb{T}_{t}$,
\begin{equation}\label{eq:Appendix_E_6}
    \begin{aligned}
    &\beta_{i, t}(\hat{v}_{i, t}, \bm{v}_{-i, t}) - \beta_{i, t}(v_{i, t}, \bm{v}_{-i, t}) \leq - \sup_{\tau\in\mathbb{T}_{t}}\rho_{i}(\tau)\\
    &+ \inf_{\tau\in \mathbb{T}_{t}}\Big\{ -\mathbb{E}_{\hat{v}_{i,t}}^{\sigma}\Big[\sum_{s=t}^{\tau}-\ell_{i_{s}}(\tilde{v}_{i, s}, \sigma_{s}(\bm{\tilde{v}}, s)) +\sum_{s=t+1}^{\tau-1}\beta_{i, s}(\bm{\tilde{v}}_{s}) + \theta_{i, T}(\bm{\tilde{v}}_{T})\Big| v_{i,t} \Big] \Big\} \\
    &+ \mathbb{E}^{\sigma}\Big[\sum_{s=t}^{\tau'} -\ell(\tilde{v}_{i, s}, \sigma_{s}(\bm{\tilde{v}}_{s })) + \sum_{s=t+1}^{\tau'-1} \beta_{i_{s}}( \bm{\tilde{v}}_{s}) + \theta_{i, \tau'}( \bm{\tilde{v}}_{\tau'}) + \rho_{i}(\tau')\Big| v_{i,t}\Big]\\
    \leq& \sup_{\tau'\in\mathbb{T}_{t}}\Big\{ \mathbb{E}^{\sigma}\Big[\sum_{s=t}^{\tau'} -\ell(\tilde{v}_{i, s}, \sigma_{s}(\bm{\tilde{v}}_{s })) + \sum_{s=t+1}^{\tau'-1} \beta_{i, s}( \bm{\tilde{v}}_{s}) + \theta_{i, \tau'}( \bm{\tilde{v}}_{\tau'}) + \rho_{i}(\tau')\Big| v_{i,t}\Big]  \Big\}\\
    &+ \inf_{\tau\in \mathbb{T}_{t}}\Big\{ -\mathbb{E}_{\hat{v}_{i,t}}^{\sigma}\Big[\sum_{s=t}^{\tau}-\ell_{i, s}(\tilde{v}_{i, s}, \sigma_{s}(\bm{\tilde{v}}_{s})) +\sum_{s=t+1}^{\tau-1}\beta_{i, s}(\bm{\tilde{v}}_{s}) + \theta_{i_{\tau}}(\bm{\tilde{v}}_{\tau})+\rho_{i}(\tau) \Big| v_{i,t}\Big] \Big\} \\
    =& \sup_{\tau'\in\mathbb{T}_{t}}\Big\{ \mathbb{E}^{\sigma}\Big[\sum_{s=t}^{\tau'} -\ell(\tilde{v}_{i, s}, \sigma_{s}(\bm{\tilde{v}}_{s })) + \sum_{s=t+1}^{\tau'-1} \beta_{i, s}( \bm{\tilde{v}}_{s}) + \theta_{i, \tau'}( \bm{\tilde{v}}_{\tau'}) + \rho_{i}(\tau')\Big| v_{i,t} \Big]  \Big\}\\
    &- \sup_{\tau\in \mathbb{T}_{t}}\Big\{ \mathbb{E}_{\hat{v}_{i,t}}^{\sigma}\Big[\sum_{s=t}^{\tau}-\ell_{i, s}(\tilde{v}_{i, s}, \sigma_{s}(\bm{\tilde{v}}_{s})) +\sum_{s=t+1}^{\tau-1}\beta_{i, s}(\bm{\tilde{v}}_{s}) + \theta_{i, \tau}(\bm{\tilde{v}}_{\tau})+\rho_{i}(\tau) \Big| v_{i,t}\Big] \Big\}.
    \end{aligned}
\end{equation}

Hence, (\ref{eq:Appendix_E_2_2}) and (\ref{eq:Appendix_E_6}) show that the market model with $\beta_{i, t}$, $\theta_{i, t}$, and $\rho_{i}$ constructed in (\ref{eq:thm_beta})-(\ref{eq:thm_rho}), respectively, is DIC.

Next, we prove that the formulation of $\Lambda^{\sigma}_{i}$ in (\ref{eq:integral_envelope}) is valid. Substituting $\beta_{i, t}$, $\theta_{i, t}$, and $\rho_{i}$ constructed in (\ref{eq:thm_beta})-(\ref{eq:thm_rho}), respectively, with $\Lambda^{\sigma}_{i}$ given in (\ref{eq:integral_envelope}), yields:
\begin{equation}\label{eq:Appendix_E_7}
    \begin{aligned}
   J^{\Gamma}_{i, t}(v_{i, t}, h_{t},\tau) =& \Lambda^{\sigma}_{i}(v_{i, t}, v'_{i, t}; \tau)\\
=&\int^{v_{i, t}}_{\bar{v}_{i,t} }
    \mathbb{E}^{\sigma}\Big[  \sum\limits_{s=t}^{\tau}\big(1 - \exp(\sigma_{s}(\tilde{\bm{v}}_{s}))\mathcal{G}^{s}_{t}(\tilde{v}^{s}_{i,t}|\sigma) \big)\Big| x, h_{t}\Big] dx.
\end{aligned}
\end{equation}
From (\ref{eq:lemma_envelope}) of Lemma (\ref{lemma:envelope_necessary_condition}), we can see that (\ref{eq:Appendix_E_7}) satisfy the envelope condition.

\hfill$\square$

\section{Proof of Proposition \ref{prop:necessary_condition} }\label{app:prop:necessary_condition}

We divide the proof into to parts: \textit{(i)} $v_{i, t}\geq \kappa^{l}_{i}(t)$ and \textit{(ii)} $v_{i, t}\leq \kappa^{l}_{i}(t)$.
Let $\bm{v}_{-i, t}\in \bm{V}_{-i,t}$ denote the period-$t$ instrumental preference of owners other than owner $i$, for any $t\in\mathbb{T}$.

\subsection*{ \textit{(i)} $v_{i, t}\geq \kappa^{l}_{i}(t)$}

Let $\hat{v}_{i, t}\geq v_{i, t}\geq \kappa^{l}_{i}(t)$. For the distance $d^{S}_{i, t}(v_{i, t},\hat{v}_{i, t})$ we have
\begin{equation*}
    \begin{aligned}
    d^{S}_{i_{t}}(v_{i, t},\hat{v}_{i, t}) = & -\ell\big(v_{i, t}, \sigma_{t}(v_{i, t},  \bm{v}_{-i, t})\big) + \ell\big(\hat{v}_{i, t}, \sigma_{t}(v_{i, t},  \bm{v}_{-i, t})\big)\\
    =& -\ell\big(v_{i, t}, \sigma_{t}(v_{i, t},  \bm{v}_{-i, t})\big)+ \ell\big(\hat{v}_{i, t}, \sigma_{t}(\hat{v}_{i, t},  \bm{v}_{-i, t})\big)\\ &-\ell\big(\hat{v}_{i, t}, \sigma_{t}(\hat{v}_{i, t},  \bm{v}_{-i, t})\big) + \ell\big(\hat{v}_{i, t}, \sigma_{t}(v_{i, t},  \bm{v}_{-i, t})\big)\\
    =& -\ell\big(v_{i, t}, \sigma_{t}(v_{i, t},  \bm{v}_{-i, t})\big)+ \ell\big(\hat{v}_{i, t}, \sigma_{t}(\hat{v}_{i, t},  \bm{v}_{-i, t})\big)\\
    &-\ell\big(\hat{v}_{i,t}, \sigma_{t}(\hat{v}_{i, t},  \bm{v}_{-i, t})\big)+ \theta_{i, t}(\hat{v}_{i, t},  \bm{v}_{-i, t}) + \ell\big(\hat{v}_{i, t}, \sigma_{t}(v_{i, t},  \bm{v}_{-i, t})\big) \\
    &- \theta_{i_{t}}(v_{i, t},  \bm{v}_{-i, t})\\
    &+ \theta_{i, t}(v_{i, t},  \bm{v}_{-i, t}) - \theta_{i, t}(\hat{v}_{i, t},  \bm{v}_{-i, t})\\
    (\text{DIC } \rightarrow)\geq& -\ell\big(v_{i, t}, \sigma_{t}(v_{i, t},  \bm{v}_{-i, t})\big)+ \ell\big(\hat{v}_{i, t}, \sigma_{t}(\hat{v}_{i, t},  \bm{v}_{-i, t})\big)\\
    &+ \theta_{i, t}(v_{i, t},  \bm{v}_{-i, t}) - \theta_{i, t}(\hat{v}_{i, t},  \bm{v}_{-i, t})\\
    =& \Lambda^{\sigma}_{i}(v_{i, t}, \bar{v}_{i,t};t) - \Lambda^{\sigma}_{i}(\hat{v}_{i, t}, \bar{v}_{i,t};t).
    \end{aligned}
\end{equation*}
Hence, the condition (\ref{eq:necessary_C1}) is satisfied.

\subsection*{\textit{(ii)} $v_{i,t}\leq \kappa^{l}_{i}(t)$}

Define, 
\begin{equation*}
    \begin{aligned}
    Z^{m}_{i}(v_{i, t})\equiv \sup_{\tau\in\mathbb{T}_{t+1}}\Bigg\{ \mathbb{E}^{\sigma}\Big[\sum_{s=t}^{\tau}-\ell_{s}(\tilde{v}_{i, s}, \sigma_{i, s}(\bm{\tilde{v}}_{s})) + \sum_{s=t}^{\tau-1} \beta_{i, s}(\bm{\tilde{v}}_{s})) + \theta(\bm{\tilde{v}}_{\tau})) +\rho_{i}(\tau)  \Big|v_{i,t} \Big] \Bigg\} .
    \end{aligned}
\end{equation*}
From the definition of $\Lambda^{\sigma}_{i}$ in (\ref{eq:integral_envelope}), we have
\begin{equation}\label{eq:Appendix_F_1}
    \begin{aligned}
    &\sup_{\tau\in\mathbb{T}_{t+1}}\Lambda^{\sigma}_{i}(\hat{v}_{i,t}, \bar{v}_{i,t}; \tau) - \sup_{\tau\in\mathbb{T}_{t}}\Lambda^{\sigma}_{i}(v_{i, t}, \bar{v}_{i,t}; \tau)\\
    =& Z^{m}_{i}(\hat{v}_{i, t}) - Z^{m}_{i}(v_{i, t})\\
    =& \sup_{\tau\in\mathbb{T}_{t+1}}\Bigg\{ \mathbb{E}^{\sigma}\Big[\sum_{s=t}^{\tau}-\ell_{s}(\tilde{v}_{i, s}, \sigma_{i, s}(\bm{\tilde{v}}_{s})) + \sum_{s=t}^{\tau-1} \beta_{i, s}(\bm{\tilde{v}}_{s})) + \theta(\bm{\tilde{v}}_{\tau})) +\rho_{i}(\tau) \Big| \hat{v}_{i_{t}}  \Big] \Bigg\} \\
    &- \sup_{\tau\in\mathbb{T}_{t+1}}\Bigg\{ \mathbb{E}^{\sigma }\Big[\sum_{s=t}^{\tau}-\ell_{s}(\tilde{v}_{i, s}, \sigma_{i, s}(\bm{\tilde{v}}_{s})) + \sum_{s=t}^{\tau-1} \beta_{i, s}(\bm{\tilde{v}}_{s})) + \theta(\bm{\tilde{v}}_{\tau})) +\rho_{i}(\tau) \Big| v_{i, t}  \Big] \Bigg\}\\
    =& \sup_{\tau\in\mathbb{T}_{t+1}}\Bigg\{ \mathbb{E}^{\sigma}\Big[\sum_{s=t}^{\tau}-\ell_{s}(\tilde{v}_{i,s}, \sigma_{i,s}(\bm{\tilde{v}}_{s})) + \sum_{s=t+1}^{\tau-1} \beta_{i,s}(\bm{\tilde{v}}_{s})) + \theta(\bm{\tilde{v}}_{\tau})) +\rho_{i}(\tau) \Big| \hat{v}_{i,t}  \Big] \Bigg\} \\
    &- \sup_{\tau\in\mathbb{T}_{t+1}}\Bigg\{ \mathbb{E}^{\sigma}\Big[\sum_{s=t}^{\tau}-\ell_{s}(\tilde{v}_{i,s}, \sigma_{i, s}(\bm{\tilde{v}}_{s})) + \sum_{s=t+1}^{\tau-1} \beta_{i,s}(\bm{\tilde{v}}_{s})) + \theta(\bm{\tilde{v}}_{\tau})) +\rho_{i}(\tau) \Big| v_{i,t}  \Big] \Bigg\}\\
    &+ \beta_{i, t}(\hat{v}_{i, t}, \bm{v}_{-i, t}) - \beta_{i, t}(v_{i, t}, \bm{v}_{-i, t}).
    \end{aligned}
\end{equation}
From the optimality of truthful reporting in DIC market model, we have
\begin{equation*}
    \begin{aligned}
    &\text{RHS of (\ref{eq:Appendix_F_1})} \\
    \leq&  \sup_{\tau\in\mathbb{T}_{t+1}}\Bigg\{ \mathbb{E}^{\sigma}\Big[\sum_{s=t}^{\tau}-\ell_{s}(\tilde{v}_{i, s}, \sigma_{i_{s}}(\bm{\tilde{v}}_{s})) + \sum_{s=t+1}^{\tau-1} \beta_{i, s}(\bm{\tilde{v}}_{s})) + \theta(\bm{\tilde{v}}_{\tau})) +\rho_{i}(\tau)   \Big| \hat{v}_{i,t}\Big] \Bigg\} \\
    &-\sup_{\tau\in\mathbb{T}_{t+1}}\Bigg\{ \mathbb{E}_{\hat{v}_{i,t}}^{\sigma}\Big[\sum_{s=t}^{\tau}-\ell_{s}(\tilde{v}_{i, s}, \sigma_{i, s}(\bm{\tilde{v}}_{s})) + \sum_{s=t+1}^{\tau-1} \beta_{i, s}(\bm{\tilde{v}}_{s})) + \theta(\bm{\tilde{v}}_{\tau})) +\rho_{i}(\tau)  \Big| v_{i,t} \Big] \Bigg\}\\
    =& \sup_{\tau\in\mathbb{T}_{t}}\Big\{\bar{J}^{\Gamma}_{i, t}(\hat{v}_{i,t}, \hat{v}_{i,t},h_{t}, \tau)\Big\} - \sup_{\tau\in\mathbb{T}_{t}}\Big\{\bar{J}^{\Gamma}_{i, t}(v_{i,t}, \hat{v}_{i,t},h_{t}, \tau)\Big\}\\
\leq& \sup_{\tau\in\mathbb{T}_{t}}\Big\{d^{-S}_{i,t}(\hat{v}, v_{i,t}; \tau)\Big\}.
    \end{aligned}
\end{equation*}
Hence, the condition (\ref{eq:necessary_C2}) is satisfied.

\hfill$\square$

\section{Proof of Proposition \ref{prop:delta_DIC} }\label{app:prop:delta_DIC}

Fix $\bm{v}_{-i, t}$ as the instrumental preferences of owners other than owner $i$. Let $v_{i,t}$, $\hat{v}_{i,t}\in V_{i,t}$.
From the formulation of $\beta_{i,t}$ in (\ref{eq:thm_beta}), we have, for any two $\tau', \tau''\in\mathbb{T}_{t}$,
\begin{equation}\label{eq:Appendix_G_1}
    \begin{aligned}
    &\beta_{i, t}(\hat{v}_{i, t}, \bm{v}_{-i, t}) - \beta_{i,t}(v_{i, t}, \bm{v}_{-i, t})
    = \sup_{\tau\in\mathbb{T}_{t}} \Lambda^{\sigma}_{i}(\hat{v}_{i, t}, \bar{v}_{i,t}; \tau)- \sup_{\tau\in\mathbb{T}_{t}} \Lambda^{\sigma}_{i}(v_{i, t}, \bar{v}_{i,t}; \tau)\\
    &+ \mathbb{E}^{\sigma}\Big[\sum_{s=t}^{\tau'} -\ell(\tilde{v}_{i, {t}}, \sigma_{s}(\bm{\tilde{v}}_{s })) + \sum_{s=t+1}^{\tau'-1} \beta_{i,s}( \bm{\tilde{v}}_{s}) + \theta_{i, \tau'}( \bm{\tilde{v}}_{\tau'})\Big| v_{i,t}\Big]\\
    &-\mathbb{E}^{\sigma}\Big[\sum_{s=t}^{\tau''} -\ell(\tilde{v}_{i, t}, \sigma_{s}(\bm{\tilde{v}}_{s })) + \sum_{s=t+1}^{\tau''-1} \beta_{i, s}( \bm{\tilde{v}}_{s}) + \Lambda^{\sigma}_{i}(\tilde{v}_{i, \tau''}, \bar{v}_{i,t};\tau'')\Big| \hat{v}_{i,t}\Big].
    \end{aligned}
\end{equation}
From the definition of $h^{-S}_{i, t}$ in (\ref{eq:H_bar_S}), (\ref{eq:Appendix_G_1}) becomes
\begin{equation}\label{eq:Appendix_G_2}
    \begin{aligned}
    &\beta_{i,t}(\hat{v}_{i, t}, \bm{v}_{-i, t}) - \beta_{i, t}(v_{i, t}, \bm{v}_{-i, t})\\
    \leq & \Xi^{-S}_{i, t} +\sup_{\tau\in\mathbb{T}_{t}}\Big\{\bar{J}^{\Gamma}_{i, t}(\hat{v}_{i,t}, \hat{v}_{i,t},h_{t}, \tau)\Big\} - \sup_{\tau\in\mathbb{T}_{t}}\Big\{\bar{J}^{\Gamma}_{i, t}(v_{i,t}, \hat{v}_{i,t},h_{t}, \tau)\Big\}\\
    &+ \mathbb{E}^{\sigma}\Big[\sum_{s=t}^{\tau'} -\ell(\tilde{v}_{i, t}, \sigma_{s}(\bm{\tilde{v}}_{s })) + \sum_{s=t+1}^{\tau'-1} \beta_{i, s}( \bm{\tilde{v}}_{s}) + \theta_{i,\tau'}( \bm{\tilde{v}}_{\tau'})\Big| v_{i,t}\Big]\\
    &-\mathbb{E}^{\sigma}\Big[\sum_{s=t}^{\tau''} -\ell(\tilde{v}_{i, t}, \sigma_{s}(\bm{\tilde{v}}_{s })) + \sum_{s=t+1}^{\tau''-1} \beta_{i,s}( \bm{\tilde{v}}_{s}) + \Lambda^{\sigma}_{i}(\tilde{v}_{i, \tau''}, \bar{v}_{i};\tau'')\Big| \hat{v}_{i,t}\Big].
    \end{aligned}
\end{equation}
Since $\Lambda^{\sigma}_{i}(v_{i, t}, \bar{v}_{i};\tau)\geq \Lambda^{\sigma}_{i}(v_{i, t}, \bar{v}_{i,t};t)$, for any $v_{i, t}\in V_{i}$, $t\in\mathbb{T}$, $\tau\in\mathbb{T}_{t}$, we have $\Lambda^{\sigma}_{i}(v_{i, t}, \bar{v}_{i};\tau)\geq -\ell_{t}(v_{i, t}, \sigma_{t}(\bm{v}_{t})) + \theta_{i, t}(\bm{v}_{t}) )$.
Then, (\ref{eq:Appendix_G_2}) becomes
\begin{equation}\label{eq:Appendix_G_3}
    \begin{aligned}
    &\beta_{i, t}(\hat{v}_{i, t}, \bm{v}_{-i,t}) - \beta_{i,t}(v_{i,t}, \bm{v}_{-i, t})\leq -\sup_{\tau\in\mathbb{T}_{t}}\Big\{\bar{J}^{\Gamma}_{i, t}(v_{i,t}, \hat{v}_{i,t},h_{t}, \tau)\Big\}+\sup_{\tau\in\mathbb{T}_{t}}\Big\{\rho_{t}(\tau)\Big\}\\
     &+ \Xi^{-S}_{i, t} + \mathbb{E}^{\sigma}\Big[\sum_{s=t}^{\tau'} -\ell(\tilde{v}_{i, t}, \sigma_{s}(\bm{\tilde{v}}_{s })) + \sum_{s=t+1}^{\tau'-1} \beta_{i, s}( \bm{\tilde{v}}_{s}) + \theta_{i, \tau'}( \bm{\tilde{v}}_{\tau'})\Big| v_{i,t}\Big].
    \end{aligned}
\end{equation}
From the formulation of $\rho_{i}$ in (\ref{eq:thm_rho}), we can find the upper bound of (\ref{eq:Appendix_G_3}) as follows:
\begin{equation*}
    \begin{aligned}
    &\beta_{i, t}(\hat{v}_{i, t}, \bm{v}_{-i, t}) - \beta_{i, t}(v_{i, t}, \bm{v}_{-i, t})\\
    \leq& \Xi^{-S}_{i, t} + \sup_{\tau\in\mathbb{T}_{t}}\Big\{\rho_{t}(\tau)\Big\}\\
    &+ \sup_{\tau\in\mathbb{T}_{t}}\Big\{\mathbb{E}^{\sigma}\Big[\sum_{s=t}^{\tau'} -\ell(\tilde{v}_{i, t}, \sigma_{s}(\bm{\tilde{v}}_{s })) + \sum_{s=t+1}^{\tau'-1} \beta_{i, s}( \bm{\tilde{v}}_{s}) + \theta_{i, \tau'}( \bm{\tilde{v}}_{\tau'}) + \rho_{i}(\tau)\Big| v_{i,t}\Big]\Big\}\\
    &\sup_{\tau\in\mathbb{T}_{t}}\Big\{\mathbb{E}_{\hat{v}_{i,t}}^{\sigma}\Big[\sum_{s=t}^{\tau'} -\ell(\tilde{v}_{i, t}, \sigma_{s}(\bm{\tilde{v}}_{s })) + \sum_{s=t+1}^{\tau'-1} \beta_{i, s}( \bm{\tilde{v}}_{s}) + \theta_{i, \tau'}( \bm{\tilde{v}}_{\tau'}) + \rho_{i}(\tau)\Big| v_{i,t}\Big]\Big\},
    \end{aligned}
\end{equation*}
which implies that 
\begin{equation*}
    \sup_{\tau\in\mathbb{T}_{t}}\Big\{\bar{J}^{\Gamma}_{i, t}(v_{i,t}, v_{i,t},h_{t}, \tau)\Big\} + \Xi^{-S}_{i, t} + \sup_{\tau\in\mathbb{T}_{t}}\Big\{\rho_{t}(\tau)\Big\} \geq \sup_{\tau\in\mathbb{T}_{t}}\Big\{\bar{J}^{\Gamma}_{i, t}(v_{i,t}, \hat{v}_{i,t},h_{t}, \tau)\Big\}.
\end{equation*}
Then, it is straightforward to see that the market model is $\Xi^{-S}_{i, t} + \sup_{\tau\in\mathbb{T}_{t}}\Big\{\rho_{t}(\tau)\Big\}$-DIC. Similar procedures can be applied to prove the case when the optimal stopping calls for stopping.

\hfill$\square$

\bibliographystyle{unsrt}  
\bibliography{references}

\end{document}